\documentclass[11pt,a4paper]{article}
\usepackage[margin=28mm]{geometry}
\usepackage[british]{babel}
\usepackage[utf8]{inputenc}
\usepackage[T1]{fontenc}
\usepackage{lmodern,csquotes}
\usepackage{amsmath,amssymb,amsthm,mathtools,braket}
\usepackage{makecell,xcolor,graphicx,float,subcaption,tikz,varwidth}
\usepackage[skip=0.75\baselineskip plus 2pt]{parskip}
\usepackage[nobreak]{cite}
\usepackage[bottom,hang]{footmisc}
\usepackage[normalem]{ulem}

\usepackage{hyperref,bookmark}

\numberwithin{equation}{section}

\definecolor{dark-blue}{rgb}{0.3,0.3,0.7}
\definecolor{dark-red}{rgb}{0.7,0,0}
\definecolor{dark-green}{rgb}{0,0.7,0}
\definecolor{KulLogo1}{RGB}{82, 189, 236}
\definecolor{KulLogo2}{RGB}{0, 64, 122}
\definecolor{KulGrid1}{RGB}{31, 171, 231}
\definecolor{KulGrid2}{RGB}{29, 141, 176}
\definecolor{KulGrid3}{RGB}{17, 110, 138}
\definecolor{linkcolor}{rgb}{0,0,0.8}

\hypersetup{
  hidelinks,
  linktoc       = page,
  colorlinks    = true,
  allcolors     = linkcolor,
  bookmarksopen = false,
}

\newcommand{\+}{{\mkern2mu}}
\newcommand*{\dd}{\mathop{}\! \mathrm{d}}
\newcommand*{\DD}{\mathop{}\! \mathcal{D}}
\newcommand*{\ii}{i}
\newcommand*{\vol}{\mathrm{vol}}
\newcommand*{\Vol}{\operatorname{Vol}}
\newcommand*{\Diff}{\mathrm{Diff}}
\newcommand{\abs}[1]{\mathopen{|}#1\mathclose{|}}
\renewcommand*{\Re}{\operatorname{Re}}

\newcommand*{\cL}{c_\mathrm{L}}
\newcommand*{\tvol}{\mathrm{v\tilde{o}l}}
\newcommand{\hypgeom}{{\+{}_2F_1}}
\newcommand{\scheme}{B}
\newcommand{\widesim}[2][1.5]{
  \mathrel{\overset{#2}{\scalebox{#1}[1]{$\sim$}}}
}

\makeatletter
  \def\dlmfeqlink#1.#2.#3\@nil{(\href{https://dlmf.nist.gov/#1.#2.E#3}{#1.#2.#3})}
  \newcommand{\dlmfeqcite}[1]{\cite[\dlmfeqlink#1\@nil]{NIST:DLMF}}
\makeatother

\makeatletter
\newcommand{\footnoteref}[1]{\protected@xdef\@thefnmark{\ref{#1}}\@footnotemark}
\makeatother

\newcommand*{\textdash}{\unskip\nobreak\space\textemdash\nobreak\space\ignorespaces}

\begingroup
\global\mathcode`/=\numexpr "4000 + \count2\relax
\endgroup

\newcommand{\figurewidth}{240pt}
\graphicspath{{./figures/}}

\title{{\huge Quantum Liouville Cosmology}\vspace*{1em}}

\author{%
  Dionysios Anninos\textsuperscript{1,2},
  Thomas Hertog\textsuperscript{2,3} and
  Joel Karlsson\textsuperscript{2,3}
  \vspace{0.5em}%
}

\date{%
  \textsuperscript{1}{\small\slshape Department of Mathematics, King's College London, Strand, London WC2R 2LS, UK} \\
  \textsuperscript{2}{\small\slshape Institute for Theoretical Physics, KU Leuven, 3001 Leuven, Belgium} \\
  \textsuperscript{3}{\small\slshape Leuven Gravity Institute, KU Leuven, 3001 Leuven, Belgium}%
  \vspace*{3em}
}

\begin{document}
\maketitle
\begin{abstract}
  \noindent
  We provide a detailed analysis of the disk path integral of timelike Liouville theory, conceived as a tractable and precise toy-model quantum cosmology in two dimensions.
  Disk path integrals with the insertion of matter field operators, taken along a judiciously chosen complex contour, yield states akin to the Hartle--Hawking wavefunction.
  Working in the fixed $K$-representation, where $K$ is the trace of the extrinsic curvature, we compute the one-loop wavefunctions and put forward a conjecture for the all-loop expressions.
  A suitable pairing of Liouville disk path integrals yields a $K$-independent quantity that may form the basis for a well-defined inner product on the space of Euclidean histories.
  We also consider other ensembles, including one with fixed area, and provide a static patch perspective with a timelike feature.
\end{abstract}
\thispagestyle{empty}
\clearpage
\tableofcontents
\pagenumbering{arabic}

\section{Introduction}

Ample evidence indicates that cosmological theory must ultimately be based on quantum gravity.
Indeed, the quantum dynamics of the early universe is the arena par excellence where our theories of quantum gravity are put to the test.
Current models of quantum cosmology, however, remain on somewhat shaky grounds, both conceptually and computationally.
This in turn hinders one's ability to derive precise predictions for cosmological observables, let alone verify them.
The measure problem in inflation and, relatedly, the quantum physical modelling of the observer as part of the theory, provide a case in point.

Hence, there is a need for tractable toy models of quantum cosmology that are on solid theoretical footing and from which certain core principles could be gleaned or sharpened.
Those principles concern first and foremost whether and how a quantum mechanical framework is applicable to cosmology \cite{DeWitt:1967yk,DeWitt:1967ub}.
This matters both in the ultraviolet, where a big bang singularity may wipe out the very notion of classical spacetime, and in the infrared, where continued accelerated expansion can give rise to puzzling if not outright paradoxical phenomena \cite{Guth:2011ie}.
Further, if the universe is a quantum system, it must have a state, which may or may not reside in a Hilbert space, and which enters at least at some level in any detailed predictions \cite{Hartle:2010dq,Maldacena:2024uhs}.

In this paper, we consider a precise toy model that, we aim to demonstrate, elucidates some of the theoretical underpinnings of quantum cosmology.
To make progress, we forgo \textdash for now \textdash phenomenological constraints.
On the other hand, inspired by earlier work \cite{CarneirodaCunha:2003mxy,Bautista:2019jau,Anninos:2021ene,Anninos:2024iwf,Muhlmann:2022duj}, we retain a number of realistic features.
These include the presence of a large number of locally propagating degrees of freedom, a tunable parameter that allows us to take a semiclassical limit where quantum fluctuations of the geometry are small, the existence of classical big bang/big crunch cosmologies (as well as bouncing solutions), an analogue of the conformal mode problem in Euclidean signature, a systematically calculable Euclidean gravitational path integral, and a Wheeler--DeWitt equation with a wrong sign kinetic term that supports a large number of solutions.
The price we pay is that we work in two spacetime dimensions.
That being said, we take the perspective that the richness of the above list compensates for this.

Given this considerable list of requirements, one might wonder whether such a model even exists.
Fortunately, it turns out there is one that is rather straightforward to construct and describe.
The model we consider is given by quantum mechanically coupling a unitary two-dimensional conformal field theory, with a discrete spectrum of states, to a fluctuating metric field governed by an action that contains a cosmological term with $\Lambda > 0$.
We take the central charge $c$ of the conformal field theory \textdash which is a measure of local degrees of freedom \textdash to be positive and large, but finite.
The semiclassical limit of the theory corresponds to a large-$c$ expansion, whereby the small parameter $1/c$ suppresses the size of gravitational quantum fluctuations.

We work in conformal gauge \cite{David:1988hj,Distler:1988jt}, in which the physical metric takes the form $g_{\mu\nu} = e^{2 \varphi} \tilde{g}_{\mu\nu}$, with the reference metric $\tilde{g}_{\mu\nu}$ some fixed metric in an open neighbourhood of a point.
Upon fixing the conformal gauge and further integrating out the conformal matter, the problem takes the form of a specific Liouville theory known as the timelike Liouville conformal field theory.
Though Liouville theory is often employed in the context of the noncritical string worldsheet, here we view it as part of a quantum gravitational theory in its own right \cite{Polchinski:1989fn}.
What makes the Liouville theory timelike is the fact that the kinetic term governing the Liouville field $\varphi$ has the \enquote{wrong} sign, directly mimicking the unbounded nature of the conformal mode in Euclidean gravity \cite{Gibbons:1978ac}.
Though nonunitary when viewed as a theory in and of itself, timelike Liouville theory can be analysed with standard conformal bootstrap methods \cite{Ribault:2015sxa,Muhlmann:2025ngz,Roussillon:2025tmv}.
Despite the \enquote{wrong} sign, recent mathematical work \cite{Chatterjee:2025yzo,Usciati:2025cdn} has made progress toward a rigorous construction of the timelike Liouville path integral.
A crucial element in this construction is the complexification of the path integral contour for the timelike Liouville field \cite{Anninos:2021ene,Anninos:2024iwf,Usciati:2025cdn}.

Our main purpose is to analyse the quantum mechanical structure of disk one-point functions of timelike Liouville theory.
We take the perspective that this disk path integral produces a solution to the Wheeler--DeWitt equation of the underlying gravitational theory.
The disk partition function of spacelike Liouville theory was computed at one loop in a particular limit in \cite{Mahajan:2021nsd}, and more generally in \cite{Chaudhuri:2024yau}.
Here, we focus on the disk path integral of timelike Liouville theory \cite{Bautista:2021ogd,Anninos:2024iwf,Usciati:2025cdn} and mainly consider a disk decorated, in a gauge invariant manner, with the insertion of matter field operators.
We find that by carefully choosing the complexified contour, such disk path integrals produce states reminiscent of the Hartle--Hawking no-boundary wavefunction \cite{Hartle:1983ai}.
In particular, we recover the elegant condition that the wavefunction is well-behaved and decaying in the small spatial-volume limit.
In our toy model, however, the states are definable and calculable at all orders in the quantum loop expansion.
Exploiting this, we analyse the disk integral with a range of methods and choices of boundary conditions \cite{Fateev:2000ik,Zamolodchikov:2001ah_,Teschner:2000md}, to reveal its detailed structure in various forms.

The paper is structured as follows.
In §\ref{sec:gravity_and_Liouville}, we review how timelike Liouville theory arises from two-dimensional quantum gravity coupled to conformal matter.
We also introduce the wavefunctions we study as disk one-point functions and review \cite{Anninos:2024iwf,Bautista:2021ogd,Fateev:2000ik,Teschner:2000md} to put forward a conjecture for the exact, all-loop wavefunctions.
In §\ref{sec:PI}, we formulate the path integrals computing the wavefunctions and analyse their semiclassical saddle points.
We reinterpret the FZZT \cite{Fateev:2000ik,Teschner:2000md} boundary conditions geometrically as fixing the trace of the extrinsic curvature of the boundary ($K$).
In addition, we demonstrate how the conformal structure of the Liouville one-point functions arises.
This is known to stem from the integral over the moduli space of saddles when the insertion is light and does not backreact on the geometry \cite{Fateev:2000ik}, but we elaborate also on the case of a heavy insertion, which does backreact and has no moduli space.
One-loop quantum fluctuations of the disk are discussed in §\ref{sec:fluct}.
Due to the boundary conditions, which we treat with care throughout, the Liouville field fluctuates both in the bulk and on the boundary of the disk.
Moreover, we analyse the number of negative and zero modes, which turns out to be subtler in timelike Liouville theory as compared to the spacelike case.
In §\ref{sec:ensembles}, we assemble the pieces discussed in the previous sections into the one-loop wavefunction and compare with the expectation from §\ref{sec:gravity_and_Liouville}.
We analyse various ensembles, fixing the trace of the extrinsic curvature or the boundary length, the latter also at fixed area, and discuss their relations, how the overall phase differs between them and their relative advantages and uses.
Lastly, in §\ref{sec:outlook}, we comment on a remarkable property of the wavefunction in the $K$-representation that may further clarify the cosmological Hilbert space and provide an outlook on how this might generalize to higher dimensions.
We also comment on a thermal perspective, in contrast to the wavefunction point of view we employ throughout most of the paper.

\section{Gravity as a Liouville theory} \label{sec:gravity_and_Liouville}
We study Euclidean two-dimensional gravity with a positive cosmological constant~$\Lambda$ coupled to a matter conformal field theory (CFT).
On a fixed manifold $M$ at fixed genus, the partition function reads
\begin{equation} \label{eq:start}
  Z^{M}_\textnormal{grav} = \int_{}\ \frac{\DD g}{\Vol(\Diff)}\+ e^{\vartheta \chi_M - \Lambda A} Z_\mathrm{CFT}[g]\+,
\end{equation}
where $\chi_M$ is the Euler characteristic of $M$, and $A \equiv \int_{D}\! \vol$ the area of the spacetime metric $g_{\mu\nu}$.
The real parameter $\vartheta$ is a coupling constant that controls the contributions of higher-genus topologies.
We consider $\Lambda > 0$ and large $\vartheta$, and take the matter CFT to be unitary with a discrete spectrum and a normalizable vacuum state.
Taking $\Lambda>0$, in addition to being of physical interest, has the effect of suppressing large area configurations.
Note that, although contributions from higher-genus topologies are formally suppressed at large $\vartheta$, they exhibit divergences related to the integral over radii of the contractible cycles \cite{Anninos:2022ujl}.
(The situation might be improved by introducing supersymmetry \cite{Anninos:2023exn,Muhlmann:2025ngz} or other deformations \cite{Allameh:2025gsa}.)
Here, as a first step, we consider fixed topology, focusing on the disk.

In this section, we review how Liouville theory arises by gauge fixing the above path integral \cite{Polyakov:1981rd} (see also e.g.\ \cite{Ginsparg:1993is_}).
Taking a unitary matter CFT, one is led to consider timelike Liouville theory \cite{Polchinski:1989fn}.
As in spacelike Liouville theory \cite{Chaudhuri:2024yau}, Dirichlet boundary conditions are inconsistent with conformal symmetry but it is possible to fix the physical length of the boundary.
Alternatively, one can introduce a boundary cosmological constant as a chemical potential for the boundary length \cite{Fateev:2000ik,Bautista:2021ogd}.
This differs from the treatment of semiclassical timelike Liouville theory on the disk in \cite{Anninos:2024iwf}.
We elaborate further on these points in §\ref{sec:PI} and §\ref{sec:ensembles}.
Here, we also review exact results in spacelike Liouville theory on the disk \cite{Fateev:2000ik,Teschner:2000md} as well as previous analyses of the timelike theory in the same setting \cite{Anninos:2024iwf,Bautista:2021ogd}, on the basis of which we put forward a conjecture for the semiclassical all-loop one-point function in timelike Liouville theory, which we will verify at one loop in subsequent sections.

\subsection{The Weyl gauge and timelike Liouville theory}
The path integral can be gauge fixed to the Weyl gauge by inserting
\begin{equation}
  1 = \Delta_\mathrm{FP}[g] \int_{}\ \DD f \DD\varphi\, \delta(g - e^{2\varphi_f} \tilde{g}_{f})\+,
\end{equation}
into \eqref{eq:start}, where $\Delta_\mathrm{FP}$ is the associated Faddeev--Popov determinant.
Here, there would be an additional integral over modular parameters for higher-genus topologies.
Above, we have chosen a fiducial background metric $\tilde{g}$, acted on by a Weyl transformation parametrized by $\varphi$ and a diffeomorphism $f$.
With this insertion, the integral over physical metrics $g$ is readily performed due to the delta functional.
Since $\Delta_\mathrm{FP}$ and all quantities in \eqref{eq:start} are diffeomorphism invariant, the integral over $f$ cancels against $\Vol(\Diff)$.
The Faddeev--Popov determinant is the same as in bosonic string theory \cite{Polchinski:1998rq} and hence given by the standard $bc$ ghost CFT of central charge $-26$, $\Delta_\mathrm{FP}[g] = Z_{bc}[g]$.
Upon this gauge fixing, and recalling that the physical metric is $g = e^{2\varphi} \tilde{g}$, there is a redundancy under Weyl transformations
\begin{equation} \label{eq:Weyl_transformation}
  \tilde{g} \mapsto e^{2\omega} \tilde{g}\+,\qquad\quad
  \varphi \mapsto \varphi - \omega\+,
\end{equation}
with $\omega$ some smooth function.
The observables of the gravitational theory have to respect this, since they should not depend on the chosen gauge.

Although there is no kinetic term for the metric in \eqref{eq:start}, dynamics for the conformal mode $\varphi$ is generated by the conformal anomaly \cite{Polyakov:1981rd}
\begin{equation} \label{eq:conf_anomaly}
  Z_\mathrm{CFT}[e^{2\varphi} \tilde{g}] = e^{\frac{c}{24\pi} S_\textnormal{anom}[\varphi; \tilde{g}]} Z_\mathrm{CFT}[\tilde{g}]\+,\qquad
  S_\textnormal{anom} = \int_{\mathrlap{M}}\ \tvol\+ \bigl(\tilde{g}^{\mu\nu} \partial_\mu \varphi\+ \partial_\nu \varphi + \tilde{R} \varphi\bigr) + 2 \int_{\mathrlap{\partial M}}\ \tvol\+ \tilde{K} \varphi\+,
\end{equation}
where $\tilde{R}$ is the Ricci scalar and $\tilde{K}$ the trace of the extrinsic curvature of the fiducial background.
Putting the above together, the theory, on fixed topology, factorizes into the ghost CFT, the matter CFT and Liouville theory
\begin{equation} \label{eq:geometric_Liouville}
  Z_\textnormal{grav}^M = e^{\vartheta \chi_M} Z_{bc}[\tilde{g}] Z_\mathrm{CFT}[\tilde{g}] Z_\mathrm{L}[\tilde{g}]\+,\qquad
  Z_\mathrm{L} = \int_{}\ \DD \varphi\+ e^{\frac{c-26}{24\pi} S_\textnormal{anom}[\varphi; \tilde{g}] - \Lambda A[e^{2\varphi} \tilde{g}]}\+.
\end{equation}
In this path integral, only the physical length of the boundary, $\ell \equiv \int_{\partial M}\! \vol = \int_{\partial M}\! \tvol\+ e^\varphi$, is fixed.
We discuss the boundary conditions in more detail below.
Since $Z_\textnormal{grav}^M$ is independent of the choice of fiducial background $\tilde{g}$, it follows that Liouville theory is itself a CFT with central charge $\cL$ such that
\begin{equation}
  \cL + c - 26 = 0\+.
\end{equation}

The path integral measure, $\DD\varphi$, for the Liouville field is background independent, diffeomorphism invariant and ultralocal.
As such, it can be taken as proportional to the volume form associated to
\begin{equation} \label{eq:geometric_measure}
  \| \delta\varphi \|_g^2 = \int_{\mathrlap{M}}\ \vol\+ (\delta \varphi)^2 = \int_{\mathrlap{M}}\ \tvol\+ e^{2 \varphi} (\delta \varphi)^2\+.
\end{equation}
The relation between this field space metric and the continuum limit of a matrix model is discussed in \cite{Ferrari:2011we}.
The above metric is not flat; it depends on the field value $\varphi$ when written in terms of the fiducial background $\tilde{g}$ on which the Liouville theory is defined.
This leads to complications, e.g.\ in the semiclassical expansion beyond the one-loop order, but also has its virtues, as we will see.
For instance, at the formal level it follows immediately that Liouville theory satisfies \eqref{eq:conf_anomaly} with central charge $\cL = 26 - c$ from the above measure and
\begin{equation} \label{eq:anom_consistency}
  S_\textnormal{anom}[\varphi; e^{2\omega}\tilde{g}] + S_\textnormal{anom}[\omega; \tilde{g}] = S_\textnormal{anom}[\varphi + \omega; \tilde{g}]\+.
\end{equation}
The latter is a consistency relation that follows from \eqref{eq:conf_anomaly} and can be verified using the explicit form of the anomaly action.

The above shows that, in contrast to many familiar CFTs, the classical action of Liouville theory is not Weyl invariant but contributes to the conformal anomaly.
Moreover, in the above formulation, the conformal anomaly is completely captured by the classical action with no contribution from the measure.
Note also that, although conformally invariant,%
\footnote{The Liouville action is diffeomorphism invariant and the anomaly action vanishes for Weyl transformations such that $e^{2\omega} \tilde{g}$ is related to $\tilde{g}$ by a diffeomorphism.}
the action contains a dimensionful parameter $\Lambda$.
This is possible due to the affine transformation \eqref{eq:Weyl_transformation} of $\varphi$ under Weyl transformations.

Liouville theory admits two semiclassical limits $\cL \to \pm\infty$ with $\Lambda/\cL$ kept finite, in which quantum fluctuations are suppressed.
Here, we restrict to real values of the central charge, although allowing for complex values can also be useful \cite{Harlow:2011ny,Collier:2024kmo,Blommaert:2025eps}.
From \eqref{eq:geometric_Liouville}, we see that the sign of the kinetic term of $\varphi$ differs between the two limits.
When $\cL \to + \infty$, it has the correct sign for the path integral over real field configurations to converge.
This is the semiclassical limit of \emph{spacelike} Liouville theory.
When $\cL \to - \infty$, the kinetic term has the opposite sign and the integration cycle has to be complex for the path integral to converge \cite{Gibbons:1978ac,Anninos:2021ene,Usciati:2025cdn}.
Since $\varphi$ is the conformal mode of the physical metric, this is analogous to the conformal mode problem of Einstein gravity.
This flavour of the theory is known as \emph{timelike} Liouville theory.
Although timelike Liouville theory is a nonunitary CFT, this does not a priori imply that the gravity theory is nonunitary.
Indeed, the Weyl redundancy \eqref{eq:Weyl_transformation} only appears after fixing the Weyl gauge and is not present in \eqref{eq:start}.

Spacelike and timelike Liouville theory, though clearly related, are different theories in the sense that the path integration cycles are different and, hence, observables cannot directly be analytically continued between the two \cite{Harlow:2011ny}.
With unitary matter, only the timelike semiclassical limit can be realized in gravity.
Relatedly, timelike Liouville theory is more similar to Einstein gravity in that the semiclassical geometries have positive curvature for $\Lambda > 0$ \cite{Polchinski:1989fn}.

A longstanding conjecture \cite{David:1988hj,Distler:1988jt} posits that Liouville theory can, alternatively, be formulated with a flat but background-dependent measure $\tilde{\DD}\varphi$, associated to the field-space metric without the exponential factor $e^{2\varphi}$ on the right-hand side of \eqref{eq:geometric_measure}, as
\begin{equation} \label{eq:canonical_Liouville}
  Z_\mathrm{tL} = \int_{}\ \tilde{\DD} \varphi\+ e^{-S_\mathrm{tL}[\phi; \tilde{g}]}\+,\quad\
  S_\mathrm{tL} = \frac{1}{4\pi} \int_{\mathrlap{M}}\ \tvol\+ \bigl(-\tilde{g}^{\mu\nu} \partial_\mu \phi\+ \partial_\nu \phi - q \tilde{R} \phi + 4\pi \Lambda e^{2\beta\phi}\bigr) - \frac{1}{2\pi} \int_{\mathrlap{\partial M}}\ \tvol\+ q \tilde{K} \phi\+.
\end{equation}
Here, we have focused on the timelike flavour of the theory.
The action for spacelike Liouville theory is recovered by $\phi \to \pm\ii \phi$, $\beta \to \mp\ii b$ and $q \to \pm\ii Q$.
In \eqref{eq:canonical_Liouville}, the Liouville field has been rescaled $\varphi = \beta\phi$ to normalize the kinetic term.
The parameters $q$ and $\beta$ are fixed by requiring conformal invariance and that the central charge is $\cL = 26 - c$, as above.
This fixes, see e.g.\ \cite{Ginsparg:1993is_},
\begin{equation}
  q = \beta^{-1} - \beta\+,\qquad
  \cL = 1 - 6 q^2.
\end{equation}
In this formulation, the conformal anomaly receives contributions both from the action and the measure.
Note that, for real $\beta$, $q$ can take any real value and $\cL \leq 1$.
In the spacelike theory, $\cL = 1 + 6 Q^2$, $Q = b^{-1} + b \geq 2$ for real $b$ and, hence, $\cL \geq 25$.

Both formulations above are useful and we will distinguish them by using $\varphi$ in the first and $\phi$ in the second, related by $\varphi = \beta \phi$.
Similarly, abusing notation, the vertex operator $e^{2a\varphi} = e^{2\alpha\phi}$ will be denoted by either $V_a$ or $V_\alpha$ depending on the context.

\subsection{Wavefunctions of the Liouville universe} \label{sec:Liouville_wfn}
Having introduced the model, we turn to the object of interest: the wavefunction of the universe.
Adopting the proposal by Hartle and Hawking \cite{Hartle:1983ai}, we are prompted to consider a gravitational path integral over metrics on a manifold with a single $S^1$ boundary.
We may generalize slightly by exciting the matter sector through an insertion.
More concretely, after fixing the Weyl gauge, we are led to consider the following class of wavefunctions:
\begin{equation} \label{eq:wfu}
  \Psi^\textnormal{grav}_O = \biggl\langle\int_{\mathrlap{D}}\ \tvol\+ V_\alpha O\biggr\rangle _{\! \textnormal{grav}}\+,
\end{equation}
since we fix the topology to that of a disk, $D$.
Here, the expectation value is evaluated in the full gravity theory, $O$ is a scalar primary of the matter CFT and $V_\alpha = e^{2\alpha\phi}$ a Liouville vertex operator such that
\begin{equation} \label{eq:Delta_consistency}
  \Delta_\alpha + \Delta_O = 1\+,
\end{equation}
to ensure Weyl invariance.%
\footnote{Under a Weyl transformation $\tilde{g} \mapsto e^{2\omega} \tilde{g}$, a primary transforms as $O \mapsto e^{-2 \Delta_O \omega} O$, following the notation of e.g.\ \cite{Fateev:2000ik}.
  Hence, $\Delta$ denotes half the scaling dimension.}

This integrated one-point function should be interpreted as the wavefunction of the state created by the insertion in the basis provided by the boundary conditions.
Such an interpretation is customary for so-called \enquote{physical states} with $\alpha = Q/2 + \ii P$ for real $P$ in spacelike Liouville theory \cite{Fateev:2000ik}.
The physical states of timelike Liouville theory have $\alpha = -q/2 + P$ with $P \in \mathbb{R}$, such that $\Delta_\alpha = -q^2/4 + P^2$ \cite{Ribault:2015sxa}.
In the gravity theory, on the other hand, we are led to consider
\begin{equation} \label{eq:alpha_from_O}
  \alpha = -\frac{q}{2} \pm \sqrt{\frac{q^2}{4} + 1 - \Delta_O}\+,
\end{equation}
since $\Delta_\alpha = \alpha (q+\alpha)$ \cite{Ribault:2015sxa,Bautista:2019jau}.
For $\Delta_O \leq q^2/4 + 1$, this corresponds to a physical state with real $\alpha \in (-\beta^{-1}, \beta)$ since the matter CFT is unitary ($\Delta_O \geq 0$).
However, the gravity theory also allows for $\alpha = -q/2 + \ii p$, with $p \in \mathbb{R}$, when $\Delta_O > q^2/4 + 1$.

To be interpreted as a wavefunction, the one-point function in \eqref{eq:wfu} should not be normalized.
That is, it is given simply by a path integral with an insertion, \emph{not} divided by the zero-point function $Z_\textnormal{grav}$.
In Liouville theory, this is typically how one-point functions are defined but in the matter CFT sector, one should multiply the normalized one-point function by the disk partition function.

For the theory to factorize as in \eqref{eq:geometric_Liouville}, it is important that the boundary conditions of the matter CFT are conformal.
Hence, conformal boundary conditions should be used for the Liouville sector as well.
As will be explained in §\ref{sec:conf}, this does not allow for Dirichlet boundary conditions for $\varphi$.
Instead, one can fix the physical length of the boundary, $\ell = \int_{\partial D}\! \tvol\+ e^\varphi$.
This respects conformal invariance since the physical metric $g = e^{2\varphi} \tilde{g}$ is Weyl invariant.
Alternatively, one can introduce a boundary cosmological constant $\Lambda_\mathrm{b}$, acting as a chemical potential for $\ell$ \cite{Fateev:2000ik}.
We will see below that, similarly to how $\Lambda$ (the chemical potential for area) fixes the curvature, $\Lambda_\mathrm{b}$ fixes the extrinsic curvature of the boundary of the semiclassical geometry.

With conformal boundary conditions, the physical Hilbert space, and the wavefunction, factorizes into the matter CFT and Liouville sectors.
We will focus on the Liouville sector but first comment on the integral in \eqref{eq:wfu}.
Choosing a flat fiducial background of radius $\tilde{L}$ and using a complex coordinate with $\abs{z} \leq 1$, we have
\begin{equation} \label{eq:fiducial_bkg_complex_coord}
  \dd \tilde{s}^2 = \tilde{L}^2 \dd \bar{z} \dd z\+,\qquad\quad
  \langle O(z) \rangle \propto (1 - \bar{z} z)^{-2\Delta_O}\+.
\end{equation}
Hence, the integral in \eqref{eq:wfu} is proportional to
\begin{equation} \label{eq:wfu_integral}
  \int \frac{\dd^2 z}{(1 - \bar{z} z)^2} \propto \Vol(\mathbb{H}^2)\+,
\end{equation}
which diverges due to the contribution close to the boundary.
However, there are zero modes in the ghost sector associated with the conformal group of the background, $\operatorname{Conf}(D) = \mathrm{PSL}(2,\mathbb{R})$, the volume of which should, therefore, be divided out.
As we recall in §\ref{sec:conf}, the two noncompact directions of $\mathrm{PSL}(2,\mathbb{R})$ also span $\mathbb{H}^2$, so the divergences cancel.
Any remaining finite factor that can arise from differently normalized measures depends only on $\tilde{L}$ and the UV scale needed to define ultralocal path integral measures.
This dependence is captured by the $\tilde{L}^2$ from $\tvol$ and the conformal anomaly of the ghost sector.
We will not be concerned with the remaining finite normalization.

The wavefunction of the full theory takes the form
\begin{equation}
  \ket{\Psi^\textnormal{grav}_O} = \ket{\Psi_\alpha} \otimes \ket{O} \otimes \ket{\textnormal{ghost}}\+,
\end{equation}
where we recall that $\alpha$ should be chosen as to satisfy \eqref{eq:Delta_consistency}.
Going forward, we focus on the Liouville sector and the associated wavefunction $\ket{\Psi_\alpha}$.
If we fix the physical length of the boundary,
\begin{equation} \label{eq:one-pt_wfn}
  \langle V_\alpha(z_0) \rangle_\ell = \frac{\Psi_\alpha(\ell)}{(1 - \bar{z}_0 z_0)^{2 \Delta_\alpha}}\+,
\end{equation}
where the expectation value now only refers to the Liouville sector and $z_0$ is a point in the interior of the disk.
This completely captures the dependence of $\Psi^\textnormal{grav}_O(\ell)$ on $\ell$ and $\Lambda$.
Even though the dependence of the one-point function on $z_0$ is fixed by conformal invariance, we will use it as a consistency check of the semiclassical computation.

The wavefunctions $\Psi_\alpha(\ell)$ of timelike Liouville theory were studied in \cite{Anninos:2024iwf}.
Motivated by the functional Wheeler--DeWitt (WDW) equation, upon gauge fixing the semiclassical field $\varphi(\theta) = \varphi_0$, a gauge-fixed WDW equation,
\begin{equation} \label{eq:WDW}
  \biggl[\frac{\beta^2}{2} \biggl(\ell \frac{\dd}{\dd \ell}\biggr)^2 + \frac{\Lambda}{2\pi} \ell^2 - \frac{1}{2} (q+2\alpha)^2\biggr] \Psi_\alpha = 0
  \quad\implies\quad
  \Psi_\alpha^\pm = J_{\pm(q+2\alpha)/\beta}\Bigl(\sqrt{\tfrac{\Lambda}{\pi \beta^2}}\+ \ell\Bigr)\+,
\end{equation}
is obtained, here expressed in terms of the boundary length $\ell = 2\pi e^{\varphi_0}$.
Although not derived rigorously, it is expected that this WDW equation is correct in the semiclassical limit $\beta \to 0$ since this suppresses quantum fluctuations of the conformal mode $\varphi = \beta \phi$.

The sign of the first term is the Lorentzian version of the conformal mode problem and is responsible for the oscillatory behaviour of the wavefunction at large $\ell$.
When $(q+2\alpha)/\beta$ is an integer, the two solutions above are linearly dependent and one can use $Y_{(q+2\alpha)/\beta}$ as the second solution.
It was stressed in \cite{Anninos:2024iwf} that with mildly excited matter fields, such that $\alpha$ in \eqref{eq:alpha_from_O} is real, one of the wavefunctions behaves like the Hartle--Hawking wavefunction, decaying at small $\ell$.
When the matter fields are highly excited, such that $\alpha$ gets an imaginary part, $q+2\alpha$ is purely imaginary and the wavefunctions oscillate more and more rapidly with finite amplitude as $\ell \to 0$.
Relatedly, the classical solutions bounce at a finite $\ell$, much like de Sitter space, in the former case but have a big bang singularity, and thus arbitrarily small values of $\ell$, in the latter \cite{Anninos:2024iwf}.

The two linearly independent solutions are related by $\alpha \leftrightarrow -q-\alpha$.
It is no coincidence that $\Delta_\alpha = \alpha (q+\alpha) = \Delta_{-q-\alpha}$.
This raises the question whether the two wavefunctions are associated with different vertex operators $V_\alpha$ and $V_{-q-\alpha}$.
In spacelike Liouville, the corresponding vertex operators $V_\alpha$ and $V_{Q-\alpha}$ are identified up to a normalization captured by a reflection coefficient \cite{Fateev:2000ik}.
In this case, the solutions to the equation analogous to \eqref{eq:WDW} are modified Bessel functions and the one that grows exponentially at large $\ell$ is discarded.
The decaying modified Bessel function satisfies $K_{(Q-2\alpha)/b} = K_{-(Q-2\alpha)/b}$, consistent with the identification of the vertex operators.

In the timelike case the question is more subtle.
A proposal for the timelike DOZZ formula implies that $V_\alpha$ and $V_{-q-\alpha}$ are similarly identified \cite{Zamolodchikov:2005fy,Kostov:2005kk,Kostov:2005av,Kostov:2006zp}.
Assuming this, the one-point function of timelike Liouville was bootstrapped in \cite{Bautista:2021ogd}, resulting in a particular linear combination%
\footnote{The integral transform that changes the boundary conditions from fixed $\ell$ to fixed $\Lambda_\mathrm{b}$ converges on $\mathbb{R}_+$ only for the Bessel function $J_{\nu}$ with $\Re(\nu) > 0$.
  To compare with \cite{Bautista:2021ogd}, we have continued the result to $\Re(\nu) < 0$.}
\begin{equation} \label{eq:Psi_linear_combination}
  \Psi_\alpha = \ii^{(q + 2\alpha)/\beta} J_{(q + 2\alpha)/\beta}\biggl(\sqrt{\tfrac{\Lambda}{\sin(\pi\beta^2)}}\+ \ell\biggr)
  + \ii^{-(q + 2\alpha)/\beta} J_{-(q + 2\alpha)/\beta}\biggl(\sqrt{\tfrac{\Lambda}{\sin(\pi\beta^2)}}\+ \ell\biggr)\+.
\end{equation}
In addition, \cite{Bautista:2021ogd} concludes that their one-point function is only valid for certain rational values of $\beta^2$ when $\Lambda > 0$ and $\Lambda_\mathrm{b} > 0$.
The conclusion that only this linear combination is allowed would be striking, if true, since it is in conflict with the semiclassical analysis that allows for any solution to the WDW equation.
Instead, it picks out a particular linear combination of the Hartle--Hawking and Vilenkin wavefunctions and suffers from a serious divergence as $\ell \to 0$, which is only avoided by the Hartle--Hawking wavefunction \cite{Anninos:2024iwf}.

As mentioned above, one expects that the WDW equation \eqref{eq:WDW} is valid as $\beta \to 0$.
In spacelike Liouville theory, this gives the Bessel function $K_{(Q-2\alpha)/b}(\sqrt{\Lambda/(\pi b^2)}\+ \ell)$ whereas the exact one-point function is \cite{Fateev:2000ik,Teschner:2000md}
\begin{equation}
  \langle V_\alpha(z_0) \rangle_\ell
  = \frac{2 \Gamma(2 b \alpha - b^2)}{b\+ \Gamma(b^{-2} - 2 b^{-1} \alpha + 1)} \frac{\bigl[\pi \Lambda \Gamma(b^2)/\Gamma(1-b^2) \bigr]^{(Q-2\alpha)/2b}}{(1 - \bar{z}_0 z_0)^{2 \alpha(Q-\alpha)}} K_{(Q-2\alpha)/b}\biggl(\sqrt{\tfrac{\Lambda}{\sin(\pi b^2)}}\+ \ell\biggr)\+.
\end{equation}
Hence, the solution to the gauge fixed WDW equation is exact up to a rescaling of the argument of the Bessel function, and the overall normalization which the WDW equation is insensitive to.
The intuition for this is that $\ell$ captures all diffeomorphism invariant data of the boundary metric.
As such, the uncontrolled minisuperspace truncation of Einstein gravity, in this model, is replaced by a gauge fixing procedure that captures the functional form of the exact result.

In the timelike theory, without the identification of $V_\alpha$ and $V_{-q-\alpha}$, any linear combination of the two Bessel functions solves the shift equations of \cite{Bautista:2021ogd}.%
\footnote{This assumes that the relation $\cosh(\pi \beta s)^2 = - \sin(\pi\beta^2) \Lambda_\mathrm{b}^2/\Lambda$ continues to hold.
  The parameter $s$ enters in the bootstrap analysis both in the spacelike \cite{Fateev:2000ik} and timelike \cite{Bautista:2021ogd} theories.
  Upon $s \to \pm \ii s$ and, as before, $ \beta \to \mp \ii b$, the relation here turns into the well-established spacelike one.}
Vertex operators identified or not, this suggests that the wavefunctions in \eqref{eq:WDW} are exact up to rescalings analogous to those in the spacelike theory, which would severely restrict the form of the semiclassical loop expansion.
We will verify this conjecture at one loop in §§\ref{sec:PI}--\ref{sec:ensembles}.

\section{Path integral for Liouville one-point functions} \label{sec:PI}
We now turn to the semiclassical one-loop analysis of the Liouville one-point functions, related to the wavefunction by \eqref{eq:one-pt_wfn}.
Though ultimately equivalent, using the geometric, background-independent formulation of Liouville theory is useful at the one-loop level since, then, the fluctuations are captured by a free, massive scalar on the semiclassical geometry independent of the fiducial background.
At this order, one does not have to confront the complications of the curved measure and can substitute it for a flat saddle-point measure.

For the purpose of this analysis, we will keep the sign of $\cL$ unspecified, capturing both the spacelike and timelike semiclassical limits $\cL \to \pm \infty$, since this does not introduce any additional complications.
Our primary interest is the timelike theory; the spacelike theory will be used as a consistency check since the exact result is well-established in that setting \cite{Fateev:2000ik,Teschner:2000md}.
We will consider both \enquote{light} operators that do not backreact and \enquote{heavy} ones that do.

In this section, we will formulate the problem and analyse the saddle points.
The disk partition function of spacelike Liouville theory is studied semiclassically in \cite{Mahajan:2021nsd,Chaudhuri:2024yau}.
Although we consider one-point functions and focus on timelike Liouville theory, our set-up is similar to that of \cite{Chaudhuri:2024yau}.
The saddle points of timelike Liouville theory are discussed in \cite{Anninos:2024iwf} but, here, we consider different boundary conditions.
In particular, we will see that fixing the boundary cosmological constant fixes the normal derivative of the Liouville field \cite{Mahajan:2021nsd}, which we reinterpret geometrically as fixing the trace of the extrinsic curvature of the boundary.
In addition, we refine the analysis of the moduli space of saddles in \cite{Fateev:2000ik}, with important consequences for the phase of the path integral once the measure on the moduli space is accounted for.
We also demonstrate how the dependence of the saddle point on the point of insertion leads to the conformal structure of the one-point function of a heavy vertex operator.

\subsection{One-loop path integral} \label{sec:one-loop_PI}
Using the path integral measure induced by \eqref{eq:geometric_measure}, the disk one-point function of the vertex operator $V_a = e^{2a\varphi}$ is
\begin{equation} \label{eq:one-pt_PI}
  \langle V_a(z_0) \rangle = \int \DD \varphi\+e^{2a \varphi(z_0) - \Lambda \hat{A} - \Lambda_\mathrm{b} \hat{\ell} - \frac{\cL}{24\pi} \int_{D}\! \tvol\+ (\tilde{g}^{\mu\nu} \partial_\mu \varphi\+ \partial_\nu \varphi + \tilde{R} \varphi) - \frac{\cL}{12\pi} \int_{\partial D}\! \tvol\+ \tilde{K} \varphi}\+,
\end{equation}
where $z_0$ is a point in the interior of the disk and
\begin{equation}
  \hat{A} \equiv \int_{\mathrlap{D}}\ \tvol\+ e^{2\varphi}\+,\qquad\quad
  \hat{\ell} \equiv \int_{\mathrlap{\partial D}}\ \tvol\+ e^\varphi\+,
\end{equation}
are the physical area and boundary length, respectively.
Here, we use FZZT \cite{Fateev:2000ik,Teschner:2000md} boundary conditions and have, hence, introduced a boundary cosmological constant.
This implies that the Liouville field is allowed to fluctuate without restriction both in the bulk and on the boundary.
The one-point function above is the Laplace transform of the one with fixed boundary length.
We elaborate on this in §\ref{sec:ensembles}.

Writing $\varphi = \varphi_\ast + \delta\varphi$, we find the saddle point equations
\begin{equation} \label{eq:Liouville_eom}
  0 = -2 \tilde{\square} \varphi_\ast + \tilde{R} + \frac{48\pi \Lambda}{\cL} e^{2\varphi_\ast} - \frac{48\pi a}{\cL} \tilde{\delta}(z-z_0)\+,\quad
  0 = \tilde{n}^\mu \partial_\mu \varphi_\ast + \tilde{K} + \frac{12 \pi \Lambda_\mathrm{b}}{\cL} e^{\varphi_\ast} \bigg|_{\partial M}\+,
\end{equation}
where $\tilde{n}$ is the outward-pointing normal with respect to the fiducial background metric.
Note that we have not imposed any boundary conditions on $\varphi$ when deriving these; they both follow from demanding that the action is stationary.
The saddle point equations admit a geometric formulation in terms of the saddle point metric $g_\ast = e^{2\varphi_\ast} g$ as
\begin{equation} \label{eq:geom_eom}
  R_\ast = -\frac{48\pi \Lambda}{\cL} + \frac{48\pi a}{\cL} \delta_\ast(z-z_0)\+,\qquad\quad
  K_\ast = - \frac{12\pi \Lambda_\mathrm{b}}{\cL}\+.
\end{equation}
Hence, the saddle point geometry has constant curvature away from the insertion at $z = z_0$ and its boundary has constant extrinsic curvature.
We keep $\Lambda/\cL$ and $\Lambda_\mathrm{b}/\cL$ finite in the semiclassical limit to have finite curvature saddles.
We also see that if $a$ is kept fixed as $\abs{\cL} \to \infty$ (light insertion), the delta function can be dropped and the geometry is smooth, whereas if $a/\cL$ is kept finite (heavy insertion), there is a conical singularity at $z_0$.%
\footnote{We use scalar delta functions, i.e.\ the scalar-density delta is $\delta(z) = \sqrt{\det \tilde{g}}\+ \tilde{\delta}(z) = \sqrt{\det g_\ast}\+ \delta_\ast(z)$.}

The semiclassical path integral around a saddle $\varphi_\ast$ is
\begin{equation} \setlength{\arraycolsep}{0pt}
  \begin{array}[b]{>{\displaystyle}l >{\displaystyle}r >{\displaystyle}l}
    \langle V_a \rangle_\ast = e^{2a\varphi_\ast(z_0) - S_\ast} \int \DD \delta\varphi \exp\Biggl[
      & - \frac{\cL}{24\pi} \int_{\mathrlap{D}}\ \vol_\ast \biggl[g_\ast^{\mu\nu} \partial_\mu \delta \varphi\+ \partial_\nu \delta \varphi - \frac{R_\ast}{2} \bigl(e^{2\delta\varphi} - 1 - 2 \delta\varphi\bigr) \biggr] & \\
      & {} + \frac{\cL}{12\pi} \int_{\mathrlap{\partial D}}\ \vol_\ast\+ K_\ast \bigl(e^{\delta\varphi} - 1 - \delta\varphi\bigr)
      & \Biggr]\+,
  \end{array}
\end{equation}
where by $R_\ast$ we refer to the constant curvature away from the insertion and $S_\ast$ is the on-shell action.
Note that the fluctuation $\delta\varphi$ effectively lives on the semiclassical geometry, not the fiducial background.
When there are zero modes, these should be excluded from the perturbative path integral and, instead, be integrated over globally.
We will elaborate on this in §\ref{sec:conf}.

The higher-point interactions, both bulk and boundary, are suppressed when $\abs{\cL} \to \infty$ since the region in field space that dominates the path integral is $\delta\varphi \sim \mathcal{O}(1/\sqrt{\cL})$.
Equivalently, this can be seen by rescaling $\delta\varphi$ to make the kinetic term canonical.
For this to be valid in the timelike theory, one has to implement a Gibbons--Hawking--Perry rotation \cite{Gibbons:1978ac} and integrate over imaginary fluctuations to make the fluctuations suppressed.
By power counting, we see that $S_\ast = \mathcal{O}(\cL)$ and that the one-loop correction is $\mathcal{O}(\cL^0)$.
Hence, it is consistent to drop the delta function in \eqref{eq:geom_eom} when the insertion is light since $a\+ \delta\varphi \sim \mathcal{O}(1/\sqrt{\cL})$ can be neglected at one loop.

The boundary conditions for the Liouville field have to respect conformal invariance, as discussed in §\ref{sec:Liouville_wfn}.
Since this excludes Dirichlet boundary conditions, the Liouville field fluctuates both in the bulk and on the boundary.
Hence, it is useful to split the quantum fluctuation around the saddle point into a bulk piece and a boundary piece, as in \cite{Chaudhuri:2024yau}.
To make this unique, we impose a boundary condition on the bulk fluctuation and a bulk condition on the boundary fluctuation:%
\footnote{In the timelike case, when $R_\ast > 0$, the Klein--Gordon problem $(-\square_\ast - R_\ast) \delta\varphi_\mathrm{b} = 0$ with Dirichlet boundary conditions $\delta\varphi_\mathrm{b} |_{\partial M} = \delta\varphi|_{\partial M}$ can be ill-posed since the Klein--Gordon operator with homogeneous boundary conditions might admit a nonvanishing solution.
  There is then an associated bulk zero-mode.
  We will return to this issue below.}
\begin{align} \label{eq:bulk_bdy_split}
  &\delta\varphi = \delta\varphi_\mathrm{B} + \delta\varphi_\mathrm{b}\+,
  &&\delta\varphi_\mathrm{B} |_{\partial D} = 0\+,
  &&(-\square_\ast - R_\ast) \delta\varphi_\mathrm{b} = 0\+.
\end{align}
These conditions are chosen precisely such that there is no quadratic term coupling $\delta\varphi_\mathrm{B}$ and $\delta\varphi_\mathrm{b}$ in the action, which are thus orthogonal.

In addition to neglecting the interaction terms, which contribute at higher loop orders, the background-independent measure in \eqref{eq:geometric_measure} can be replaced by a saddle point measure by $\vol \to \vol_\ast$ at the one-loop level, again because $\delta\varphi \sim \mathcal{O}(1/\sqrt{\cL})$.
Hence, the one-loop path integral takes the form
\begin{equation} \label{eq:1-pt_at_one_loop}
  \begin{aligned}[b]
    \langle V_a(z_0) \rangle_\textnormal{1-loop}
    = e^{2 a \varphi_\ast(z_0) - S_\ast}
    & \int \DD_\ast \delta\varphi_\mathrm{B}\+ \exp\Biggl[
      - \frac{\cL}{24\pi} \int_{\mathrlap{D}}\ \vol_\ast \Bigl(g_\ast^{\mu\nu} \partial_\mu \delta \varphi_\mathrm{B}\+ \partial_\nu \delta \varphi_\mathrm{B} - R_\ast \delta\varphi_\mathrm{B}^2 \Bigr)
      \Biggr]
    \\ & \mathllap{{}\times{}}
    \int \DD_\ast \delta\varphi_\mathrm{b}\+ \exp\Biggl[
      - \frac{\cL}{24\pi} \int_{\mathrlap{\partial M}}\ \vol_\ast\+ \delta\varphi_\mathrm{b} \bigl(n_\ast^\mu \partial_\mu - K_\ast\bigr) \delta\varphi_\mathrm{b}
      \Biggr]\+,
  \end{aligned}
\end{equation}
where we have integrated the action for the boundary fluctuation by parts and used the bulk condition above.

To define dimensionless ultralocal path integral measures, one has to introduce a UV (energy) scale $\Lambda_\mathrm{UV}$.
The one-loop path integral measures for the fluctuations can then be normalized as
\begin{equation} \label{eq:PI_measure_normalization}
  1 = \int \DD_\ast \delta\varphi_\mathrm{B}\+ \exp\Biggl[- \frac{\cL \Lambda_\mathrm{UV}^2}{24\pi} \int_{\mathrlap{D}}\ \vol_\ast\+ \delta\varphi_\mathrm{B}^2 \Biggr],\quad
  1 = \int \DD_\ast \delta\varphi_\mathrm{b}\+ \exp\Biggl[- \frac{\cL \Lambda_\mathrm{UV}}{24\pi} \int_{\mathrlap{\partial D}}\ \vol_\ast\+ \delta\varphi_\mathrm{b}^2 \Biggr],
\end{equation}
where we have, in addition, included factors $\frac{\cL}{24\pi}$ due to $\delta\varphi$ not being canonically normalized.
In the timelike theory, this implies a Gibbons--Hawking--Perry \cite{Gibbons:1978ac} rotation in field space to deal with the conformal mode problem since, in that case, $\cL < 0$.
This means that the integral is performed over purely imaginary fluctuations $\delta\varphi$.
We will see that this prescription is almost correct but that a finite number of modes will have to be rotated back, as in the case of the sphere in higher dimensions \cite{Polchinski:1988ua} and timelike Liouville theory on the two-sphere \cite{Anninos:2021ene}.
We discuss this in more detail in §\ref{sec:fluct}.

\subsection{Saddle points} \label{sec:saddle}
Next, we turn to the saddle points of the path integral with the vertex operator inserted.
Recall the saddle point equations \eqref{eq:geom_eom}
\begin{equation}
  R_\ast = -\frac{48\pi \Lambda}{\cL} + \frac{48\pi a}{\cL} \delta_\ast(z-z_0)\+,\qquad\quad
  K_\ast = - \frac{12\pi \Lambda_\mathrm{b}}{\cL}\+.
\end{equation}
With a light insertion, such that the delta function is not present, or a heavy insertion located at the origin, $z_0 = 0$, we may write an ansatz
\begin{equation} \label{eq:Liouville_simple_saddle}
  \dd s_\ast^2 = \frac{4 L^2}{(1 + \eta \rho^2)^2} (\dd \rho^2 + \rho^2 \dd \theta^2)\+,
\end{equation}
where $0 \leq \rho \leq \gamma$, $\theta \sim \theta + 2\pi (1+k)$, $k > -1$, and $L > 0$ is a length scale.
Here, the origin $z = 0$ corresponds to $\rho = 0$; we elaborate on the precise relation between $\rho$, $\theta$ and $z$ in \eqref{eq:rho_theta_tilde}--\eqref{eq:varphi_saddle} below.
As we will see below, $k$, which controls the conical singularity at $\rho = 0$, is determined solely by $a$, the label of the inserted operator $V_a = e^{2a\varphi}$.
Without loss of generality, we may take $\eta \in \{0, \pm 1\}$ by simultaneously rescaling $\rho$, $L$ and $\gamma$.
We will not consider the flat disk, $\eta = 0$, further and will return to the case of a heavy insertion away from the origin in §\ref{sec:conf}.
Note that there is a unique real semiclassical geometry with these boundary conditions, in contrast to what happens with positive curvature and fixed boundary length.
This does not imply that there is a unique saddle point for the Liouville field.
We elaborate on this point in §\ref{sec:conf}.

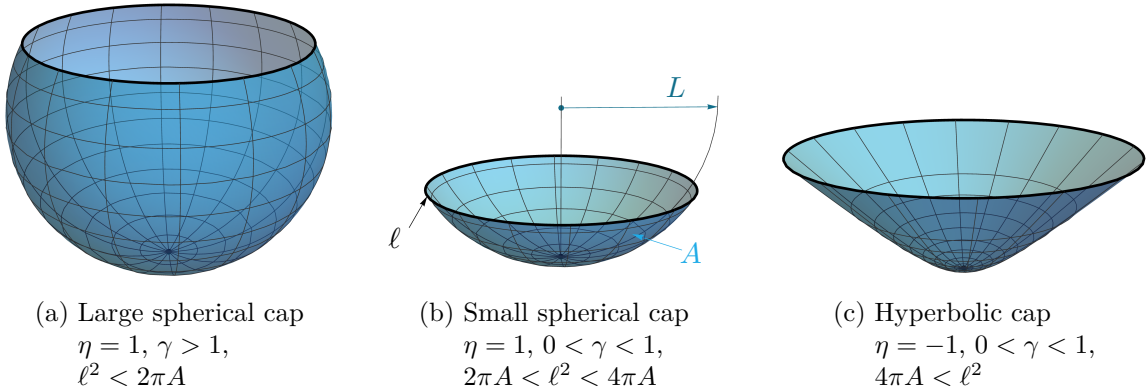
\begin{figure}[H]
  \centering
  \renewcommand{\figurewidth}{140pt}
  \captionsetup[subfigure]{justification=centering}
  \begin{subfigure}{\figurewidth}
    \centering
    \includegraphics[width=\figurewidth,trim={0 42pt 0 0},clip]{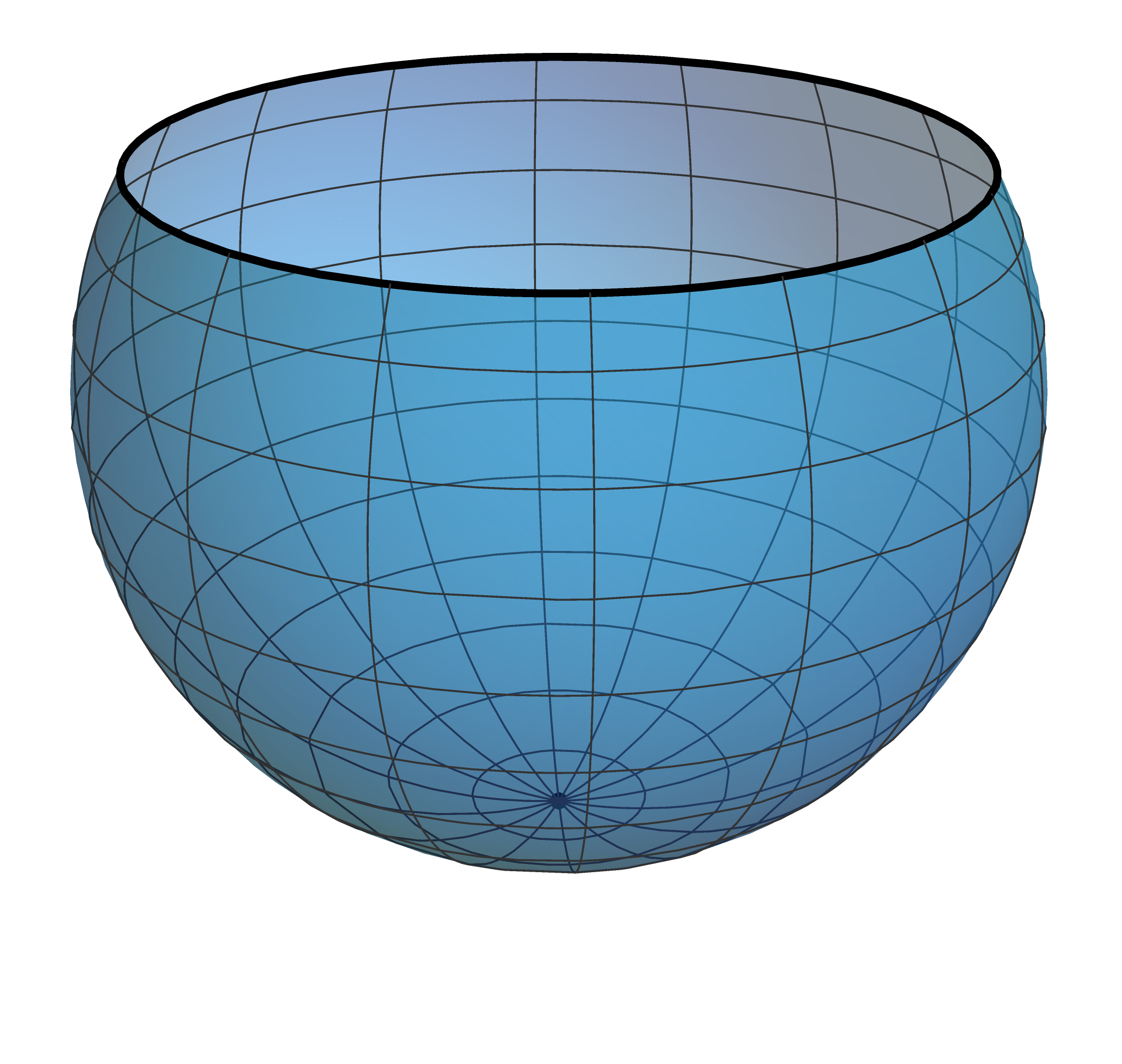}
    \caption{%
      \begin{varwidth}[t]{\figurewidth}
        Large spherical cap \\
        $\eta = 1$, $\gamma > 1$,\\
        $\ell^2 < 2\pi A$
      \end{varwidth}}
  \end{subfigure}
  \hfil
  \begin{subfigure}{\figurewidth}
    \centering
    \begin{tikzpicture}
      \node[inner sep=0, outer sep=0] at (0,0) {\includegraphics[width=\figurewidth,trim={0 26pt 0 0},clip]{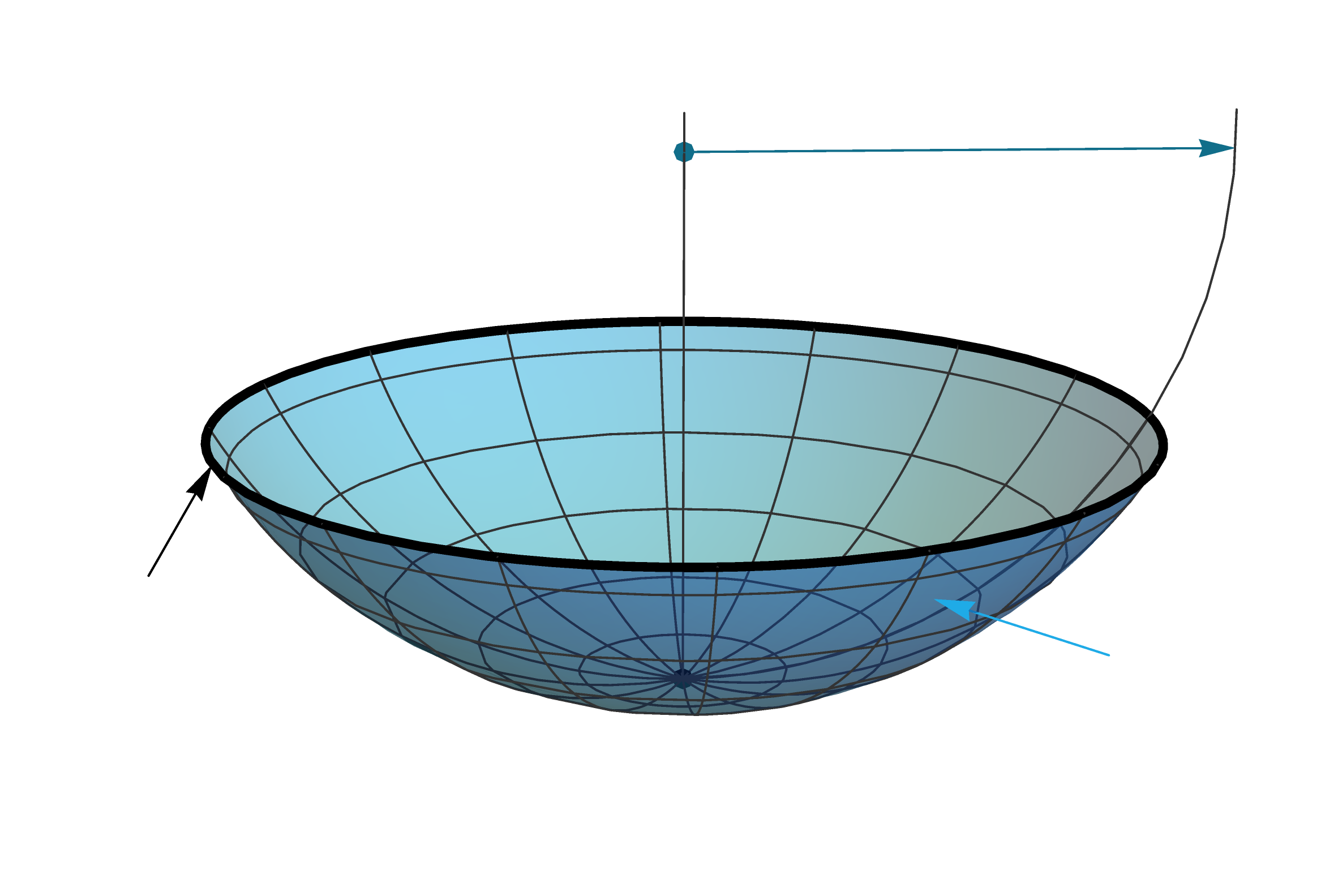}};
      \node[color=KulGrid3] at (1.6,1.15) {$L$};
      \node[color=KulGrid1] at (1.8,-1.05) {$A$};
      \node[color=black] at (-2.1,-0.85) {$\ell$};
    \end{tikzpicture}
    \caption{%
      \begin{varwidth}[t]{\figurewidth}
        Small spherical cap \\
        $\eta = 1$, $0 < \gamma < 1$,\\
        $2\pi A < \ell^2 < 4\pi A$
      \end{varwidth}}
  \end{subfigure}
  \hspace*{6pt}
  \hfil
  \begin{subfigure}{\figurewidth}
    \centering
    \includegraphics[width=\figurewidth,trim={0 72pt 0 0},clip]{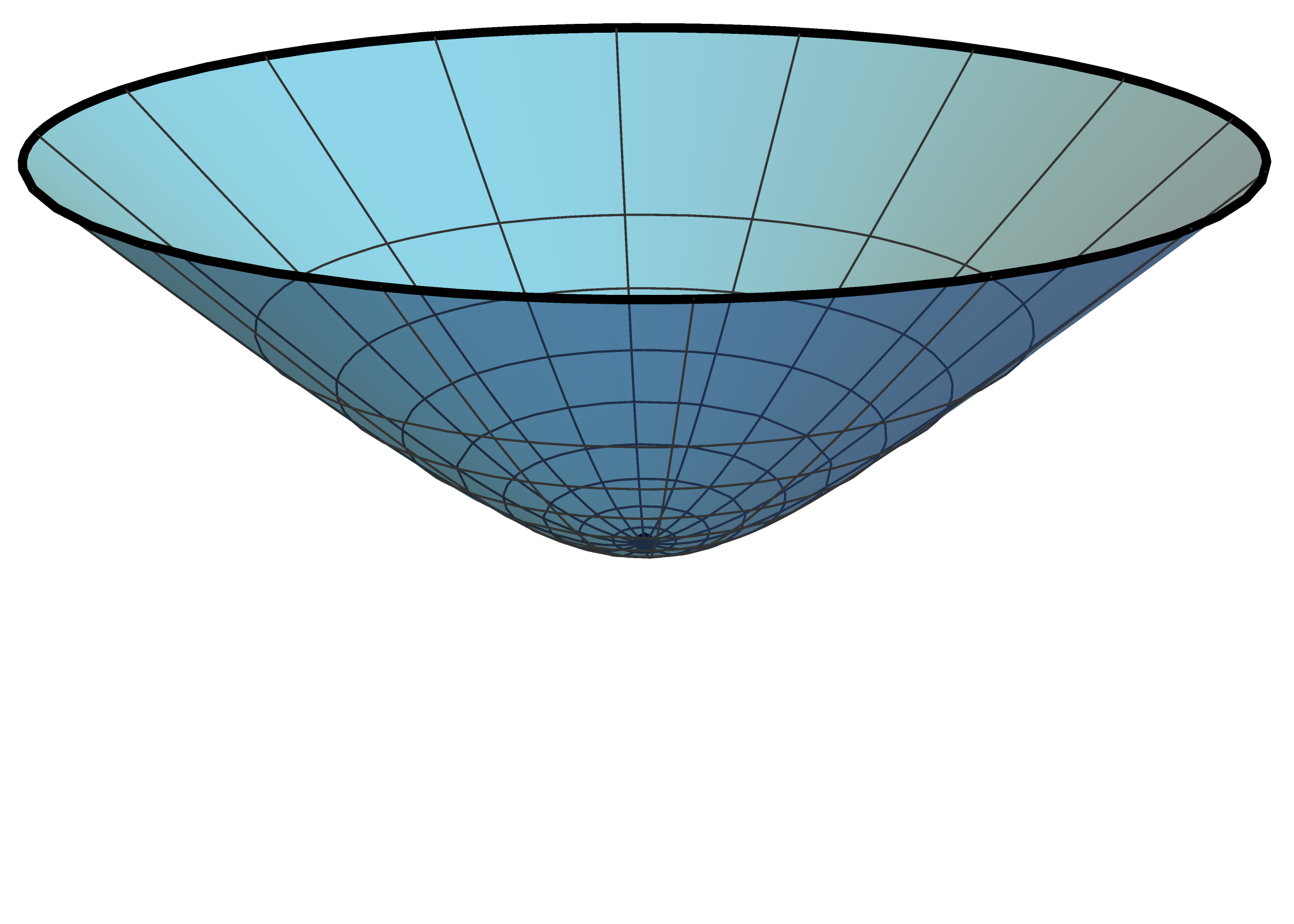}
    \caption{%
      \begin{varwidth}[t]{\figurewidth}
        Hyperbolic cap\\
        $\eta = -1$, $0 < \gamma < 1$,\\
        $4\pi A < \ell^2$
      \end{varwidth}}
  \end{subfigure}
  \caption{Illustrations of semiclassical saddle geometries \eqref{eq:Liouville_simple_saddle}, without conical singularities ($k=0$), and the corresponding ranges for the parameter $\gamma$ and $\ell^2/A$.
    The geometries have been isometrically embedded in $\mathbb{R}^3$ (a, b) and $\mathbb{R}^{2,1}$ (c).}
  \label{fig:saddles}
\end{figure}

The above geometry is a constant curvature disk of finite size, with $\eta$ determining the sign of the curvature.
The parameters $L$ and $\gamma$ fix the curvature scale and how big a portion of the sphere or hyperbolic plane the disk covers.
For positive curvature, $\gamma$ may take any real positive value, whereas it is bounded by $0 < \gamma < 1$ for negative curvature.
When the curvature is positive, $\gamma < 1$ and $\gamma > 1$ correspond to \enquote{small} and \enquote{large} spherical caps relative to the hemisphere ($\gamma = 1$), respectively.
The geometry has, evaluating $R_\ast$ away from the origin,
\begin{equation} \label{eq:saddle_properties}
  R_\ast = \frac{2\eta}{L^2}\+,\quad
  K_\ast = \frac{1 - \eta \gamma^2}{2 L \gamma}\+,\quad
  A_\ast = 4 \pi L^2 \frac{(1+k) \gamma^2}{1 + \eta\gamma^2}\+,\quad
  \ell_\ast = 4 \pi L \frac{(1+k) \gamma}{1 + \eta \gamma^2}\+.
\end{equation}
Note that, with a positive cosmological constant, the curvature is positive in the timelike theory ($\eta = +1$) and negative in the spacelike theory ($\eta = -1$).
The magnitude of $\Lambda$ fixes the curvature scale $L$.
From the second equation above,
\begin{equation} \label{eq:saddle_gamma}
  \gamma = -\eta L K_\ast + \eta \sqrt{\eta + L^2 K_\ast^2}\+,
\end{equation}
which can be used to write the geometry complete in terms of the cosmological constants.
Hence, any real $\Lambda_\mathrm{b}$ admits a real saddle in the timelike theory but only $\Lambda_\mathrm{b}/\cL < -\sqrt{\Lambda/(6\pi \cL)}$ does so in the spacelike theory.
The saddles are illustrated in Figure~\ref{fig:saddles}.

To fix $k$ in terms of $a$, we choose a fiducial background metric $\dd \tilde{s}^2$.
The metric $\dd s_\ast^2$ then defines the semiclassical Liouville field configuration through $\dd s_\ast^2 = e^{2\varphi_\ast} \dd \tilde{s}^2$, which will also be used to compute the on-shell action.
To this end, we first rewrite the saddle geometry in coordinates that are independent of $k$ and $\gamma$:
\begin{equation} \label{eq:rho_theta_tilde}
  \rho = \gamma \tilde{\rho}^{1+k}\+,\quad
  \theta = (1+k) \tilde{\theta}
  \quad\implies\quad
  \dd s_\ast^2 = \frac{4 L^2 (1+k)^2 \gamma^2 \tilde{\rho}^{2k}}{(1 + \eta \gamma^2 \tilde{\rho}^{2(1+k)})^2} \bigl(\dd \tilde{\rho}^2 + \tilde{\rho}^2 \dd \tilde{\theta}^2\bigr)\+,
\end{equation}
where $0 \leq \tilde{\rho} \leq 1$ and $\tilde{\theta} \sim \tilde{\theta} + 2\pi$ are related to the complex coordinate $z$ from \eqref{eq:fiducial_bkg_complex_coord} by $z = \tilde{\rho}\+ e^{\ii \tilde{\theta}}$.
Choosing a flat disk of radius $\tilde{L}$ as the fiducial background, we have%
\footnote{\label{footnote:complex_saddle}%
  In addition, there are complex saddles with $\varphi_\ast \to \varphi_\ast + 2\pi \ii n$ with $n \in \mathbb{Z}$.
  Note that shifting by just $\pi \ii$ is prevented by the boundary conditions even though the physical metric in the bulk does not change.
  The on-shell action of these saddles acquires an additional term $2\pi \ii n \cL \chi_M/6$.}
\begin{equation} \label{eq:varphi_saddle}
  \dd \tilde{s}^2 = \tilde{L}^2 \bigl(\dd \tilde{\rho}^2 + \tilde{\rho}^2 \dd \tilde{\theta}^2\bigr)\+,\qquad
  \varphi_\ast = \log \frac{2 L (1+k) \gamma \tilde{\rho}^{k} }{\tilde{L} (1 + \eta \gamma^2 \tilde{\rho}^{2(1+k)})} \+.
\end{equation}

We can now determine $k$ by integrating the equation of motion for $\varphi_\ast$ over a small disk $0 \leq \tilde{\rho} \leq \epsilon$, including the contribution from the delta function in \eqref{eq:Liouville_eom}.
This gives
\begin{equation} \label{eq:fixing_k}
  - \frac{24 \pi}{\cL} a + \mathcal{O}(\epsilon^2)
  = \int_{\mathrlap{D}}\ \tvol\+ \tilde{\square} \varphi_\ast
  = \int_{\mathrlap{\rho = \epsilon}}\ \tvol\+ \tilde{n}^\mu \partial_\mu \varphi_\ast \overset{\epsilon \to 0}{\longrightarrow}\+ 2 \pi k
  \quad \implies \quad
  k = - \frac{12}{\cL} a\+.
\end{equation}
Here we see again that the conical singularity can be neglected when the insertion is light ($a = \mathcal{O}(1)$) whereas the conical deficit is finite in the semiclassical limit when it is heavy ($a = \mathcal{O}(\cL)$).
Recall that $k > -1$, corresponding to
\begin{equation}
  a < \frac{\cL}{12} \quad \text{(spacelike)}\+,\qquad\quad
  a > \frac{\cL}{12} \quad \text{(timelike)}\+.
\end{equation}
The bound on $a$ in the spacelike theory corresponds to the Seiberg bound \cite{Seiberg:1990eb,Polchinski:1990mh_} in the semiclassical limit.

The on-shell action diverges due to the insertion.
To regularize it, we excise a small disk $0 \leq \tilde{\rho} < \epsilon$ centred at $z_0$, following \cite{Zamolodchikov:1995aa}.
The contribution from the insertion is then accounted for by smearing it over the circle at $\tilde{\rho} = \epsilon$:%
\footnote{One can smear the insertion over a small circle already off-shell.
  The additional boundary condition at $\tilde{\rho} = \epsilon$ is $\tilde{n}^\mu \partial_\mu \varphi_\ast = - 48\pi a / (\cL \Vol(S_\epsilon^1))$, consistent with \eqref{eq:fixing_k} and satisfied by our saddle.
  The fluctuation path integral remains unmodified if one imposes boundary conditions such that $\delta\varphi(\tilde{\rho} = \epsilon)$ goes to a constant as $\epsilon \to 0$, see §\ref{sec:bulk_fluct} and, in particular, footnote~\ref{footnote:inner_bdy_terms}.}
\begin{equation}
  -2a \varphi_\ast(0) \ \to \ -\frac{2a}{\Vol(S^1_\epsilon)} \int_{\mathrlap{\tilde{\rho}=\epsilon}}\ \tvol\+ \varphi_\ast\+.
\end{equation}
This is related to the UV scale appearing in the ultralocal path integral measures by $\epsilon = (\Lambda_\mathrm{UV} \tilde{L})^{-1}$.
Dropping terms that vanish in the $\epsilon \to 0$ limit, we find
\begin{equation} \label{eq:tree-level_centred_insertion}
  \begin{aligned}[b]
    S_\ast - 2 a \varphi_\ast(0)
    & = \Lambda A + \Lambda_\mathrm{b} \ell + \frac{\cL}{6} \biggl[-\frac{k^2}{2} \log \Lambda_\mathrm{UV} \tilde{L} + \frac{\ell^2}{4\pi A} + (1+k)\biggl(- 1 + \log\frac{2(1+k)A}{\tilde{L} \ell}\biggr)\biggr]\+.
  \end{aligned}
\end{equation}
Note that the $k$ in the coefficient of the second logarithm comes from the insertion whereas the $1$ is related to the conformal anomaly.%
\footnote{Recall that the conformal anomaly is captured completely at the classical level in the formulation with a background-independent measure.}
Although it is possible to express the result in terms of only the cosmological constants $\Lambda$ and $\Lambda_\mathrm{b}$, this and several formulas that will appear later simplify by writing them in terms of the area, $A = A_\ast$, and the boundary length, $\ell = \ell_\ast$.
Two particularly useful relations between these and the parameters that enter in the metric are
\begin{equation}
  \frac{\ell^2}{A} = \frac{4\pi (1+k)}{1+\eta\gamma^2}\+,\qquad\qquad
  \frac{A}{\ell} = L \gamma\+.
\end{equation}
The final result is expressed in different ways and compared to exact results in §\ref{sec:ensembles}.

\subsection{Conformal structure of the one-point functions} \label{sec:conf}
Having discussed the saddle points, we turn to the conformal structure of the one-point functions.
This comes about in slightly different ways depending on whether the insertion is light or heavy.
In the former case, it is related to zero modes while in the latter it comes from how the saddle geometry depends on $z_0$.
In both cases, it has to do with the conformal group of the disk, $\mathrm{PSL}(2, \mathbb{R})$ (see Appendix~\ref{app:disk_conf} for a brief review).

A similar zero-mode analysis can be found in \cite{Fateev:2000ik}.
Here, we refine that analysis slightly by integrating only over the moduli space $\mathrm{PSL}(2, \mathbb{R})/\mathrm{U}(1)$ of saddles instead of the full $\mathrm{PSL}(2, \mathbb{R})$.
The distinction is important to be able to normalize the measure by comparing with the normalization of the infinitesimal zero modes \eqref{eq:PI_measure_normalization}, as well as to keep track of finite factors that will play a role for the consistency check of the results.
In addition, we perform the analysis on the disk instead of the upper half-plane.

Liouville theory is invariant under conformal transformations since the anomaly action $S_\textnormal{anom}[\omega; \tilde{g}]$ vanishes for $\omega$ such that $e^{2\omega} \tilde{g} = \tilde{g}'$ is related to $\tilde{g}$ by a diffeomorphism.
Explicitly, with $\dd \tilde{s}^2 = \Omega(\bar{z}, z) \dd \bar{z} \dd z$, a conformal transformation $z \mapsto z' = g(z)$ maps%
\footnote{Note that these conventions yield a left-action since $\varphi \circ g_1^{-1} \circ g_2^{-1} = \varphi \circ (g_2 \circ g_1)^{-1}$.}
\begin{equation} \label{eq:conf_action}
  \varphi \mapsto g \cdot \varphi = \varphi' + \omega_g\+,\qquad
  \varphi' = \varphi \circ g^{-1}\+,\quad
  \omega_g =  \log \biggl(\abs{\partial g^{-1}} \frac{\Omega \circ g^{-1}}{\Omega}\biggr)\+,
\end{equation}
where, by abuse of notation, we use $g$ to refer both to the group element and the associated function.

At the infinitesimal level, a diffeomorphism $x \mapsto x' = x + \xi(x')$ is parametrized by a vector field $\xi^\mu$ with vanishing normal component at the boundary, and $\delta \tilde{g}_{\mu\nu} = -2 \tilde{\nabla}_{(\mu} \xi_{\nu)}$.
For there to be a corresponding conformal transformation, $\xi$ has to be a conformal Killing vector and we get $\omega_\xi = -\frac{1}{2} \tilde{\nabla}_\mu \xi^\mu$.
This implies that
\begin{equation}
  \delta \varphi = - \xi^\mu \partial_\mu \varphi - \frac{1}{2} \tilde{\nabla}_\mu \xi^\mu = - \frac{1}{2} \nabla_\mu \xi^\mu\+,
\end{equation}
with $\nabla_\mu$ the covariant derivative associated to $g = e^{2 \varphi}  \tilde{g}$.
Hence, $\delta\varphi = - \frac{1}{2} \nabla_{\! \ast \mu} \xi^\mu$ is a zero mode of the quadratic action in the saddle point expansion.
Note that only nonisometric conformal Killing vectors of the semiclassical background correspond to zero modes since Killing vectors, associated with isometries, have $\nabla_{\! \ast \mu} \xi^\mu = 0$.
It is of course expected that there is a zero mode for every generator of the conformal group that does not leave the saddle point $\varphi_\ast$ invariant, analogous to the phenomenon of spontaneous symmetry breaking.

A short calculation shows that, in two dimensions, $(-\square - R) \nabla_\mu \xi^\mu = 0$ for any conformal Killing vector, consistent with them corresponding to zero modes on compact manifolds without boundaries.
When there is a boundary, we see that they satisfy the bulk condition in \eqref{eq:bulk_bdy_split} and hence correspond to boundary modes.
Generically, they are nonvanishing on the boundary, which explains why Dirichlet boundary conditions break conformal invariance.
The only exception is when $(-\square - R) \delta\varphi = 0$ admits a nonvanishing solution with homogeneous Dirichlet boundary conditions, in which case our decomposition into bulk and boundary modes breaks down.
As we explain below, this only happens on the hemisphere.

\subsubsection*{Light insertion}
Due to the conformal invariance, $g \cdot \varphi_\ast$, with $\varphi_\ast$ the reference solution \eqref{eq:varphi_saddle}, is a saddle point for any $g \in \mathrm{PSL}(2, \mathbb{R})$.
However, $\varphi_\ast$ is invariant under the $\mathrm{U}(1)$ isometries $g(z) = e^{\ii \alpha} z$ of the saddle so there is a distinct Liouville saddle point only for each coset $g \mathrm{U}(1)$.
The zero modes are, thus, dealt with by introducing collective coordinates and integrating over the moduli space of saddles $\mathrm{PSL}(2, \mathbb{R})/\mathrm{U}(1)$.
Since the on-shell action is invariant, it need not be included in the integral and we focus on
\begin{equation}
  I \equiv \int_{}\ \dd\mu[g \mathrm{U}(1)]\+ e^{2a (g \cdot \varphi_\ast)(z_0)}\+,
\end{equation}
where, as above, $z_0$ is the point of insertion.
The measure on the coset space is obtained by disintegrating a bi-invariant Haar measure on $\mathrm{PSL}(2, \mathbb{R})$.
Its normalization is fixed by demanding that it agrees with the path integral measure for the fluctuations when evaluated at the identity or, equivalently, can be chosen arbitrarily as long as the corresponding Jacobian is included.
Note that the moduli space is noncompact.
Hence, without an insertion, the integral would diverge.

The integrand can be simplified by using that the measure is left-invariant and judiciously choosing a group element to act with.
A general element of $\mathrm{PSL}(2, \mathbb{R})$ acting on the disk can be parametrized as
\begin{equation} \label{eq:PSL2R_element}
  g(z) = \frac{e^{\ii \alpha} z + w}{e^{\ii \alpha} \bar{w} z + 1}\+,\qquad
  \alpha \sim \alpha + 2\pi\+,\quad
  \abs{w} < 1\+,
\end{equation}
which is a composition of a rotation parametrized by $\alpha$ and a \enquote{special conformal} transformation that maps the origin to $w$.
With this parametrization, one readily computes the bi-invariant Haar measure (up to an arbitrary normalization)
\begin{equation}
  \dd\mu(g) = \frac{\dd \alpha \dd^2 w}{(1 - \bar{w} w)^2}\+.
\end{equation}
Using the left-invariance of the measure, the integral is
\begin{equation} \label{eq:moduli_space_integral}
  \begin{aligned}[b]
    I &= \int_{}\ \dd\mu[g \mathrm{U}(1)]\+ e^{2a (g_0 \cdot g \cdot \varphi_\ast)(z_0)}
    = \biggl(\frac{2 \gamma L/\tilde{L}}{1 - \bar{z}_0 z_0}\biggr)^{2a} \int_{}\ \frac{\dd^2 w}{(1 - \bar{w} w)^2} \biggl(\frac{1 - \bar{w} w}{1 + \eta \gamma^2 \bar{w} w}\biggr)^{2a}
    \\
    &= \frac{\ell^2}{4(2a-1)A} \biggl(\frac{2 A/(\ell \tilde{L})}{1 - \bar{z}_0 z_0}\biggr)^{2a}\+,
    \qquad\text{for}\ \Re(a) > \frac{1}{2}\+,
  \end{aligned}
\end{equation}
where $g_0$ is any conformal transformation that maps the origin to $z_0$ and the condition on $a$ is needed for the integral to converge.

The integral above is not the full contribution to the one-point function since we used an arbitrary normalization of the measure.
To compensate, we should include a Jacobian.
Linearizing the group element \eqref{eq:PSL2R_element} around the identity, the holomorphic part of the vector field generating the transformation is
\begin{equation} \label{eq:infinitesimal_conf}
  \xi
  = \ii z\+ \delta\alpha + \delta w - z^2\+ \delta \bar{w}\+.
\end{equation}
The corresponding fluctuation of the Liouville field is
\begin{equation} \label{eq:delta_varphi_from_w}
  \delta \varphi = - \frac{1}{2} \nabla_{\! \ast \mu} \xi^\mu
  = \frac{1 + \eta\gamma^2}{1 + \eta\gamma^2 \bar{z} z} \bar{z}\+ \delta w - \frac{1 + \eta\gamma^2}{1 + \eta\gamma^2 \bar{z} z} z\+ \delta \bar{w}\+.
\end{equation}
We will see in §\ref{sec:bdy_fluct} that these are indeed zero modes and determine the Jacobian by matching the linearized measure on the moduli space with the normalization \eqref{eq:PI_measure_normalization}.

We already see that \eqref{eq:moduli_space_integral} has the correct kinematic structure and implies that
\begin{equation}
  \langle V_a(z_0) \rangle_\textnormal{tree}
  = \frac{\langle V_a(0) \rangle_\textnormal{tree}}{(1 - z_0 \bar{z}_0)^{2 \Delta_a^\textnormal{tree}}}\+,\qquad
  \Delta_a^\textnormal{tree} = a \+,
\end{equation}
where the conformal dimension agrees both with the naive expectation from the Weyl transformation $\varphi \mapsto \varphi + \omega$ and, at leading order in the semiclassical limit, with the exact $\Delta_\alpha = \alpha (q + \alpha)$, $\alpha = \beta a$.

\subsubsection*{Heavy insertion}
When the insertion is heavy, $a = \mathcal{O}(\cL)$, and not at the origin, the saddle point is not the one discussed in §\ref{sec:saddle}.
However, it is related to that saddle by the action \eqref{eq:conf_action} of any conformal transformation $g_0$ that maps the origin to $z_0$:
\begin{equation}
  \varphi_\ast^{g_0} = \varphi_\ast' + \omega_0\+,\qquad\quad
  \omega_0 = \log \frac{1 - z_0 \bar{z}_0}{\abs{1 - \bar{z}_0 z}^2}\+,
\end{equation}
where $\varphi_\ast$ is the saddle \eqref{eq:varphi_saddle} with the conical singularity at the origin.
Since the two saddles are related by a conformal transformation, they describe the same semiclassical geometry $\dd s_\ast^2 = e^{2 \varphi_\ast} \dd \tilde{s}^2$, albeit in different coordinates.
It follows that $\varphi_\ast^{g_0}$ solves the equations of motion \eqref{eq:geom_eom} from their geometric nature.

To compute the on-shell action we follow §\ref{sec:saddle} but excise a disk of radius $\epsilon$ centred around $z_0$ instead of the origin.%
\footnote{It is crucial that the \emph{fiducial} radius $\epsilon \tilde{L}$ of the excised disk is independent of $z_0$.}
We will show that the Liouville action can be computed as
\begin{equation} \label{eq:conf_heavy_action}
  S_\mathrm{L}^\gamma[\varphi_\ast^{g_0}; \tilde{g}]
  = S_\mathrm{L}^{g_0^{-1}(\gamma)}[\varphi_\ast; \tilde{g}]\+,
\end{equation}
where $\gamma$ is the boundary of the excised region.

Even though a small disk is excised, there is no boundary term $\tilde{K} \varphi$ there.
Hence, the Liouville action on the annulus $\epsilon \leq \abs{z} \leq 1$ differs from the anomaly action, which does include that boundary term, by%
\footnote{The Liouville action also contains the cosmological constant term but since the area is smooth as $\epsilon \to 0$ there is no need to keep track of that term.}
\begin{equation}
  \frac{24\pi}{\cL} S_\mathrm{L}^\gamma[\varphi_\ast^{g_0}; \tilde{g}]
  = S_\textnormal{anom}^\gamma[\varphi_\ast^{g_0}; \tilde{g}] - 2 \int_{\mathrlap{\gamma}}\
  \tvol\+ \tilde{K} \varphi_\ast^{g_0}\+.
\end{equation}
Since the solution on the annulus is smooth, we can use \eqref{eq:anom_consistency} to relate the anomaly action above to $S_\textnormal{anom}^\gamma[\varphi_\ast'; e^{2\omega_0}\tilde{g}]$.
As above, this anomaly action differs from the Liouville action $S_\mathrm{L}^\gamma[\varphi_\ast'; e^{2\omega_0} \tilde{g}]$ by a boundary term at $\gamma$.
Furthermore, $S_\mathrm{L}^\gamma[\varphi_\ast'; e^{2\omega_0}\tilde{g}] = S_\mathrm{L}^{g_0^{-1}(\gamma)}[\varphi_\ast; \tilde{g}]$ by a change of coordinates since $e^{2\omega_0} \dd\tilde{s}^2 = \dd\tilde{s}^{\prime\+ 2}$.
Hence, the on-shell action is given by
\begin{equation} \label{eq:heavy_on-shell_action}
  \frac{24\pi}{\cL} S_\mathrm{L}^\gamma[\varphi_\ast^{g_0}; \tilde{g}]
  = \frac{24\pi}{\cL} S_\mathrm{L}^{g_0^{-1}(\gamma)}[\varphi_\ast; \tilde{g}]
  + S_\textnormal{anom}^\gamma[\omega_0; \tilde{g}]
  + 2 \int_{\mathrlap{\gamma}}\ \tvol_{\omega_0}\+ \tilde{K}_{\omega_0} \varphi_\ast'
  - 2 \int_{\mathrlap{\gamma}}\ \tvol\+ \tilde{K} \varphi_\ast^{g_0}\+,
\end{equation}
where $\tvol_{\omega_0}$ and $\tilde{K}_{\omega_0}$ are the induced volume form and extrinsic curvature of the metric $e^{2\omega_0} \dd \tilde{s}^2$, respectively.
Note that $g_0^{-1}(\gamma)$ is a circle enclosing the origin.
That this circle is not centred exactly at the origin gives a subleading effect when $\epsilon \to 0$.
On the other hand, it is important that its radius is $\epsilon / (1 - \bar{z}_0 z_0)$ at leading order in small $\epsilon$.

The anomaly action $S_\textnormal{anom}^\gamma[\omega_0; \tilde{g}]$ would vanish were it not for the excised region centred at $z_0$.
The bulk term of it is corrected only at subleading order in $\epsilon$ since the integrand is smooth, and hence cancels the boundary term at $\abs{z} = 1$.
Hence, the only nontrivial contribution comes from the boundary term at $\gamma$.
Since $\varphi_\ast^{g_0} = \varphi_\ast' + \omega_0$ and $\tvol_{\omega_0} \tilde{K}_{\omega_0} = \tvol (\tilde{K} + \tilde{n}^\mu \partial_\mu \omega_0)$, this boundary term almost exactly cancels the two other $\gamma$ integrals in \eqref{eq:heavy_on-shell_action}:
\begin{equation}
  S_\textnormal{anom}^\gamma[\omega_0; \tilde{g}]
  + 2 \int_{\mathrlap{\gamma}}\ \tvol_{\omega_0}\+ \tilde{K}_{\omega_0} \varphi_\ast'
  - 2 \int_{\mathrlap{\gamma}}\ \tvol\+ \tilde{K} \varphi_\ast^{g_0}
  = 2 \int_{\mathrlap{\gamma}}\ \tvol\+ (\tilde{n}^\mu \partial_\mu \omega_0) \varphi_\ast' + \mathcal{O}(\epsilon)
  \ \overset{\epsilon \to 0}{\longrightarrow}\ 0\+.
\end{equation}
The right-hand side vanishes when $\epsilon \to 0$ since $\tvol = \epsilon \tilde{L} \dd \theta$ and $\tilde{n}^\mu \partial_\mu \omega_0$ is smooth at $z_0$, in contrast to $\tilde{K} = (\epsilon \tilde{L})^{-1}$, while $\varphi_\ast'$ diverges only logarithmically.
This is a manifestation of the fact that both $\dd \tilde{s}^2$ and $e^{2\omega_0} \dd \tilde{s}^2$ look locally like flat space in a neighbourhood of $z_0$, in contrast to $\dd s_\ast^2$.
Thus, \eqref{eq:conf_heavy_action} holds and the action is conformally invariant also when a small region is excised, provided one transforms the boundary of the excised region.

When computing $S^{g_0^{-1}(\gamma)}[\varphi_\ast; \tilde{g}]$, the only relevant modification to the calculation in §\ref{sec:saddle} is that the radius of the excised region is $\epsilon/(1 - \bar{z}_0 z_0)$ instead of $\epsilon$.
Similarly, since $\varphi_\ast^{g_0} = \varphi_\ast' + \omega_0$, the only modification to the contribution from the smeared insertion is that $\varphi_\ast$ is effectively smeared over a circle of radius $\epsilon/(1 - \bar{z}_0 z_0)$ and the additional contribution from $\omega_0(z_0) = -\log(1 - \bar{z}_0 z_0)$.
Combining these effects and recalling $k = -12 a/\cL$,
\begin{equation}
  S_\mathrm{L}^\gamma[\varphi_\ast^{g_0}; \tilde{g}] - 2 a \varphi_\ast^{g_0}(z_0)
  = S_\ast - 2 a \varphi_\ast(0) - \frac{\cL}{12} k(k+2) \log(1 - z_0 \bar{z}_0)\+,
\end{equation}
with $S_\ast - 2 a \varphi_\ast(0)$ given by \eqref{eq:tree-level_centred_insertion}.
We have thus found
\begin{equation}
  \langle V_a(z_0) \rangle_\textnormal{tree}
  = \frac{\langle V_a(0) \rangle_\textnormal{tree}}{(1 - z_0 \bar{z}_0)^{2 \Delta_a^\textnormal{tree}}}\+,\qquad
  \Delta_a^\textnormal{tree} = \frac{a (\cL - 6 a)}{\cL} \+.
\end{equation}

From the Weyl transformation $\varphi \mapsto \varphi + \omega$ one might naively have expected the dimension of $e^{2a \varphi}$ to be $a$, but $\Delta_a^\textnormal{tree}$ differs from this when the operator is heavy, $a = \mathcal{O}(\cL)$, due to the backreaction on the saddle.
It is a nontrivial consistency check that $\Delta_a^\textnormal{tree}$ agrees with the exact $\Delta_\alpha = \alpha (q + \alpha)$, $\alpha = \beta a$, at leading order in the semiclassical limit.
However, it disagrees at subleading order:
\begin{equation}
  \Delta_\alpha - \Delta_\alpha^\textnormal{tree} = - \beta\alpha - \frac{13}{6} (\beta\alpha)^2 + \mathcal{O}(\beta^2)\+.
\end{equation}
Note that the product $\beta\alpha$ is kept finite for a heavy insertion.
This predicts a one-loop correction $-k(12+13k)/24$ to $\Delta_\alpha$.
It would be interesting to understand how the conformal dimensions are corrected at the loop level since, naively, the one-loop determinants are only sensitive to the semiclassical geometry, which is independent of $z_0$.
In §\ref{sec:bulk_fluct}, we speculate that this might be related to boundary conditions at the conical singularity.

\section{Quantum fluctuations} \label{sec:fluct}
Having analysed the saddle points of the one-point function $\langle V_a \rangle$, we turn to the quantum fluctuations.
Recall that these are split into boundary fluctuations $\delta\varphi_\mathrm{b}$ and bulk fluctuations $\delta\varphi_\mathrm{B}$, discussed in §\ref{sec:bdy_fluct} and §\ref{sec:bulk_fluct}, respectively.

Boundary fluctuations of spacelike Liouville theory have been studied before in \cite{Chaudhuri:2024yau}.
Here, we generalize \cite{Chaudhuri:2024yau} by considering a disk saddle of arbitrary curvature \textdash positive curvature being relevant for timelike Liouville theory \textdash and allowing for a conical singularity.
We find that, although the number of negative modes is independent of $\ell^2/A$ as long as the curvature is negative, when the curvature is positive, it can change as one varies $\ell^2/A$.
In addition, we calculate the zero-mode Jacobian in the case of a light vertex operator by matching the fluctuations induced by conformal transformations, discussed in §\ref{sec:conf}, to specific boundary modes.
A similar analysis for timelike Liouville theory on the sphere can be found in \cite{Anninos:2021ene}.

The bulk fluctuation determinant was recently computed in \cite{Chaudhuri:2024rgn} in the absence of a conical singularity.
Here, we formulate the problem with the singularity present and discuss the subtleties associated with the boundary conditions at the singularity.
We also analyse how the number of negative modes depends on $\ell^2/A$ in the regime of positive curvature.
It turns out that there is always one negative mode in total but whether it is a bulk or boundary mode depends on whether the spherical cap is large or small.

\subsection{Boundary fluctuations} \label{sec:bdy_fluct}
We first turn to the boundary modes.
Recall that these are subject to the Klein--Gordon equation $(-\square_\ast - 2 \eta/L^2) \delta\varphi_\mathrm{b} = 0$.
Using the coordinates in \eqref{eq:Liouville_simple_saddle} (recall $\theta \sim \theta + 2\pi (1+k)$), the boundary modes can be expanded in a Fourier series as
\begin{equation} \label{eq:varphi_b_Fourier}
  \delta\varphi_\mathrm{b} = \sum_{n \in \mathbb{Z}} \delta\varphi_{\mathrm{b},n} f_n(\rho) e^{\ii n \theta/(1+k)}\+,
\end{equation}
where $f_n(\gamma) = 1$, and the Klein--Gordon equation becomes
\begin{equation}
  {-} \frac{(1 + \eta \rho^2)^2}{4}\biggl[\rho^{-1} \partial_\rho (\rho \partial_\rho) - \frac{n^2}{(1+k)^2 \rho^2}\biggr] f_n(\rho) - 2 \eta f_n(\rho) = 0\+.
\end{equation}
The regular%
\footnote{We discuss the regularity condition at $\rho = 0$ in presence of a conical singularity further in §\ref{sec:bulk_fluct}.}
solution with $f_n(\gamma) = 1$ is
\begin{equation} \label{eq:varphi_b_solution}
  f_n(\rho) = \frac{g_n(\rho)}{g_n(\gamma)}\+,\qquad
  g_n(\rho) = \rho^{\frac{\abs{n}}{1+k}} \biggl(\frac{\abs{n}}{1+k} + \frac{1 - \eta \rho^2}{1 + \eta \rho^2}\biggr)\+.
\end{equation}
If $g_n(\gamma) = 0$, there is no solution and the Klein--Gordon problem is ill-posed.
On the hyperbolic cap, $\eta = -1$ and $g_n(\gamma) > 0$ since $0 < \gamma < 1$.
On the spherical cap, $\eta = 1$ and $g_n(\gamma) = 0$ for
\begin{equation}
  \gamma = \sqrt{\frac{1+k+\abs{n}}{1+k-\abs{n}}}\+.
\end{equation}
In particular, this happens with a light insertion ($k = 0$) for the $n = 0$ mode on the hemisphere ($\gamma = 1$).
With a heavy insertion, all $\abs{n} < 1 + k$ give rise to such an issue for some value of $\gamma$.
Note, however, that this does not imply that the one-point function is ill-defined; it simply signals that the bulk/boundary split we employ breaks down.
Since, for a particular vertex operator, the issue appears at discrete values of $\gamma$, it can be resolved by computing the one-point function away from those values and taking limits.

The radial solutions satisfy
\begin{equation} \label{eq:lambda_b}
  (n_\ast^\mu \partial_\mu - K_\ast) f_n |_{\partial M} = \lambda_{\mathrm{b},n}\+,\qquad
  \lambda_{\mathrm{b},n}
  = \frac{2\pi}{\ell} \frac{n^2 - (1+k)^2}{\abs{n} + \frac{\ell^2}{2\pi A} - (1+k)} \+.
\end{equation}
From this, we see that there are two zero modes with $\abs{n} = 1+k$ if $k$ is an integer.
We will elaborate on this in the case of a light insertion, where these are related to conformal transformations, below.
In the spacelike theory, there are also negative modes with $\abs{n} < 1+k$.
In particular the constant mode has $\lambda_{\mathrm{b},0} < 0$.
In the timelike theory, the same modes are negative but only for some values of $\ell^2/A$ since $0 < \ell^2/A < 4\pi (1+k)$.
In particular, for a light insertion, the $n = 0$ mode is only negative on a small spherical cap ($\ell^2 > 2\pi A$).

To make the integral over the negative modes converge, their contours of integration have to be rotated relative to those of the positive modes.
This produces a phase $\pm \ii$ for each negative mode, depending on the direction of rotation.
With the measure in \eqref{eq:PI_measure_normalization}, and rotating the integration contour such that the Gaussian integrals converge, we get%
\footnote{The branch of the square root depends on the number of negative modes and the prescription for how to rotate their integration contours.}
\begin{equation} \label{eq:bdy_det}
  \begin{aligned}
    & \int \DD_\ast' \delta\varphi_\mathrm{b}\+ \exp\Biggl[
      - \frac{\cL}{24\pi} \int_{\mathrlap{\partial D}}\ \vol_\ast\+ \delta\varphi_\mathrm{b} \bigl(n_\ast^\mu \partial_\mu - K_\ast\bigr) \delta\varphi_\mathrm{b}
      \Biggr]
    = \frac{1}{\sqrt{D_\mathrm{b}'(k)}}\+,
    \\
    & D_\mathrm{b}'(k) = \sideset{}{^{\+ \prime}}\prod_{n \in \mathbb{Z}} \frac{2\pi}{\Lambda_\mathrm{UV} \ell} \frac{n^2 - (1+k)^2}{\abs{n} + \frac{\ell^2}{2\pi A} - (1+k)} \+,
  \end{aligned}
\end{equation}
where the primes indicate that zero modes are not included.
This can be computed using zeta function regularization, as detailed in Appendix~\ref{app:bdy_det_zeta}.
The result, which agrees with \cite{Chaudhuri:2024yau}, is
\begin{equation} \label{eq:bdy_det_heavy}
  D_\mathrm{b}(k)
  = -\frac{2\pi}{(1+k)^2} \biggl(\frac{\Lambda_\mathrm{UV} \ell}{2\pi}\biggr)^{-2\bigl(\frac{\ell^2}{2\pi A} - (1+k)\bigr)} \biggl(\frac{\ell^2}{2\pi A} - (1+k)\biggr) \frac{\Gamma\bigl(\frac{\ell^2}{2\pi A} - (1+k)\bigr)^2}{\Gamma(1+k)^2\Gamma(-1-k)^2}\+,
\end{equation}
where the zero modes that appear at integer $k$ are visible through the poles of $\Gamma(-1-k)$.
Here, we have written $D_\mathrm{b}(k)$ in terms of the area $A$ and boundary length $\ell$.
When computing the one-point function, it should be rewritten in terms of the appropriate variables depending on ensemble and boundary conditions by using on-shell relations \eqref{eq:saddle_properties} and the saddle point equations \eqref{eq:geom_eom}.
We do this in §\ref{sec:ensembles}.

Determinants with zero modes removed can be obtained through limits of the above expression.
In particular,
\begin{equation} \label{eq:bdy_det_light}
  D_\mathrm{b}'(0)
  = \lim_{k\to 0} \frac{\Lambda_\mathrm{UV}^2 D_\mathrm{b}(k)}{\lambda_{\mathrm{b},1} \lambda_{\mathrm{b},-1}}
  = - \frac{\pi}{2} \Biggl(\frac{\Lambda_\mathrm{UV} \ell}{2\pi}\Biggr)^{-2\bigl(\frac{\ell^2}{2\pi A} - 2\bigr)} \frac{\Gamma\bigl(\frac{\ell^2}{2\pi A} + 1\bigr)^2}{\frac{\ell^2}{2\pi A} - 1},
\end{equation}
since $\lambda_{\mathrm{b},\pm 1}$ are the zero modes at $k = 0$.
The sign here is correct since it captures the negative $n = 0$ mode on the hyperbolic cap and the small spherical cap ($\ell^2 > 2\pi A$).

\subsubsection*{Zero mode Jacobian}
Lastly, we turn to the zero mode Jacobian for the light insertions.
To fix it, we compare \eqref{eq:varphi_b_Fourier} and \eqref{eq:varphi_b_solution} with the fluctuation \eqref{eq:delta_varphi_from_w} induced by an infinitesimal conformal transformation and match the linearized measure on the moduli space with the measure \eqref{eq:PI_measure_normalization} for the fluctuations.
Since the left-invariant measure $(1-\bar{w}w)^{-2} \dd^2 w$ on $\mathrm{PSL}(2, \mathbb{R})/\mathrm{U}(1)$ gives $\dd^2 \delta w$ when linearized around the origin $w = 0$, this gives
\begin{equation}
  \delta\varphi_{\mathrm{b},-1} = \delta w\+,\quad
  \delta\varphi_{\mathrm{b},+1} = \delta \bar{w}\+,\quad
  J \dd^2 \delta w = \frac{\cL \Lambda_\mathrm{UV} \ell}{12 \pi^2} \dd^2 \delta\varphi_{\mathrm{b},1}
  \quad\implies\quad J = \frac{\cL \Lambda_\mathrm{UV} \ell}{12 \pi^2}\+.
\end{equation}
Note that the sign of the Jacobian $J$ is negative for timelike Liouville theory.
This is due to the Gibbons--Hawking--Perry rotation implied by \eqref{eq:PI_measure_normalization} and the fact that the zero modes have to be rotated back to real contours to span the $\mathrm{PSL}(2, \mathbb{R})/\mathrm{U}(1)$ moduli space of saddles, consistent with the prescription in \cite{Polchinski:1988ua}.

Multiplying by the integral \eqref{eq:moduli_space_integral}, the contribution to the one-point function is
\begin{equation} \label{eq:moduli_integral_result}
  J I = \frac{\cL}{48 \pi^2} \frac{\Lambda_\mathrm{UV} \ell^3}{(2a-1)A} \biggl(\frac{2 A/(\ell \tilde{L})}{1 - z_0 \bar{z}_0}\biggr)^{2a}\+,
\end{equation}
which is independent of the arbitrary normalization of the measure on the coset space due to the Jacobian.

\subsection{Bulk fluctuations} \label{sec:bulk_fluct}
With the measure \eqref{eq:PI_measure_normalization}, again rotating the integration contour in the timelike case, we immediately see that
\begin{equation}
  \begin{aligned}
    & \int \DD_\ast \delta\varphi_\mathrm{B}\+ \exp\Biggl[
      - \frac{\cL}{24\pi} \int_{\mathrlap{M}}\ \vol_\ast \biggl(g_\ast^{\mu\nu} \partial_\mu \delta \varphi_\mathrm{B} \partial_\nu \delta \varphi_\mathrm{B} - R_\ast \delta\varphi_\mathrm{B}^2 \biggr)\Biggr]
    = \frac{1}{\sqrt{D_\mathrm{B}(k)}}\+,
    \\
    & D_\mathrm{B}(k) = \det \biggl( \frac{- \square_\ast - 2\eta/L^2}{\Lambda_\mathrm{UV}^2}\biggr) \+,
  \end{aligned}
\end{equation}
since $\delta\varphi_\mathrm{B}$ has homogeneous Dirichlet boundary conditions.
The eigenvalue problem related to the above determinant is
\begin{equation} \label{eq:bulk_Klein--Gordon}
  (-\square_\ast - 2\eta/L^2) \delta\varphi_\mathrm{B} = \lambda_\mathrm{B}\+ \delta\varphi_\mathrm{B}\+.
\end{equation}
Using coordinates \eqref{eq:Liouville_simple_saddle} and decomposing $\delta\varphi_\mathrm{B}$ into Fourier modes
\begin{equation}
  \delta\varphi_\mathrm{B} = \sum_{n \in \mathbb{Z}} \delta\varphi_{\mathrm{B},n}\+ e^{\ii n \theta/(1+k)}\+,
\end{equation}
the eigenvalue problem becomes
\begin{equation}
  - \frac{(1+\eta\rho^2)^2}{4} \biggl(\rho^{-1} \partial_\rho (\rho \partial_\rho) - \frac{n^2}{(1+k)^2} \rho^{-2}\biggr) \delta\varphi_{\mathrm{B},n} = \lambda\+ \delta\varphi_{\mathrm{B},n}\+,
\end{equation}
with $\lambda = L^2 \lambda_\mathrm{B} + 2 \eta$.
This is a singular Sturm--Liouville problem $-\partial_\rho(p\+ \partial_\rho \delta\varphi_\mathrm{B}) + q\+ \delta\varphi_\mathrm{B} = \lambda w\+ \delta\varphi_\mathrm{B}$ on $\rho \in (0, \gamma]$ with
\begin{equation} \label{eq:bulk_Sturm--Liouville}
  p = \rho\+,\qquad
  q = \frac{n^2}{(1+k)^2} \rho^{-1}\+,\qquad
  w = \frac{4 \rho}{(1 + \eta \rho^2)^2}\+.
\end{equation}
Through a change of variables, one can also write it as a hypergeometric differential equation
\begin{equation}
  z (1 - z) \partial_z^2 f_n + \Biggl(\frac{\abs{n}}{1+k} + 1 - 2 z\Biggr) \partial_z f + \eta \lambda f_n = 0\+,\quad
  z = \frac{\eta \rho^2}{1 + \eta \rho^2}\+,\
  \delta\varphi_{\mathrm{B},n} = \rho^{\abs{n}/(1+k)} f_n\+.
\end{equation}
Close to $\rho = 0$, the two independent solutions scale as $\delta\varphi_{\mathrm{B},n} \sim \rho^{\pm \abs{n}/(1+k)}$ for $n \neq 0$.
The plus branch is always square integrable but the minus branch only for $\abs{n} < 1+k$.
When $n = 0$, the two solutions behave as $\delta\varphi_{\mathrm{B},n} \sim \rho^0$ and $\delta\varphi_{\mathrm{B},n} \sim \log \rho$ close to the origin.
Hence, for $\abs{n} \geq 1+k$, the Sturm--Liouville operator is essentially self-adjoint ($\rho = 0$ is limit-point) whereas an additional boundary condition is needed at $\rho = 0$ when $\abs{n} < 1+k$ ($\rho = 0$ is limit-circle).

When $k < 0$, the boundary condition is ambiguous only in the $n = 0$ sector;
the $n \neq 0$ behaviour $\rho^{-\abs{n}/(1+k)} \delta\varphi_{\mathrm{B},n} \to 0$ as $\rho \to 0$ is implied by square-integrability.
When there is no conical singularity, $k=0$, the $n = 0$ solution $\sim \log \rho$ is discarded by demanding regularity of $\delta\varphi_\mathrm{B}(\rho,\theta)$ at the pole of the cap.
The above are the only boundary conditions that allow for the integration by parts we used to arrive at \eqref{eq:1-pt_at_one_loop}.%
\footnote{\label{footnote:inner_bdy_terms}%
  For this one needs $\vol_\ast\+ \delta\varphi_{(1)}\+ n_\ast^\mu \partial_\mu \delta\varphi_{(2)} \to 0$ as $\epsilon \to 0$, with $\delta\varphi_{(1),(2)}$ bulk or boundary fluctuations, whereas self-adjointness only requires this expression to vanish after antisymmetrization in $\delta\varphi_{(1),(2)}$.}
However, the fact that the conformal dimensions of the heavy operators need to receive one-loop corrections suggests that the correct boundary conditions might be more involved than this, particularly in the $n = 0$ sector.
Still, we expect the $n = 0$ boundary condition to reduce to the standard regularity condition in the $k \to 0$ limit.

For now, we focus on light insertions ($k = 0$) where the boundary conditions are well-understood, as discussed above.
The solution satisfying the boundary conditions is
\begin{equation} \label{eq:bulk_mode}
  \delta\varphi_{\mathrm{B},n} = \rho^{\abs{n}} \hypgeom\biggl(\Delta, 1-\Delta; 1 + \abs{n}; \frac{\eta \rho^2}{1+\eta\rho^2}\biggr)\+,\qquad
  \Delta = \frac{1}{2} + \sqrt{\frac{1}{4} + \eta \lambda}\+,
\end{equation}
and the eigenvalue equation determining the spectrum becomes
\begin{equation}
  \hypgeom\biggl(\frac{1}{2} + \sqrt{\frac{1}{4} + \eta \lambda}, \frac{1}{2} - \sqrt{\frac{1}{4} + \eta \lambda}; 1 + \abs{n}; \frac{\eta \gamma^2}{1+\eta\gamma^2}\biggr) = 0\+.
\end{equation}
On the hyperbolic cap, there are no negative or zero modes since $-\square_\ast$ and the effective mass $m^2 = -2\eta/L^2$ are both positive in that case.

On the spherical cap, the issue is more subtle since $m^2 = -2\eta/L^2$ is negative.
It was proven in \cite{Chaudhuri:2024rgn} that the eigenvalues of $-\square_\ast$ on a spherical cap are strictly decreasing functions of $\gamma$.
Hence, we can understand the number of negative and zero modes by studying the $\gamma \to \infty$ limit, in which the saddle geometry is a sphere with the north pole excised.

Homogeneous Dirichlet boundary conditions close to the north pole coincide with the regularity condition on the full sphere for $n \neq 0$ modes.
In the $\theta$-independent sector, the Dirichlet condition is different from regularity.
Still, one can show that the spectrum converges to that on the full sphere.
Hence, $L^2 \lambda_\mathrm{B} \to l(l+1) - 2$ and there is one negative mode and three zero modes on the full sphere.

On the spherical cap with finite $\gamma$, there is thus at most one negative or zero mode since the three zero modes on the full sphere are lifted.
We have already encountered this zero mode in §\ref{sec:bdy_fluct}, where we noted that the Klein--Gordon problem with $m^2 = -2\eta/L^2$ is ill-posed on the hemisphere due to the radial function $g_n(\rho)$ vanishing on the boundary.
By the monotonicity mentioned above, this implies that the lowest bulk mode is positive on small spherical caps, crosses zero and turns negative on large spherical caps.
As above, the integration contour for the negative mode has to be rotated for the Gaussian integral to converge, producing a phase $\pm \ii$.

When there is no conical singularity, the zeta-regularized fluctuation determinant is \cite{Chaudhuri:2024rgn}
\begin{equation} \label{eq:bulk_det}
  D_\mathrm{B}(0) = 8\pi \sqrt{2\pi} e^{\frac{59}{12} - 2\zeta'(-1)} \biggl(\frac{\Lambda_\mathrm{UV} \ell}{4\pi e^{4/3}}\biggr)^{\frac{\ell^2}{\pi A}} \biggl(\frac{\Lambda_\mathrm{UV} A}{\ell}\biggr)^{-1/3} \bigl(\Lambda_\mathrm{UV}^2 A\bigr)^{-2} \frac{\frac{\ell^2}{2\pi A} - 1}{\Gamma\bigl(\frac{\ell^2}{2\pi A}\bigr)^2}\+.
\end{equation}
The fact that the lowest mode is negative on large spherical caps is reflected here by the sign of $\ell^2/(2\pi A) - 1$.
We provide a consistency check of this result in the special case of a hemisphere in Appendix~\ref{app:hemisphere_det}.
In addition, we have verified numerically that the results of \cite{Chaudhuri:2024rgn} are consistent with the standard gluing properties of the path integral for a free massive scalar field.
It would be interesting to investigate whether the methods of \cite{Chaudhuri:2024rgn} can be extended to the case $k \neq 0$.

\subsubsection*{Summary of negative and zero modes}
We have seen that, in the spacelike theory, there is one negative boundary mode for every $n \in \mathbb{Z}$ with $\abs{n} < 1 + k$, and zero modes for $\abs{n} = 1 + k$ if $k$ is integer.
In particular, there are no negative bulk modes present in the spacelike case.
The modes that become light at integer $k$, and turn negative as $k$ is further increased, manifest as simple poles of the exact result of \cite{Fateev:2000ik,Teschner:2000md} at fixed $\ell$.

The timelike case is more subtle since the same boundary modes are negative, but only for sufficiently large $\ell^2/A$ of the saddle point geometry, and there can be negative bulk modes in addition to the boundary ones.
In particular, for a light insertion ($k = 0$), there is precisely one negative mode and two zero modes, the latter corresponding to conformal transformations of the disk.
The negative mode is a boundary mode when the spherical cap saddle is small and the lowest bulk mode when the cap is large.
This is similar to the situation on the sphere, where the $\varphi = \text{const}$ mode is negative \cite{Anninos:2021ene}.
On the disk, the negative mode has a radial profile but is still invariant under the $\mathrm{U}(1)$ isometry.

As we will see in §\ref{sec:ensembles}, there can also be additional phases from the Laplace transforms relating different boundary conditions and ensembles.
Having discussed the different pieces, we now proceed to assemble them into the one-point function.

\section{Wavefunctions at one loop and beyond} \label{sec:ensembles}
In this section, we put the various pieces discussed in the previous sections together, presenting the semiclassical wavefunction at one-loop order and comparing it with the all-loop expectations discussed in §\ref{sec:Liouville_wfn} based on \cite{Fateev:2000ik,Teschner:2000md,Anninos:2024iwf,Bautista:2021ogd}.
Thus far, we have focused on FZZT boundary conditions, specified by a boundary cosmological constant $\Lambda_\mathrm{b}$.
This is discussed further in §\ref{sec:fixed_K}.
In §\ref{sec:fixed_ell} we instead fix the physical length $\ell$ of the boundary, which is conjugate to $\Lambda_\mathrm{b}$.
Lastly, in §\ref{sec:fixed_A}, we fix not only $\ell$ but also the physical area of the bulk geometry $A$.

The different ensembles and boundary conditions are related by Laplace transforms since the cosmological constants $\Lambda$ and $\Lambda_\mathrm{b}$ act as chemical potentials for $A$ and $\ell$, respectively.
At tree level, the saddle point approximation of the integral transforms leads to Legendre transforms.
At the one-loop level, we must also include a one-loop factor from the quadratic expansion around the saddle point.

To make the discussion transparent, in this section we denote by $S_\mathrm{L}$ the Liouville action \emph{without} the cosmological constant terms $\Lambda \hat{A}$ and $\Lambda_\mathrm{b} \hat{\ell}$, in contrast to §\ref{sec:PI}.
When they appear, we write these terms explicitly.
Recall that $\hat{A}$ and $\hat{\ell}$ are the area and length operators, that is, they are computed using the physical metric and hence depend on $\varphi$.
Hence, the path integral in the fixed area ensemble, for instance, contains a delta function $\delta(A - \hat{A})$, as described in more detail below.
Moreover, we focus on the timelike theory but include some comments on the spacelike case, in particular in §\ref{sec:fixed_A}.

In addition, having already discussed the conformal dimensions in §\ref{sec:conf}, we focus on the wavefunctions $\Psi_\alpha$, defined in \eqref{eq:one-pt_wfn} by
\begin{equation} \label{eq:one-pt_wfn_again}
  \langle V_\alpha(z_0) \rangle = \frac{\Psi_\alpha}{(1 - \bar{z}_0 z_0)^{2 \Delta_\alpha}}\+.
\end{equation}
Recall that this timelike Liouville wavefunction has been studied before in \cite{Bautista:2021ogd} by bootstrap methods and in \cite{Anninos:2024iwf} from a Wheeler--DeWitt (WDW) and saddle point perspective, as discussed in §\ref{sec:Liouville_wfn}.
Here, we extend the semiclassical analysis beyond the saddle point level and ensure that conformal invariance is preserved by the boundary conditions, see §\ref{sec:Liouville_wfn} and §\ref{sec:conf}.
Recall also that this wavefunction captures the gravitational part of a gravitational path integral with a diffeomorphism invariant insertion \eqref{eq:wfu}, and arises by fixing the Weyl gauge, as described in §\ref{sec:gravity_and_Liouville}.

\subsection{Fixed trace of the extrinsic curvature} \label{sec:fixed_K}
With the notation described above, the timelike Liouville one-point function we are led to consider by gauge-fixing the gravitational path integral is, with fixed cosmological constants,
\begin{equation}
  \langle V_a \rangle_{\Lambda, \Lambda_\mathrm{b}} = \int \DD\varphi\+ e^{-S_\mathrm{L} - \Lambda \hat{A} - \Lambda_\mathrm{b} \hat{\ell}} V_a\+.
\end{equation}
Recall from §\ref{sec:saddle} that there is a single real saddle point geometry contributing to this path integral.

Since we scale $\Lambda$ and $\Lambda_\mathrm{b}$ with $\cL$, it is convenient to introduce parameters that remain finite in the semiclassical limit.
To this end, we define
\begin{equation} \label{eq:L_K_defs}
  L \equiv \sqrt{\frac{\abs{\cL}}{24\pi\Lambda}}\+,\qquad\qquad
  K \equiv - \frac{12\pi \Lambda_\mathrm{b}}{\cL}\+.
\end{equation}
At the level of the saddle point geometry, these are the radius of the spherical cap and the trace of the extrinsic curvature of the boundary, respectively.
Note that $K$, even though not expressible in terms of $\varphi$, is the proper definition of the trace of the extrinsic curvature at the quantum level, analogous to how the momentum $p$ in quantum mechanics is not related to the velocity $\dot{x}$ of an off-shell path.

Rewriting \eqref{eq:tree-level_centred_insertion} in terms of the parameters above by using \eqref{eq:saddle_properties}, we find the tree-level result
\begin{equation} \label{eq:Psi_heavy_tree}
  \Psi_a(K)
  = (\Lambda_\mathrm{UV} \tilde{L})^{\frac{\cL}{12} k^2} \biggl(\frac{2(1+k) L}{\tilde{L}} \gamma_K\biggr)^{-\frac{\cL}{6}(1+k)} e^{\frac{\cL}{6}(1+k) + \mathcal{O}(\log\cL)}\+.
\end{equation}
Here,
\begin{equation} \label{eq:gamma_K_def}
  \gamma_K \equiv -L K + \sqrt{1 + L^2  K^2}\+,
\end{equation}
which corresponds to the parameter $\gamma$ of the saddle point, see \eqref{eq:saddle_gamma}.
Recall that we focus on timelike Liouville theory ($\eta = +1$), $\tilde{L}$ is the radius of the fiducial background disk \eqref{eq:varphi_saddle} and $k = -12 a/\cL$ is finite when the insertion is heavy while vanishing when it is light.
Note that the branch points at $L K = \pm \ii$ are associated with $\mathcal{I}^\pm$ \textdash the past and future asymptotic boundaries of the Lorentzian $\mathrm{dS}_2$ slice of the complexified Hartle--Hawking saddle.

The power of $L$ can be derived from the formulation \eqref{eq:canonical_Liouville} of timelike Liouville theory with a background-dependent path integral measure, following \cite{David:1988hj,Distler:1988jt}.
It follows from the shift-invariance of this path integral measure that $\langle V_\alpha \rangle_{C L, LK} = C^{(q + 2\alpha)/\beta} \langle V_\alpha \rangle_{L, LK}$, where we have used that the Euler characteristic of the disk is $\chi = 1$ and view the one-point function as depending on the length scale $L$ and the dimensionless combination $LK$.
This implies that $\langle V_\alpha \rangle_{L, LK} = L^{(q + 2\alpha)/\beta} f(LK)$ for some function $f$.
The power of $\tilde{L}$, on the other hand, should be $\cL/6 - 2\Delta_\alpha$, due to conformal invariance.
Recalling that $\Delta_\alpha = \alpha (q+\alpha)$, $a = \beta^{-1} \alpha$, $\cL = 1 - 6 q^2$ and $q = \beta^{-1} - \beta$, we have
\begin{equation} \label{eq:heavy_tree_parameter_relation}
  \frac{\cL}{6} - 2\alpha(q+\alpha) = \frac{\cL}{12} (2 + 2k + k^2) + \mathcal{O}(\beta^0)\+,\qquad
  \frac{q + 2\alpha}{\beta} = -\frac{\cL}{6} (1 + k) + \mathcal{O}(\beta^0)\+.
\end{equation}
Hence, the powers of $\tilde{L}$ and $L$ in \eqref{eq:Psi_heavy_tree} are consistent with conformal invariance and the shift-invariance of the path integral measure in \eqref{eq:canonical_Liouville}.
Note that, at the one-loop level, additional factors $L^{7/6}$ and $\tilde{L}^{(12+13k)k/12}$ are needed.
We comment on the dependence on $K$ after discussing the one-loop result with a light insertion.

Specializing to a light insertion to avoid the issues with the boundary conditions at the insertion discussed in §\ref{sec:bulk_fluct}, we get the result
\begin{equation} \label{eq:Psi_K_light_one-loop}
  \Psi_a(K) = \mp \ii \abs{\cL} \scheme(\gamma_K) \frac{(\Lambda_\mathrm{UV} \tilde{L})^{\frac{7}{6}}}{2a - 1} \biggl(\frac{2 L}{\tilde{L}} \gamma_K\biggr)^{-\frac{\cL}{6} + \frac{7}{6} + 2a} e^{\frac{\cL}{6}} [1 + \mathcal{O}(1/\cL)]\+,
\end{equation}
up to one-loop order, by using \eqref{eq:bdy_det_light}, \eqref{eq:moduli_integral_result}, and \eqref{eq:bulk_det} with the prescription that a negative mode contributes a phase $\pm \ii$.
Here, $\cL/(2a-1)$ comes from the zero mode sector, $(\Lambda_\mathrm{UV} \tilde{L})^{7/6}$ is the combined logarithmic divergence of the bulk and boundary fluctuation determinants and $\scheme(\gamma_K)$ is the positive, dimensionless, scheme-dependent factor in \eqref{eq:scheme_factor}.
The result is remarkably simple due to several cancellations between the bulk and boundary determinants.
Note that the powers of $L$ and $\tilde{L}$ are consistent with the considerations above, now at the one-loop level, since $\Delta_a^\textnormal{1-loop} = a$ and
\begin{equation} \label{eq:light_1-loop_parameter_relation}
  \frac{q + 2\alpha}{\beta} = -\frac{\cL}{6} + \frac{7}{6} + 2a + \mathcal{O}(\beta^2)\+,
\end{equation}
with $a = \mathcal{O}(\beta^0)$ for a light insertion.

In the zeta-regularization scheme that we are using,
\begin{equation} \label{eq:scheme_factor}
  \scheme(\gamma) = \frac{e^{-\frac{59}{24} + \zeta'(-1)}}{24 \cdot 2^{5/12} \pi^{1/4}} (2 e^{4/3})^\frac{2}{1+\gamma^2}\+,
\end{equation}
which follows from \eqref{eq:bdy_det_light}, \eqref{eq:moduli_integral_result} and \eqref{eq:bulk_det}.
This factor can be absorbed by renormalizing the cosmological constants, $\Lambda \to \Lambda (1 + \frac{6}{\cL} \delta_\Lambda)$ and $\Lambda_\mathrm{b} \to \Lambda_\mathrm{b} (1 + \frac{6}{\cL} \delta_{\Lambda_\mathrm{b}})$, producing the additional factor
\begin{equation}
  e^{-\frac{6}{\cL} (\Lambda A\+ \delta_\Lambda + \Lambda_\mathrm{b} \ell\+ \delta_{\Lambda_\mathrm{b}})} = e^{(\delta_{\Lambda_\mathrm{b}} - \delta_\Lambda) + \frac{1}{1+\gamma_K^2} (\delta_\Lambda - 2\delta_{\Lambda_\mathrm{b}})}\+,
\end{equation}
when evaluated on the saddle point.
Equivalently, rescaling $L \to L(1 + \frac{6}{\cL}\delta_L)$ in \eqref{eq:Psi_K_light_one-loop} while keeping $LK$ fixed produces a factor $e^{\delta_L}$, whereas $LK \to LK (1 + \frac{6}{\cL} \delta_{LK})$ gives a factor $e^{[-1 + 2/(1+\gamma_K^2)] \delta_{LK}}$, at leading order as $\cL \to -\infty$.
Demanding that this renormalization is independent of $a$ and subleading in $\cL$ explains why we have not included the factors $\cL e^{\cL/6}/(2a-1)$ in $\scheme$.
One may, of course, allow for an $a$-dependent finite renormalization of the vertex operators, but that is not necessary to produce finite results.%
\footnote{Here, we only discuss finite counterterms.
  In addition, there are divergent ones that are implicitly taken care of by the zeta function regularization.}
As we will see below, the counterterms do not spoil the Bessel function structure of the wavefunction in the length basis.

From \cite{Fateev:2000ik,Teschner:2000md,Anninos:2024iwf,Bautista:2021ogd} and the discussion in §\ref{sec:Liouville_wfn}, we expect
\begin{equation} \label{eq:fixed_K_expectation}
  \Psi_\alpha(K) \propto \gamma_K^{\pm(q+2\alpha)/\beta}\+,
\end{equation}
at the nonperturbative level, after appropriate renormalizations of $L$ and $K$.
From \eqref{eq:heavy_tree_parameter_relation} and \eqref{eq:light_1-loop_parameter_relation}, we see that \eqref{eq:Psi_heavy_tree} and \eqref{eq:Psi_K_light_one-loop} agree with \eqref{eq:fixed_K_expectation} if we choose the upper sign in the exponent of the latter.
This wavefunction is related, by a Laplace transform, to the Hartle--Hawking-like wavefunction
\begin{equation} \label{eq:Bessel_arg_prop}
  J_{(q+2\alpha)/\beta}(C \ell/L)\+,\qquad
  C = \frac{\beta^{-2}}{2\pi} - \frac{13}{24\pi} + \mathcal{O}(\beta^2)\+,
\end{equation}
as explained in §\ref{sec:fixed_ell} below.
The other, Vilenkin-like $J_{-(q+2\alpha)/\beta}(C \ell/L)$ can formally be obtained by $\alpha \to -q-\alpha$, although our one-loop analysis is not valid in that regime.
From the path integral, there seems to be no reason to take a linear combination of the two Bessel functions, in contrast to what is suggested in \cite{Bautista:2021ogd} (cf.\ \eqref{eq:Psi_linear_combination} above).
In addition, it is not clear how the phases multiplying the Bessel functions in \eqref{eq:Psi_linear_combination} would appear.%
\footnote{If we include complex saddles with $\varphi_\ast \to \varphi_\ast + 2\pi\ii n$, as in footnote~\ref{footnote:complex_saddle}, each saddle comes with a factor $\exp[2\pi\ii n(-\frac{\cL}{6}+2a)]$ but it is not clear how this could produce the phases in \eqref{eq:Psi_linear_combination}.}
Still, \eqref{eq:Psi_K_light_one-loop} is consistent with the WDW analysis of \cite{Anninos:2024iwf} and naturally selects the Hartle--Hawking-like wavefunction.

\subsection{Fixed boundary length} \label{sec:fixed_ell}
The ensemble with two cosmological constants is related to the one with fixed $\Lambda$ and boundary length $\ell$ by
\begin{equation} \label{eq:K_Laplace_transf}
  \langle V_a \rangle_{\Lambda, \Lambda_\mathrm{b}} = \int_{\mathcal{C}_\ell} \frac{\dd \ell}{\ell}\+ e^{-\Lambda_\mathrm{b} \ell} \langle V_a \rangle_{\Lambda, \ell}\+,
\end{equation}
or, inversely,
\begin{equation} \label{eq:fixed_length_inv_Laplace}
  \langle V_a \rangle_{\Lambda, \ell} = \int \DD \varphi\+ e^{-S_\mathrm{L} - \Lambda \hat{A}} \delta(1 - \hat{\ell}/\ell) V_a
  = \int_{\mathcal{C}_{\Lambda_\mathrm{b}}}\! \frac{\ell \dd \Lambda_\mathrm{b}}{2\pi\ii}\+ e^{\Lambda_\mathrm{b} \ell} \langle V_a \rangle_{\Lambda, \Lambda_\mathrm{b}}\+.
\end{equation}
To represent the delta function, the $\Lambda_\mathrm{b}$ integral should run over a contour $\mathcal{C}_{\Lambda_\mathrm{b}}$ parallel to the imaginary axis, whereas the naive contour for $\ell$ is $\mathcal{C}_\ell = \mathbb{R}_+$.
We will comment on the viability of these contours below.

With these boundary conditions, the expectation is that the dependence on $\ell$ is captured by
\begin{equation} \label{eq:fixed_ell_expectation}
  \Psi_\alpha(\ell) \propto J_{\pm(q+2\alpha)/\beta}\Bigl(\sqrt{\Lambda/\sin(\pi\beta^2)}\+ \ell\Bigr)
  = J_{\pm(q+2\alpha)/\beta}(C \ell/L)\+,
\end{equation}
or a linear combination of the two, as discussed in §\ref{sec:Liouville_wfn}.
Here, $C$ is given in \eqref{eq:Bessel_arg_prop} but the subleading terms in that equation are scheme dependent since they are sensitive to rescalings of $\Lambda$.
Explicitly performing the Laplace transform above,
\begin{equation} \label{eq:Bessel_J_Laplace}
  \begin{aligned}[b]
    & \Psi_\alpha(K) = \int_{0}^{\infty} \frac{\dd \ell}{\ell} e^{-\Lambda_\mathrm{b} \ell} \Psi_\alpha(\ell)
    \propto \pm \frac{\beta}{q+2\alpha} \biggl(-\tilde{C} L K + \sqrt{1 + \tilde{C}^2 L^2 K^2}\+\biggr)^{\pm (q+2\alpha)/\beta}\+,\\
    & \tilde{C} = 1 - \frac{13}{12} \beta^2 + \mathcal{O}(\beta^4)\+.
  \end{aligned}
\end{equation}
Recalling the definition of $\gamma_K$ in \eqref{eq:Psi_heavy_tree}, we see that this is indeed \eqref{eq:fixed_K_expectation}, up to the overall normalization and a rescaling of $L$.
Choosing $\alpha$ such that $\Re(q + 2\alpha) \geq 0$, the integral converges only for the upper sign, i.e.\ the Hartle--Hawking-like wavefunction, and when, in addition, $\Re(K) \geq 0$.
The result may be continued analytically to other regions.

\subsubsection*{Semiclassical analysis}
One readily verifies that the saddle point equations for $\langle V_a \rangle_{\Lambda, \ell}$ are the same as before, see \eqref{eq:geom_eom}, but with $\Lambda_\mathrm{b}$ replaced by the saddle point value $\Lambda_{\mathrm{b}\ast}$, viewed as a Lagrange multiplier, and supplemented by the constraint $\hat{\ell} = \ell$.
This, together with the additional factor $e^{\Lambda_{\mathrm{b}\ast} \ell}$ in \eqref{eq:fixed_length_inv_Laplace}, implements the Legendre transform and leads to
\begin{equation} \label{eq:Psi_ell_heavy_tree}
  \Psi_a(\ell)
  = (\Lambda_\mathrm{UV} \tilde{L})^{\frac{\cL}{12} k^2} \biggl[\frac{2(1+k) L}{\tilde{L}} \gamma_\ell\biggr]^{-\frac{\cL}{6}(1+k)} e^{\frac{\cL}{6} (1+k) \bigl(1 \mp \sqrt{1-\ell^{\prime\+ 2}}\bigr) + \mathcal{O}(\log\cL)}\+,
\end{equation}
where
\begin{equation} \label{eq:gamma_ell}
  \gamma_\ell \equiv \ell^{\prime\+ -1} \mp \sqrt{\ell^{\prime\+ -2} - 1}\+,\qquad\quad
  \ell' \equiv \frac{\ell}{2\pi(1+k)L}\+.
\end{equation}
Note that $\gamma_\ell$ agrees with the parameter $\gamma$ from \eqref{eq:saddle_gamma} at the saddle point level.
The upper and lower signs correspond to the small and large spherical caps of boundary length $\ell$, respectively.
The saddle points are real for $\ell \leq 2\pi(1+k) L$, corresponding to $\ell' \leq 1$.
The powers of $L$ and $\tilde{L}$ are the same as in §\ref{sec:fixed_K} and the discussion there need not be repeated.
We compare the result to the Bessel functions after writing the one-loop result for a light insertion.

In the semiclassical path integral, there is an additional factor
\begin{equation} \label{eq:deltaLambda_b}
  \exp\biggl[- \delta\Lambda_\mathrm{b} \int_{\mathrlap{\partial D}}\ \vol_\ast (e^{\delta\varphi} - 1)\biggr]\+,
\end{equation}
as compared to §\ref{sec:one-loop_PI}.
At one loop, this leads to mixing between $\delta\Lambda_\mathrm{b}$ and the constant boundary mode $\delta\varphi_{\mathrm{b}, 0}$.
Taken together, from \eqref{eq:1-pt_at_one_loop}, \eqref{eq:lambda_b} and \eqref{eq:deltaLambda_b}, the quadratic action for these fluctuations is
\begin{equation}
  \ell \delta\Lambda_\mathrm{b} \delta\varphi_{\mathrm{b},0} + \frac{\cL}{24\pi} \ell \lambda_{\mathrm{b},0} \delta\varphi_{\mathrm{b},0}^2\+.
\end{equation}
Hence, the fluctuations can be unmixed by shifting
\begin{equation} \label{eq:Lambda_ell_bdy_shift}
  \delta\varphi_{\mathrm{b},0} \quad\longrightarrow\quad \delta\varphi_{\mathrm{b},0} - \frac{12\pi}{\cL} \lambda_{\mathrm{b},0}^{-1} \delta \Lambda_\mathrm{b}\+.
\end{equation}

The shift leaves the term quadratic in $\delta\varphi$ the same, removes the mixed term $\delta\Lambda_\mathrm{b} \delta\varphi_{\mathrm{b},0}$ and produces a new term proportional to $\delta\Lambda_\mathrm{b}^2$.
Hence, the fluctuation analysis in §\ref{sec:fluct} goes through and there is an additional factor
\begin{equation} \label{eq:fixed_ell_inv_Laplace}
  \int \frac{\ell \dd\delta\Lambda_\mathrm{b}}{2\pi\ii} e^{\frac{6\pi}{\cL} \ell \lambda_{\mathrm{b},0}^{-1} \delta\Lambda_\mathrm{b}^2}
  = \sqrt{\frac{-\cL}{\pm 12\pi \sqrt{1 - \ell^{\prime\+ 2}}}}\+,
\end{equation}
contributing to $\langle V_a \rangle_{\Lambda, \ell}$.
Here, we consider a light insertion to avoid ambiguities at the insertion point, the $\pm$ sign is correlated with \eqref{eq:gamma_ell} and we have made use of \eqref{eq:lambda_b}.
Note that the region that contributes to the integral is $\delta\Lambda_\mathrm{b} = \mathcal{O}(\sqrt{\cL})$, which validates the saddle point approximation of the inverse Laplace transform since $\mathcal{O}(\delta\varphi) = \mathcal{O}(1/\sqrt{\cL})$.
As mentioned above, the standard integration contour for $\Lambda_\mathrm{b}$ runs parallel to the imaginary axis.
In the timelike theory, the integral converges on that contour only when the saddle is a small spherical cap ($\ell^2 > 2\pi A$).
When the saddle is a large spherical cap, the contour has to be rotated for the integral to converge.

Around a small spherical cap, the expression in \eqref{eq:fixed_ell_inv_Laplace} is positive, whereas, around a large spherical cap, the branch of the square root should be chosen such that the phase of the expression is the same as that contributed by a negative mode, $\pm\ii$.
This follows from using the standard orientation of the real line for convergent, Gaussian integrals of $\delta\varphi_{\mathrm{b},0}$.%
\footnote{\label{footnote:Laplace_contour_rotation}%
  Using a measure $\dd \mu(x)$ proportional to the Lebesgue measure but normalized such that $\int \dd \mu(x)\+ e^{x^2/2} = 1$ on a contour parallel to the imaginary axis (cf.\ \eqref{eq:PI_measure_normalization}), we have, along a real $x$-contour, $\pm \ii/\sqrt{2\pi} = \int \dd \mu(x)\+ \delta(x) e^{-x^2/2} = \int \dd \mu(x)\+ \frac{\dd y}{2\pi\ii} e^{-x^2/2 + yx} = \pm\ii \int\frac{\dd y}{2\pi\ii} e^{y^2/2}$, where we have shifted $x \to x+y$ to unmix the integrals.
  This is, schematically, what happens for the timelike theory around a small cap, with $x \propto \delta\varphi_{\mathrm{b},0}$ and a positive constant of proportionality.
  Note that $\delta(1-\hat{\ell}/\ell) = \delta(\delta\varphi_{\mathrm{b},0})$ at the one-loop level, that $x$ is a negative mode in the first integral (the sign is different from the one in the normalization condition) and hence contributes a phase $\pm\ii$, and that we have assumed that the contour is rotated back to $\mathbb{R}$ with its standard left-to-right orientation in $\int \dd \mu(x)\+ e^{-x^2/2}$.
  Analogous arguments can be performed in the case of a large cap and the spacelike theory.
}
If we were to, in addition, fix the orientation of the $\delta\Lambda_\mathrm{b}$ integral in the cases this is over the real line, that would fix the phase contributed by a negative mode.
The positive orientation results in a negative mode contributing $-\ii$ in timelike Liouville theory and $+\ii$ in spacelike Liouville theory.
That the phases are opposite is related to the fact that negative modes have to be rotated \emph{back} from an imaginary contour to a real one in the timelike case.

Putting the above together, the one-loop result for the wavefunction in timelike Liouville is
\begin{equation} \label{eq:Psi_ell_light_one-loop}
  \Psi_a(\ell) = -(\pm \ii)^{(3 \mp 1)/2} \frac{\abs{\cL}^{3/2}}{\sqrt{12\pi}} \scheme(\gamma_\ell) \frac{(\Lambda_\mathrm{UV} \tilde{L})^{\frac{7}{6}}}{2a - 1} \biggl(\frac{2 L}{\tilde{L}} \gamma_\ell\biggr)^{-\frac{\cL}{6} + \frac{7}{6} + 2a} \frac{e^{\frac{\cL}{6} \bigl(1 \mp \sqrt{1-\ell^{\prime\+ 2}}\bigr)}}{(1-\ell^{\prime\+ 2})^{1/4}} [1 + \mathcal{O}(1/\cL)]\+,
\end{equation}
where $\scheme(\gamma)$ is the scheme-dependent factor defined in \eqref{eq:scheme_factor}, we again use the prescription that a negative mode contributes a phase $\pm\ii$ and the $\mp$ signs in the exponents correspond to the small and large cap, as above.
Both of these saddles give rise to wavefunctions that solve the WDW equation.
Note that we focus on the regime $\ell' < 1$ where the saddles are real and the Hartle--Hawking-like wavefunction is captured by the small cap saddle.
When $\ell' = 1$, the saddles are degenerate and describe a hemisphere.
Here, the one-loop analysis breaks down, as seen from the diverging $(1-\ell^{\prime\+ 2})^{-1/4}$, stemming from \eqref{eq:fixed_ell_inv_Laplace}.
This is not inherent to the hemisphere but, rather, specific to the present ensemble; the issue with $\ell' = 1$, including the undetermined phase in \eqref{eq:Psi_ell_light_one-loop}, is not present when fixing the trace of the extrinsic curvature, as in \eqref{eq:Psi_K_light_one-loop}, or the bulk area, as in \eqref{eq:PsiA_one-loop}.

\subsubsection*{Comparison with Bessel functions}
To compare the above with the Bessel functions in \eqref{eq:fixed_ell_expectation} we use the asymptotic expansion \dlmfeqcite{10.19.3}
\begin{equation} \label{eq:Bessel_J_basic_asympt}
  J_\nu(\nu x) \sim \frac{(2\pi \nu)^{-1/2}}{(1 - x^2)^{1/4}} e^{\nu \bigl[\sqrt{1-x^2} - \log\bigl(x^{-1} + \sqrt{x^{-2} - 1}\bigr)\bigr]} [1 + \mathcal{O}(\nu^{-1})]\+,
\end{equation}
as $\nu \to +\infty$, valid for $0 < x < 1$.
From this, it follows that the Hartle--Hawking-like Bessel function in \eqref{eq:fixed_ell_expectation} has the asymptotic expansions
\begin{subequations}
  \begin{align}
    \label{eq:light_Bessel_asympt}
    & J_{(q+2\alpha)/\beta}(C \ell/L) \sim
    \frac{\beta}{\sqrt{2\pi}} \gamma_\ell^{\beta^{-2} + (2a-1)} \frac{e^{(\beta^{-2}-\frac{13}{12}) \sqrt{1-\ell^{\prime\+ 2}}}}{(1-\ell^{\prime\+ 2})^{1/4}} [1 + \mathcal{O}(\beta^2)]\+,
    \\
    & J_{(q+2\alpha)/\beta}(C \ell/L) \sim
    \frac{\beta}{\sqrt{2\pi(1+k)}} \gamma_\ell^{\frac{1+k}{\beta^2} - \frac{6+13k}{6}} \frac{e^{(\beta^{-2}-\frac{13}{12}) (1+k) \sqrt{1-\ell^{\prime\+ 2}}}}{(1-\ell^{\prime\+ 2})^{1/4}} [1 + \mathcal{O}(\beta^2)]\+,
  \end{align}
\end{subequations}
for light ($\alpha = \beta a$, $a$ finite) and heavy ($a = -\cL k/12$, $k$ finite) insertions, respectively, with $\ell'$ defined in \eqref{eq:gamma_ell} and $C$ given in \eqref{eq:Bessel_arg_prop}.
We see immediately that the $\ell$-dependence agrees with \eqref{eq:Psi_ell_heavy_tree} at tree level, i.e.\ keeping only the $\mathcal{O}(\beta^{-2})$ terms in the exponent, provided we take the upper sign in the exponent in \eqref{eq:Psi_ell_heavy_tree}, corresponding to a small spherical cap saddle.
The Hartle--Hawking-like wavefunction being captured by a small spherical cap is consistent with the higher-dimensional picture \cite{Hartle:1983ai}.

To compare the one-loop result for the light insertion \eqref{eq:Psi_ell_light_one-loop} with the asymptotic expansion of the Bessel function, we again need to consider the scheme dependence.
Rescaling $\Lambda \to \Lambda(1 + \frac{6}{\cL} \delta_\Lambda)$ and $\ell \to \ell (1 + \frac{6}{\cL} \delta_\ell)$ the counterterms evaluate to
\begin{equation}
  \frac{6}{\cL} \Lambda A\+ \delta_\Lambda = \frac{1}{2} \Bigl(-1 + \sqrt{1-\ell^{\prime\+ 2}}\Bigr) \delta_\Lambda\+,\qquad\quad
  \frac{6}{\cL} \Lambda_\mathrm{b} \ell\+ \delta_{\ell} = - \sqrt{1 - \ell^{\prime\+ 2}}\+ \delta_{\ell}\+,
\end{equation}
with on-shell values for $A$ and $\Lambda_\mathrm{b}$.
Hence, the overall normalization and the $\mathcal{O}(\beta^0)$ coefficient multiplying $\sqrt{1-\ell^{\prime\+ 2}}$ in the exponent in \eqref{eq:light_Bessel_asympt} are scheme dependent.
Taking this into account, we see that \eqref{eq:Psi_ell_light_one-loop} agrees with \eqref{eq:light_Bessel_asympt} due to the relation \eqref{eq:light_1-loop_parameter_relation} since $\beta^{-2} + (2a-1) = (q+2\alpha)/\beta$.
In addition, the relative normalization of $\Psi_a(K)$ in \eqref{eq:Psi_K_light_one-loop} and $\Psi_a(\ell)$ in \eqref{eq:Psi_ell_light_one-loop} agree, as seen from \eqref{eq:Bessel_J_Laplace} and \eqref{eq:light_Bessel_asympt}.
Note that the counterterms do not alter the Bessel function structure but merely rescale the wavefunction and its argument at subleading order in $\beta$.
In particular, from the WDW perspective, it does not affect the operator ordering.
It would be interesting to investigate this beyond the one-loop order.

Thus far, in this section, we have focused on the small spherical cap saddle.
The asymptotic expansion of $J_{-\nu}(\nu x) = \cos(\pi \nu) J_\nu(\nu x) - \sin(\pi \nu) Y_\nu(\nu x)$, with $\nu > 0$, is rapidly oscillating due to the trigonometric $\cos(\pi \nu)$ and $\sin(\pi \nu)$, which are not present in our semiclassical analysis.
However, \dlmfeqcite{10.19.3}
\begin{equation}
  Y_\nu(\nu x) \sim -\frac{(\pi \nu/2)^{-1/2}}{(1 - x^2)^{1/4}} e^{-\nu \bigl[\sqrt{1-x^2} - \log\bigl(x^{-1} + \sqrt{x^{-2} - 1}\bigr)\bigr]} [1 + \mathcal{O}(\nu^{-1})]\+,
\end{equation}
which is identical to the asymptotic expansion \eqref{eq:Bessel_J_basic_asympt} up to a factor $-2$ and, crucially, the overall sign of the exponent.
This is precisely what we expect from the semiclassical analysis since $\gamma_\ell^\textnormal{(large cap)} = 1/\gamma_\ell^\textnormal{(small cap)}$, see e.g.\ \eqref{eq:Psi_ell_heavy_tree} where the signs of the exponents of $e^{\sqrt{1-\ell^{\prime\+ 2}}}$ are opposite but those of $\gamma_\ell$ the same.
Hence, the two saddle points generate the two linearly independent solutions to the WDW equation.
Semiclassically, the only difference between the two wavefunctions is which $\Lambda_{\mathrm{b}\ast}$ saddle contributes.
Correspondingly, we expect there to be an alternative integration contour for the inverse Laplace transform \eqref{eq:fixed_length_inv_Laplace} that produces a Vilenkin-like wavefunction.
Note that this is different from the analytic continuation $\alpha \to -q-\alpha$ discussed below \eqref{eq:Bessel_J_Laplace} since, here, the $K$-representation is the same for both wavefunctions.
It would be interesting to pursue the two-loop calculation to verify that the $\mathcal{O}(\beta^2)$ contributions around the two saddles come with opposite signs, as predicted by the asymptotic expansion of the Bessel functions.
It would also be interesting to understand whether there exists a physical principle that selects a particular integration contour, and how such a prescription would compare with the bootstrap-based proposal of \cite{Bautista:2021ogd}, which effectively leads to a specific linear combination of the two WDW solutions.
We leave these questions to future work.

The asymptotic expansions above are valid in the regime where the saddles are real ($\ell' \leq 1$).
Notably, the power-law decay of the Hartle--Hawking-like wavefunction as $\ell \to 0$ is not only a feature at the saddle point level but persists at the full quantum level, although the power receives quantum corrections.
This follows immediately from the small $z$ expansion $J_\nu(z) \sim z^\nu$ of the Bessel function.
When $\ell' \geq 1$, the Hartle--Hawking-like wavefunction $J_{+(q+2\alpha)/\beta}$ is still real but oscillating.
At the saddle point level, it receives contributions from two complex-conjugate saddle geometries, as familiar from higher dimensions.
To leading order at large volume and in the semiclassical expansion, the oscillatory behaviour is captured by $e^{\ii \cL \ell/(12\pi L)}$, analogous to the $(d+1)$-dimensional result where the exponent is proportional to $\ii \Vol_d/(G_N L)$.

\subsection{Fixed area and boundary length} \label{sec:fixed_A}
Fixing not only the boundary length but also the area of the bulk, the one-point functions are related by
\begin{equation} \label{eq:Laplace_Lambda_A}
  \langle V_a \rangle_{\Lambda, \ell} = \int_{\mathcal{C}_A} \frac{\dd A}{A} e^{-\Lambda A} \langle V_a \rangle_{A, \ell}\+,
\end{equation}
or
\begin{equation}
  \langle V_a \rangle_{A, \ell} = \int \DD\varphi\+ e^{-S_\mathrm{L}} \delta(1 - \hat{A}/A) \delta(1 - \hat{\ell}/\ell) V_a
  = \int_{\mathcal{C}_\Lambda} \frac{A \dd\Lambda}{2\pi\ii} \int_{\mathcal{C}_{\Lambda_\mathrm{b}}}\! \frac{\ell \dd\Lambda_\mathrm{b}}{2\pi\ii}\+ e^{\Lambda A + \Lambda_\mathrm{b} \ell} \langle V_a \rangle_{\Lambda, \Lambda_\mathrm{b}}\+.
\end{equation}
Below, we present first the semiclassical analysis and then discuss the expected exact result and integration contours.

\subsubsection*{Semiclassical analysis}
At the saddle point level, the Legendre transform simply removes the $\Lambda A$ and $\Lambda_\mathrm{b} \ell$ terms in \eqref{eq:tree-level_centred_insertion}.
The result is
\begin{equation} \label{eq:PsiA_tree}
  \Psi_a^{(A)}(\ell) = (\Lambda_\mathrm{UV} \tilde{L})^{\frac{\cL}{12} k^2} \biggl(\frac{2(1+k) A}{\tilde{L} \ell}\biggr)^{-\frac{\cL}{6}(1+k)} e^{\frac{\cL}{6}\bigl(1+k - \frac{\ell^2}{4\pi A}\bigr) + \mathcal{O}(\log\cL)}\+,
\end{equation}
where we use a superscript to indicate that the area is fixed and have, again, stripped off the kinematic dependence on the insertion point to extract the wavefunction, as in \eqref{eq:one-pt_wfn_again}.
Note that, in contrast to §\ref{sec:fixed_ell}, there is only one real saddle geometry.
The power of $\tilde{L}$ is the same as in §\ref{sec:fixed_K} and hence consistent with conformal invariance.

At the one-loop level, in addition to \eqref{eq:deltaLambda_b}, there is now an additional factor
\begin{equation}
  \exp\biggl[- \delta\Lambda \int_{\mathrlap{D}}\ \vol_\ast (e^{2 \delta\varphi} - 1)\biggr]\+,
\end{equation}
in the semiclassical path integral.
Unmixing $\delta\Lambda$ and $\delta\varphi$ at the quadratic level now requires a constant shift of $\delta\varphi$ in the bulk.
This shift is best performed before splitting $\delta\varphi$ into bulk and boundary pieces since, otherwise, it would disrupt the homogeneous Dirichlet boundary conditions of the bulk fluctuations.
The bulk action is unmixed by
\begin{equation}
  \delta\varphi \quad\longrightarrow\quad \delta\varphi + \frac{24\pi}{\cL} R_\ast^{-1} \delta\Lambda\+,
\end{equation}
which produces a term $\delta\Lambda^2$ and mixed boundary terms $\delta\Lambda\+ \delta\Lambda_\mathrm{b}$ and $\delta\Lambda\+ \delta\varphi_\mathrm{b}$.
Because of this, after splitting into bulk and boundary fluctuations, the shift of $\delta\varphi_\mathrm{b}$ is modified to (cf.\ \eqref{eq:Lambda_ell_bdy_shift})
\begin{equation}
  \delta\varphi_\mathrm{b} \quad\longrightarrow\quad \delta\varphi_\mathrm{b} - \frac{12\pi}{\cL} \lambda_{\mathrm{b},0}^{-1} \biggl(\delta\Lambda_\mathrm{b} - \frac{2 K_\ast}{R_\ast} \delta\Lambda\biggr)\+.
\end{equation}

The shifts leave the quadratic action for $\delta\varphi$ invariant, so the analysis in §\ref{sec:fluct} again goes through, but produces an additional factor
\begin{equation}
  \int \frac{A \dd\delta\Lambda}{2\pi\ii} \int \frac{\ell \dd\delta\Lambda_\mathrm{b}}{2\pi\ii}\+ e^{
      - \frac{24\pi}{\cL} \Bigl[\bigl(\frac{A}{R_\ast} - \frac{K_\ast \ell}{R_\ast^2} - \frac{K_\ast^2 \ell}{R_\ast^2} \lambda_{\mathrm{b},0}^{-1} \bigr)\delta\Lambda^2 + \bigl(\frac{\ell}{R_\ast} + \frac{K_\ast \ell}{R_\ast} \lambda_{\mathrm{b},0}^{-1}\bigr) \delta\Lambda_\mathrm{b} \delta\Lambda - \frac{\ell}{4} \lambda_{\mathrm{b},0}^{-1} \delta\Lambda_\mathrm{b}^2\Bigr]
    } = \pm \ii \frac{\abs{\cL}}{12 \pi^2}\+,
\end{equation}
which, remarkably, is independent of $A$ and $\ell$.
Here, we focus on a light insertion and have used \eqref{eq:lambda_b} for $\lambda_{\mathrm{b},0}$.
Note that the quadratic form in the exponent is of indefinite signature, i.e.\ precisely one of the two integrals converges along the imaginary contour.
We have already seen in \eqref{eq:fixed_ell_inv_Laplace} that the integration contour for $\delta\Lambda_\mathrm{b}$ has to be rotated when the saddle is a large spherical cap, but not when it is a small cap.
Performing that integral first, the opposite holds for the integration contour for $\delta\Lambda$.
Relatedly, as shown in §\ref{sec:fluct}, the negative mode is a boundary mode on the small cap but a bulk mode on the large cap.
In the timelike theory, the contour should be rotated to produce the same phase as a negative mode, $\pm\ii$, as explained in §\ref{sec:fixed_ell} (see, in particular, footnote~\ref{footnote:Laplace_contour_rotation}).

In this ensemble, including the factor discussed above, the result for a light insertion, at one-loop order, is
\begin{equation} \label{eq:PsiA_one-loop}
  \Psi_a^{(A)}(\ell) = \frac{\cL^2}{12\pi^2} \scheme(\gamma_A) \frac{(\Lambda_\mathrm{UV} \tilde{L})^{\frac{7}{6}}}{2a-1} \biggl(\frac{2 A}{\tilde{L} \ell}\biggr)^{-\frac{\cL}{6} + \frac{7}{6} + 2a} e^{\frac{\cL}{6}\bigl(1 - \frac{\ell^2}{4\pi A}\bigr)} [1 + \mathcal{O}(1/\cL)]\+,
\end{equation}
where $\scheme$ is the scheme-dependent factor in \eqref{eq:scheme_factor} and
\begin{equation}
  \gamma_A \equiv \sqrt{\frac{4\pi A}{\ell^2} - 1}\+,
\end{equation}
for a light insertion in the timelike theory.
Note that $\Psi_a^{(A)}$ is real and positive when $a > 1/2$ since the phases from the negative mode and Laplace transforms cancel those from the zero modes.

\subsubsection*{Exact expectation}
We expect that the dependence on $\ell$ and $A$ is captured by
\begin{equation} \label{eq:Psi_A_ell_expectation}
  \Psi_\alpha^{(A)}(\ell)
  \propto (2A/\ell)^{(q+2\alpha)/\beta} \exp\biggl[\frac{\ell^2}{4 \sin(\pi \beta^2) A}\biggr]\+.
\end{equation}
This result is obtained by analytic continuation of the corresponding formula in spacelike Liouville theory \cite{Fateev:2000ik}.
Although such a continuation is in general not valid \cite{Harlow:2011ny}, we will argue that it is in the present ensemble.

To compare this with the semiclassical result, note that it follows from \eqref{eq:heavy_tree_parameter_relation} that \eqref{eq:Psi_A_ell_expectation} agrees with the saddle point result \eqref{eq:PsiA_tree} since $\cL = -6\beta^{-2} + \mathcal{O}(\beta^0)$.
At the one-loop level, the power of $2A/\ell$ in \eqref{eq:Psi_A_ell_expectation} agrees with \eqref{eq:PsiA_one-loop} by \eqref{eq:light_1-loop_parameter_relation} but the $\mathcal{O}(\beta^0)$ coefficients of $\ell^2/A$ in the exponentials differ between the expressions.
This is again due to scheme dependence since this coefficient, as well as the overall normalization, is sensitive to small rescalings of $\ell$ and $A$.
Explicitly, the counterterms produced by $\ell \to \ell (1 + \frac{6}{\cL} \delta_\ell)$ and $A \to A (1 + \frac{6}{\cL} \delta_A)$, evaluated at the on-shell values for $\Lambda$ and $\Lambda_\mathrm{b}$, are
\begin{equation}
  \frac{6}{\cL} \Lambda A\+ \delta_A = \frac{\ell^2 - 4\pi A}{4\pi A} \delta_A\+,\qquad\quad
  \frac{6}{\cL} \Lambda_\mathrm{b} \ell\+ \delta_\ell = - \frac{\ell^2 - 2\pi A}{2\pi A} \delta_\ell\+,
\end{equation}
which correspond precisely to the overall normalization and a factor $e^{\frac{\ell^2}{4\pi A} \mathcal{O}(\beta^0)}$.

\begin{figure}[H]
  \centering
  \includegraphics[width=\figurewidth]{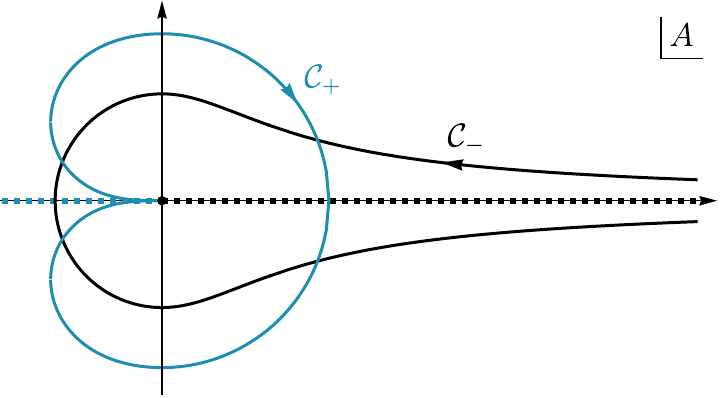}
  \caption{Illustration of the complexified contours for the area producing the Bessel functions in \eqref{eq:Bessel_contours}.
    The corresponding branch cuts for the integrand run along the real axis and are marked by dashed lines.}
  \label{fig:A_contours}
\end{figure}

In spacelike Liouville theory, the Laplace transform \eqref{eq:Laplace_Lambda_A} of \eqref{eq:Psi_A_ell_expectation} (with $\alpha \to \pm\ii \alpha$, $\beta \to \pm\ii b$ and $q \to \mp\ii Q$) converges on the positive half-line, reproducing the Bessel function $K_{(Q-2\alpha)/b}$.
In the timelike theory, in contrast, there is a severe small-area divergence due to the sign of the argument of the exponential.
Hence, a different integration contour has to be used.
To this end, we may use Schläfli's integral \dlmfeqcite{10.9.19}
\begin{equation}
  J_\nu(z) = \frac{(z/2)^\nu}{2\pi\ii} \int_\mathcal{C} \frac{\dd t}{t^{1+\nu}} e^{t - z^2/4t}\+,
\end{equation}
where the principal branch is used for $t^{1+\nu}$ and $\mathcal{C}$ is a Hankel contour wrapping the negative half-line counterclockwise (from $-\infty -\ii\epsilon$, encircling the origin and returning to $-\infty + \ii\epsilon$ without crossing the branch cut at $\mathbb{R}_-$).
From this, we see that
\begin{equation} \label{eq:Bessel_contours}
  \int_{\mathcal{C}_\pm}\! \frac{\dd A}{A} (2A/\ell)^{\nu} e^{-\Lambda A + \frac{\ell^2}{4\sin(\pi\beta^2)A}}
  = 2\pi \ii^{\mp 1} (\pm 1)^\nu \bigl[\sin(\pi\beta^2) \Lambda\bigr]^{-\nu/2} J_{\pm\nu}\Bigl(\sqrt{\Lambda/\sin(\pi\beta^2)}\+ \ell\Bigr)\+,
\end{equation}
through changes of variables $t = \ell^2/[4\sin(\pi\beta^2)A]$ and $t = -\Lambda A$ for the upper and lower signs, respectively.
With $\nu = (q+2\alpha)/\beta$ this is the Laplace transform \eqref{eq:Laplace_Lambda_A} of the fixed area wavefunction \eqref{eq:Psi_A_ell_expectation}, and reproduces both solutions~$J_{\pm(q+2\alpha)/\beta}$ to the WDW equation depending on the choice of contour $\mathcal{C}_A = \mathcal{C}_\pm$.
The contour $\mathcal{C}_+$ is the inversion of the contour $\mathcal{C}$ described above and $\mathcal{C}_-$ is the negative of $\mathcal{C}$, as apparent from the changes of variables.
They are depicted in Figure~\ref{fig:A_contours}.
For the Hartle--Hawking-like wavefunction $J_{+(q+2\alpha)/\beta}$, the principal branch is used for $A^\nu$, whereas the branch $\arg(A) \in [0, 2\pi)$ with the cut along the positive real axis is used for $J_{-(q+2\alpha)/\beta}$.
Note also that the power of $\Lambda \sim L^{-2}$ in \eqref{eq:Bessel_contours} agrees with the expectation discussed in §\ref{sec:fixed_K} from the formulation with the shift-invariant path integral measure \cite{David:1988hj,Distler:1988jt}.

The contours $\mathcal{C}_\pm$ are consistent with the semiclassical analysis.
The fact that the $\delta \Lambda$ integral has to be rotated to a real contour around the small cap is reflected in $\mathcal{C}_+$ crossing the positive real axis in the imaginary direction.
Around the large cap, instead, the $\delta \Lambda$ integral runs in the imaginary direction whereas $\mathcal{C}_-$ runs parallel to the positive real axis.
Moreover, the fact that \eqref{eq:Psi_A_ell_expectation} agrees with the one-loop result and reproduces the expected Bessel functions, upon choosing an appropriate contour $\mathcal{C}_A$, indicates that it is valid exactly, even though obtained by analytic continuation from the spacelike formula.
This relation between spacelike and timelike Liouville theory is somewhat reminiscent of the relation between AdS and dS in higher dimensions, used in \cite{Hertog:2011ky}.

\subsubsection*{Comment on spacelike Liouville theory and heavy insertions at the loop level}
The results in \eqref{eq:PsiA_tree} and \eqref{eq:PsiA_one-loop} are valid in spacelike Liouville theory as well.
The expression for $\gamma_A$ that agrees on-shell with the parameter $\gamma$ in \eqref{eq:saddle_gamma} is different in this case, but \eqref{eq:PsiA_one-loop} is valid after writing $\scheme(\gamma_A)$ explicitly in terms of $A$ and $\ell$.
In the spacelike case, the phase from the negative boundary mode cancels with the rotation of the $\delta\Lambda_\mathrm{b}$ contour.

The result can be compared to the exact one-point function coefficient
\begin{equation} \label{eq:FZZ_U_A}
  U_\alpha = \frac{\Gamma(2 b \alpha - b^2)}{b\+ \Gamma(b^{-2} - 2 b^{-1} \alpha + 1)} \Biggl(\frac{\ell\+ \Gamma(b^2)}{2 A}\Biggr)^{(Q-2\alpha)/b} \exp\Biggl(- \frac{\ell^2}{4 \sin(\pi b^2) A}\Biggr)\+.
\end{equation}
from \cite{Fateev:2000ik}, where we have omitted factors of $\tilde{L}$ and $\Lambda_\mathrm{UV}$ that follow from conformal invariance and dimensional analysis.%
\footnote{Analytically continuing the gamma functions in \eqref{eq:FZZ_U_A} to imaginary $b = \pm \ii \beta$ requires care due to the poles of the denominator, cf.\ \cite{Harlow:2011ny}, and, relatedly, the additional exponential term that becomes relevant in the asymptotic expansion of $\Gamma(x)$ as $x \to -\infty$.
  At least at the semiclassical level, it seems the issue can be avoided by employing an $\ii\epsilon$ prescription $b \to (\epsilon \pm \ii) \beta$.}
Upon expanding in small $b$, one sees that this agrees with \eqref{eq:PsiA_one-loop} at the one-loop level, up to the scheme-dependent factor
\begin{equation} \label{eq:1-pt_discrepancy}
  \frac{(4 \sqrt{e})^{\frac{\ell^2}{4\pi A}}}{2^\frac{53}{12} \pi^{9/4} e^{\frac{7}{24} - \zeta'(-1)}}
  \qquad\text{vs}\qquad
  \frac{1}{2^{3/2} \sqrt{\pi} e^{\gamma_\mathrm{E}}}\+,
\end{equation}
where the former comes from \eqref{eq:PsiA_one-loop}, the latter from \eqref{eq:FZZ_U_A} and $\gamma_\mathrm{E}$ denotes the Euler--Mascheroni constant.

In addition, using \eqref{eq:FZZ_U_A}, we can obtain a prediction for the one-loop correction to the one-point function of a heavy insertion with finite $v \equiv b \alpha = -k/2 + \mathcal{O}(\beta^2)$ as
\begin{equation} \label{eq:1-pt_heavy_1-loop_prediction}
  \frac{\langle V_\alpha(z_0) \rangle}{e^{2a\varphi_\ast(z_0)-S_\ast}}
  = b^{-2} \frac{2^{2/3} e^{\frac{13}{6} \frac{\ell^2}{4\pi A}}}{\sqrt{\pi}} \frac{(1 - 2 v)^{5/3} \Gamma(2v)}{e^{(13 + 6 \gamma_\mathrm{E}) (1-2v)/6}} \biggl(\frac{\Lambda_\mathrm{UV} A}{\ell}\biggr)^\frac{7}{6} \biggl(\frac{(\Lambda_\mathrm{UV} \tilde{L})^{-1}}{1 - \bar{z}_0 z_0}\biggr)^{\frac{v}{3} (6 - 13 v)} [1 + \mathcal{O}(b^2)]\+.
\end{equation}
Note that $\tilde{L}$ appears only in the factor correcting the conformal dimension.
This indicates that the fiducial background enters the one-loop analysis only through the excised region containing $z_0$, as anticipated in §\ref{sec:PI} and §\ref{sec:fluct}.
Further evidence for this is provided by the fact that, except this factor, $\Lambda_\mathrm{UV}$ comes with the same power, $7/6$, as in the case of a light insertion.

Recall from §\ref{sec:bulk_fluct} that there are subtleties with the boundary conditions at the conical singularity produced by the heavy insertion.
Assuming that the boundary conditions approach the standard regularity conditions as $v \to 0$, consistent with the vanishing correction to $\Delta_\alpha$ in \eqref{eq:1-pt_heavy_1-loop_prediction} in this limit, we obtain from \eqref{eq:bdy_det_heavy} and \eqref{eq:bulk_det} the one-loop factor
\begin{equation}
  \frac{\mp\ii \cL}{12\pi^2} \frac{1}{\sqrt{D_\mathrm{B}(k)}} \frac{1}{\sqrt{D_\mathrm{b}(k)}}
  \ \widesim{v \to 0}\
  b^{-2}
  \frac{(4 \sqrt{e})^{\frac{\ell^2}{4\pi A}}}{2^\frac{53}{12} \pi^{9/4} e^{\frac{7}{24} - \zeta'(-1)}}
  \frac{2^\frac{13}{6} e^{\frac{13}{6} \bigl(\frac{\ell^2}{4\pi A} -1\bigr)}}{2v} \biggl(\frac{\Lambda_\mathrm{UV} A}{\ell}\biggr)^\frac{7}{6}\+,
\end{equation}
where the $\mp\ii$ cancels the phase from the negative mode.
The result agrees with the $v\to 0$ limit of \eqref{eq:1-pt_heavy_1-loop_prediction} up to precisely the same scheme-dependent discrepancy as in \eqref{eq:1-pt_discrepancy}.
This is a nontrivial consistency check due to the order of limits; there is no moduli space of saddles when the insertion is heavy.

It would be interesting to investigate the case of finite $v$ further.
Based on the discussion in §\ref{sec:bulk_fluct}, this likely involves modifying the boundary conditions at the conical singularity in the rotationally invariant sector.
It seems plausible that this could lead to a correction to the conformal dimension since the second solution to the Klein--Gordon equation behaves as $\log \rho$ close to the origin ($\rho = 0$) and the radius of the excised disk, after mapping the insertion to the origin, is $(\Lambda_\mathrm{UV} \tilde{L})^{-1}/(1- \bar{z}_0 z_0)$.

\section{Outlook} \label{sec:outlook}

We have thoroughly and systematically analysed the disk path integral of timelike Liouville theory.
At one loop, we have obtained the expression \eqref{eq:Psi_K_light_one-loop}, for fixed $K$ boundary conditions.
Further to this, we have provided a variety of arguments indicating that the all-loop expression should be given by
\begin{equation} \label{eq:Psi_K_exact}
  \Psi_\alpha(K) = \mp\ii \beta^{-2} \bigl(\Lambda_\mathrm{UV} \tilde{L}\bigr)^{\frac{\cL}{6} - 2\alpha(q+\alpha)} \bigl(2 \Lambda_\mathrm{UV} L \gamma_K /e\bigr)^{(q+2\alpha)/\beta}\+,
\end{equation}
with $\gamma_K$ given in \eqref{eq:gamma_K_def} and $L$ in \eqref{eq:L_K_defs}, up to a subleading rescaling in the small-$\beta$ limit.
Here, we have dropped an $\mathcal{O}(\beta^0)$ and $\alpha$-dependent normalization.
Apart from this rescaling and normalization, \eqref{eq:Psi_K_exact} is scheme independent.

We note that although both spacelike and timelike Liouville theory on a disk without heavy insertions exhibit a phase in the fixed $K$ ensemble, the origins of the phases differ.
In the former, it stems from a single universal negative boundary mode.
In the latter, it follows from the one-loop calculation upon rotating the Liouville mode to an appropriate complex contour and can be due to either a bulk or boundary mode.
The unbounded nature of real boundary fluctuations \eqref{eq:bdy_det} in timelike Liouville theory is a boundary counterpart of the conformal mode problem of Euclidean gravity.

The gravitational disk path integrals can be related to solutions of the Wheeler--DeWitt equation of two-dimensional quantum gravity with $\Lambda>0$, coupled to a conformal field theory with large and positive central charge.
The full Wheeler--DeWitt wavefunction, with the matter in a nonspinning conformal primary state $\ket{O_\alpha}$, is given by
\begin{equation}
  \Psi^\textnormal{grav}_\alpha(K) = \Psi_\alpha(K) \otimes \ket{O_\alpha} \otimes \ket{\textnormal{ghost}}\+,
\end{equation}
with $\alpha$ such that $\Delta_\alpha + \Delta_{O_\alpha} = 1$, recall \eqref{eq:alpha_from_O}.
To label the gravitational wavefunction by $\alpha$, for simplicity, we assume that the matter primaries all have different scaling dimensions.
The matter CFT Hilbert space has a positive definite norm, and the ghost vacuum $\ket{\textnormal{ghost}}$ is taken to have unit norm.
Generally, one can build superpositions of the above states.
One may also generalize this by allowing spinning matter primaries, dressed by appropriate descendants of the Liouville vertex operators.
Since one-point functions of descendants are fixed by those of the primaries, our results can be straightforwardly adapted to this case.

It is worth emphasizing that many of the features of our wavefunctions bear a strong resemblance to properties encountered in higher spacetime dimensions.
For example, there exist both Hartle--Hawking-like and Vilenkin-like states when $\Delta_{O_\alpha}$ is not too large \cite{Anninos:2024iwf}.
The former has vanishing amplitude in the limit of small spatial size and oscillates in a WKB-like fashion at large size, in line with Hartle and Hawking's defining principle \cite{Hartle:1983ai}.
In our setup, however, these states can be systematically explored to all loops, and perhaps even nonperturbatively, without resorting to a minisuperspace approximation.
Indeed, in many ways the minisuperspace approximation becomes exact in the lower-dimensional theories we consider \cite{Hertog:2021jyd,Fateev:2000ik}.

What is less understood in the Wheeler--DeWitt literature is the gravitational inner product governing the $\Psi_\alpha(K)$ sector and, hence, what the cosmological Hilbert space is.
We now discuss some potentially interesting structures that have appeared along the way in our analysis that may shed some light on this.

\subsection{Cosmological pairing}

First, we note that the following pairing of the timelike Liouville wavefunctions,
\begin{equation}\label{eq:product}
  \mathcal{N}_\alpha \equiv \Psi^\textnormal{grav}_\alpha(-K)^\dag \Psi^\textnormal{grav}_\alpha(K)\+,
\end{equation}
is independent of $K$.
Here, the (nonstandard) dagger operation maps ket states to bras and acts as complex conjugation on $c$-numbers when $\alpha \in \mathbb{R}$.
The independence of $K$ continues to hold if both factors on the right-hand side are continued analytically to complex $\alpha$ \emph{after} acting with the dagger.
Hence, we take $\Psi_\alpha(K)^\dag = \bar{\Psi}_{\bar{\alpha}}(K)$ when $\alpha \in \mathbb{C}$.
This is only consistent with $\Delta_\alpha + \Delta_{O_\alpha} = 1$ when $\Delta_{\alpha} = \Delta_{\bar{\alpha}}$, which holds precisely for the values of $\alpha$ corresponding to real $\Delta_{O_\alpha}$.
For the expression to be $K$-independent, it is crucial that the $K$-dependent functions, at all loops, take the specific form \eqref{eq:Psi_K_exact}.
To leading order in the semiclassical limit, this is simply the statement that one is smoothly gluing two portions of a two-sphere together (see \cite{Anninos:2022hqo,Anninos:2024wpy} for related discussions).
We note, however, that the one-loop correction in \eqref{eq:Psi_K_light_one-loop} explicitly depends on $K$, and it is a priori less clear that the combination of fluctuating disks combined with equal and opposite $K$ should also yield a $K$-independent result.
Nonetheless, the dependence on $K$ drops out of the combination \eqref{eq:product} both for the scheme-independent one-loop correction and the scheme-dependent factor $\scheme(\gamma_K)$  in \eqref{eq:Psi_K_light_one-loop}.

More generally, we have the pairing
\begin{equation}\label{producti}
  \bigl(\alpha', \alpha\bigr) \equiv \Psi^\textnormal{grav}_{\alpha'}(-K)^\dag \Psi^\textnormal{grav}_\alpha(K)
  = \mathcal{N}_\alpha  \delta_{\alpha' \alpha}\+.
\end{equation}
The fact that $(\alpha', \alpha)$ is independent of $K$ is somewhat reminiscent of proposals for the inner product in Wheeler--DeWitt quantum cosmology \cite{DeWitt:1967yk,DeWitt:1967ub},%
\footnote{See also the excellent discussion in Section~VIII of \cite{Weinberg:1988cp}.}
especially in its formulation based on decoherent histories quantum mechanics \cite{Halliwell:1990qr}, where a Klein--Gordon inner product on a hypersurface in the space of all metrics specifies conserved probabilities of entire histories \cite{Hartle:2008ng,Halliwell:2018ejl}.

In this picture, the York time variable $K$ \cite{York:1972sj} acts like a cosmological clock in superspace.
Incidentally, this was also discussed by Hartle and Hawking in the construction of their wavefunction \cite{Hartle:1983ai}.
Note that, for $\alpha\in \mathbb{R}$, the pairing is positive definite and sesquilinear, by construction.
What is not entirely clear is whether it obeys a standard conjugation property.
Perhaps a more complete notion of conjugation will require that $K$ maps to $-K$, mimicking the reflection of Euclidean time under Euclidean conjugation.
We note that, in this spirit, the corresponding formula in the Lorentzian regime $K = -\ii K_L$ is $\mathcal{N}_\alpha = \Psi^\textnormal{grav}_\alpha(K_L)^\dag \Psi^\textnormal{grav}_\alpha(K_L)$, without a minus sign in the first argument.

It is natural, then, to consider what the analogous situation might look like in $(d+1)$ spacetime dimensions for a gravitational theory with $\Lambda>0$.
In this setting, we might consider excising some closed $d$-dimensional manifold, $M_d$, from a $(d+1)$-sphere and consider the Euclidean path integral over a portion of the sphere with boundary conditions that hold fixed the trace of the extrinsic curvature, $K(y)$, on $M_d$.
Here $y$ labels a point on $M_d$.
Interestingly, in addition to $M_d$ having a sphere topology, it may allow other topologies such as that of $S^1 \times S^{d-1}$.
In three-spacetime dimensions, for instance, one often considers a Heegaard decomposition of $S^3$ into two manifolds with a toroidal boundary \cite{Witten:1988hf}.
At least for topological quantum field theories, the Heegaard splitting can yield insight into the physical meaning of the $S^3$ partition function in terms of an entanglement entropy \cite{Kitaev:2005dm,Levin:2006zz}.
Additionally, one can have other $(d+1)$-dimensional Einstein manifolds that could fill in $M_d$, and these may also play a role \cite{Anninos:2025ltd}.

In any case, given some manifold $M_d$, we must fix additional local data on $M_d$ as compared to the two-dimensional problem.
Following recent considerations of gravitational theories with finite size boundaries \cite{Anderson:2006lqb,Anninos:2023epi}, one natural possibility is to further fix the conformal class $[\gamma]$ of the induced metric $\gamma_{ij}(y)$ on $M_d$, along with a quantum mechanical notion of $K(y)$.
The gravitational path integral thus yields a functional $Z_\textnormal{grav}^M([\gamma], K)$.
More generally, we could decorate $Z_\textnormal{grav}^M$ with insertions.
For pure Einstein gravity with $\Lambda>0$, we are thus led to consider the pairing
\begin{equation}
  \mathcal{N}_M \equiv \int \frac{\DD \gamma}{\Vol(\Diff \ltimes \mathrm{Weyl})} Z_\textnormal{grav}^M([\gamma], -K)^\dag Z_\textnormal{grav}^M([\gamma], K)\+.
\end{equation}
In the above expression, we are dividing by the volume of $\Diff \ltimes \mathrm{Weyl}$ on $M_d$ (since each conformal class $[\gamma]$ should be counted only once) or some real form of its complexification \cite{Gibbons:1978ac,Polchinski:1988ua}.
At the saddle point level, it is straightforward to confirm that the above expression holds, at least for constant $K$, see for instance (4.31) of \cite{Anninos:2024wpy}.
To leading order, and for trivial infilling topology, one indeed recovers the sphere path integral such that $\mathcal{N}_M \approx \exp {\mathcal{S}}$ with $\mathcal{S}$ the tree-level Gibbons--Hawking entropy of the de Sitter horizon \cite{Gibbons:1978ac,Gibbons:1977mu}.
This is, again, the statement that one is smoothly gluing the round sphere from two pieces.
In the case of three spacetime dimensions, whether or not we have a nontrivial integral over the conformal class depends on the topology of the two-manifold $M_2$.
For instance, if $M_2 = T^2$, we should integrate over the fundamental domain of the complex structure.
In four spacetime dimensions and higher, there is a nontrivial path integral over conformal metrics even for $M_d=S^d$.

It will be interesting to understand whether $\mathcal{N}_M$ is $K$-independent at one-loop and beyond.%
\footnote{Along these lines, and somewhat remarkably, a recent analysis of one-loop contributions on a four-sphere for an infinite tower of higher-spin Fronsdal fields reveals such a structure \cite{Anninos:2025mje}.}
Perturbing around a conformal class $[\gamma]$ containing the round $d$-sphere, the one-loop path integral must be divided by the volume of the residual conformal isometry group of $S^d$, namely $\mathcal{G}_d = \mathrm{SO}(d+1,1)$, normalized in units of $\mathcal{S}$, leading to a slightly improved expression $\mathcal{N}_{S^d} \approx \mathcal{S}^{-\frac{\dim \mathcal{G}_d}{2}} \exp{\mathcal{S}}$.
This provides an alternative description of the logarithmic corrections to the Gibbons--Hawking entropy computed in \cite{Anninos:2020hfj}.
Depending on the choice of complexified contour, there may also be a phase prefactor.
The special case $K \to -\ii d$, corresponding to the asymptotic future of $\mathrm{dS}_{d+1}$, of the above pairing has been considered previously in \cite{Maldacena:2002vr,Anninos:2017eib_,Collier:2025lux,Cotler:2025gui}.

When further developed, these lines of reasoning may provide a novel perspective on the inner product of a cosmological Hilbert space intimately tied to the notion of a York time variable \cite{York:1972sj}, but in a Euclidean setting.
This, in turn, may yield a route to include an observer or an observatory in Euclidean quantum cosmology.

\subsection{Static patch perspective}

A different perspective to the one we have taken throughout our discussion is that the gravitational disk path integral, at least in the absence of any insertions, is computing a thermodynamic partition function for the Euclidean static patch of $\mathrm{dS}_2$ in the presence of a Euclideanized timelike boundary.
This follows from the fact that the two-sphere metric can also be viewed as the Euclidean continuation of the static patch metric, if we Wick rotate the static patch time coordinate.

At the boundary, we further imagine placing the matter CFT in some Cardy state.
If we imagine excising a thin \enquote{worldtube} of an observer, the boundary conditions of the matter CFT would presumably encode the observer's state, and different Cardy states would correspond to different such choices.
It is interesting that the FZZT boundary states associated with extended branes in spacelike Liouville theory \cite{Seiberg:2003nm} are now associated with a timelike feature in $\mathrm{dS}_2$.

Here again, the problem can be considered for a variety of boundary conditions at the worldline boundary, which one can view as different choices of thermodynamic ensembles.
The fixed $\ell$ ensemble would naturally correspond to the canonical partition function, with $\ell$ the inverse temperature.
We thus have a thermal partition function that takes the form
\begin{subequations} \label{eq:Z_ell}
  \begin{equation} \label{eq:Z_ell_small}
    Z_\textnormal{grav}(\ell) = \mp \ii Z_\mathrm{CFT} \beta^{-4} \frac{(2\Lambda_\mathrm{UV} L/e)^{q/\beta}}{\ell/L} J_{1/\beta^2}(C \ell/L)\+,
  \end{equation}
  if only the small spherical cap contributes, or
  \begin{equation} \label{eq:Z_ell_large}
    Z_\textnormal{grav}(\ell) = - Z_\mathrm{CFT} \beta^{-4} \frac{(2\Lambda_\mathrm{UV} L/e)^{q/\beta}}{\ell/L} Y_{1/\beta^2}(C \ell/L)\+,
  \end{equation}
\end{subequations}
for the large cap, at least up to nonperturbative terms.
Here, the matter CFT partition function on a disk of radius $\tilde{L}$ is $(\Lambda_\mathrm{UV} \tilde{L})^{c/6} Z_\mathrm{CFT}$, the overall dependence on $\tilde{L}$ has cancelled due to $\cL + c - 26 = 0$, $L \sim (\beta^2 \Lambda)^{-1/2}$ is defined in \eqref{eq:L_K_defs}, and $C \sim \beta^{-2}$ in \eqref{eq:Bessel_arg_prop}.
We have dropped scheme-dependent contributions corresponding to rescaling $Z$ by a positive $\mathcal{O}(\beta^0)$ factor and $L \to L [1 + \mathcal{O}(\beta^2)]$.

Taking a derivative of $Z_\textnormal{grav}(\ell)$ with respect to $\Lambda$ brings down $\hat{A} = \int_{D}\! \tvol\+ e^{2\beta\phi}$ from the action.
Hence, the above result can be derived by integrating $\dd Z_\textnormal{grav}/ \dd \Lambda = -\Psi_\beta^\textnormal{grav}$ (recall \eqref{eq:wfu}).
At the one-loop level, the analysis from the previous sections can be adapted to compute the partition function instead of the one-point functions.
In the Liouville sector, the only difference from $\langle V_0 \rangle$ is that the zero modes now give $I = \int \dd^2 w\+ (1-\bar{w}w)^{-2}$ instead of \eqref{eq:moduli_space_integral}.
This infinity cancels against the ghost zero modes, analogous to \eqref{eq:wfu_integral} in the case of the one-point function.
The one-loop result is seen to agree with the exact form in \eqref{eq:Z_ell} upon using the asymptotic expansion \eqref{eq:light_Bessel_asympt} of the Bessel function.

The factor $1/\ell$ multiplying the Bessel functions in \eqref{eq:Z_ell} stems from the additional zero modes of the unpunctured disk.
So does the fact that the order of the Bessel functions is $1/\beta^2$ instead of $q/\beta$, the latter appearing in $\Psi_0(\ell) \propto J_{q/\beta}(C \ell/L)$.
It is worth noting that as a function of $\ell$, $Z_\textnormal{grav}(\ell)$ does not exhibit the standard monotonicity properties of an ordinary thermodynamic partition function \textdash it is oscillatory and not positive definite.
This may tie to the question of whether or not a static patch admits a description in terms of a closed unitary quantum mechanics.

Going to the fixed $K$ ensemble is akin to changing thermodynamic ensemble.
From this perspective, the Laplace transform \eqref{eq:Bessel_J_Laplace} can be viewed as integrating a quantum mechanical partition function on the Euclidean circle over a worldline metric \cite{Anninos:2017hhn}.
In doing so, one is implementing a Euclidean counterpart of the Hamiltonian constraint on the quantum mechanical worldline theory.
Concretely, \eqref{eq:Z_ell_small} gives
\begin{equation} \label{eq:Z_K}
  \begin{aligned}[b]
    Z_\textnormal{grav}(K)
    & = \int_0^\infty \frac{\dd \ell}{\ell} e^{- \frac{\abs{\cL}}{12\pi} K \ell} Z_\textnormal{grav}(\ell) \\
    &= \mp \ii Z_\mathrm{CFT} \beta^{-2} (2\Lambda_\mathrm{UV} L/e)^{q/\beta}
    \gamma_K^{1/\beta^2} \Bigl(\sqrt{1 + L^2 K^2} + \beta^2 L K\Bigr)\+,
  \end{aligned}
\end{equation}
with $\gamma_K$ defined in \eqref{eq:gamma_K_def}.
We have computed the integral for $K > 0$ and continued to $K < 0$.
Note that this expression is exact, modulo scheme dependence.
It is also worth noting that \eqref{eq:Z_K} roughly resembles the higher dimensional counterparts explored to leading order in \cite{Anninos:2024wpy} (see, in particular, (1.1) and (1.2) therein).
Alternatively, one can compute $Z_\textnormal{grav}$ by integrating $\dd Z_\textnormal{grav}/ \dd \Lambda = -\Psi_\beta^\textnormal{grav}$ directly in the $K$ ensemble, with the same result.
The formula \eqref{eq:Z_K} also agrees with our one-loop analysis, adapted to the partition function as above.
These methods, working in the $K$ ensemble, are directly applicable to all real $K$ without relying on analytic continuations.
The $\mathcal{O}(\beta^2)$ term in the last factor is a prediction for the two-loop correction.
It may also be worth pointing out that, except for the overall phase, $Z_\textnormal{grav}(K)$ is positive, with $\gamma_K$ decreasing monotonically from infinity to zero as $K$ ranges across the real line.

The Laplace transform of \eqref{eq:Z_ell_large} does not converge on the original contour, as anticipated from the negative mode analysis.
However, it has a real saddle with suppressed fluctuations for $K < 0$ with a contour $\mathcal{C}_\ell$ running in the imaginary direction.
Analogous to the relation between the $K$- and $\ell$-representations of the wavefunctions in §\ref{sec:fixed_K} and §\ref{sec:fixed_ell}, $Z_\textnormal{grav}(K)$ is independent of using \eqref{eq:Z_ell_small} or \eqref{eq:Z_ell_large}.
This can be verified explicitly at one loop and is manifest from the direct calculation in the $K$ ensemble, where there is only one real saddle.

Perhaps, then, $Z_\textnormal{grav}(K)$ is related to a microcanonical degeneracy of states for a portion of the $\mathrm{dS}_2$ static patch, at least in the semiclassical regime.
The difference in the phase of $Z_\textnormal{grav}(K)$ and $Z_\textnormal{grav}(\ell)$ with a contour that picks the large spherical cap is reminiscent of the phase that appears when going from the canonical to the microcanonical ensemble \cite{Anninos:2017hhn,Maldacena:2024spf}.
Remarkably, there is no such phase difference between $Z_\textnormal{grav}(K)$ and $Z_\textnormal{grav}(\ell)$ with a contour such that the small cap dominates.

Ultimately, confirming \textdash or refuting \textdash such a thermodynamic interpretation will require a more microscopic understanding of the problem.

\subsubsection*{Acknowledgements}

We thank Tarek Anous, Chiara Baracco, Nikolay Bobev, Soumyadeep Chaudhuri, Lorenz Eberhardt, Frank Ferrari, Beatrix M\"uhlmann, and Jesse van Muiden for useful comments and discussions.
This research was supported in part by KU Leuven grant C16/25/010.
D.A.\ is also funded by the Royal Society under the grant ``Concrete Calculables in Quantum de Sitter'', and the STFC consolidated grant ST/X000753/1.
J.K.\ is grateful to King's College London for their hospitality during the initial stages of this project and acknowledges support from the Research Foundation - Flanders (FWO) travel grant V458824N that enabled this visit.
J.K.\ is further supported by the FWO doctoral fellowships 1171823N and 1171825N.

\appendix
\section{Conformal transformations of the disk} \label{app:disk_conf}
Here, we briefly review the conformal transformations of the disk and derive the corresponding fluctuations of the Liouville field.
As is well known, using a complex coordinate $z$, a vector field $\vec{\xi} = \xi {\partial} + \bar{\xi} \bar{\partial}$ is a \emph{local} conformal Killing vector if $\bar{\partial}\xi = 0$.
On the sphere, $\xi_n = z^{n+1}$ gives a global conformal Killing vector only for $n \in \{0, \pm 1\}$ with complex coefficients, since vector fields with smaller $n$ are singular at the south pole $z = 0$ and those with greater $n$ are singular at the north pole $z = \infty$.
This gives rise to the familiar six-dimensional conformal group $\mathrm{PSL}(2,\mathbb{C})$.

On the disk, $n > 1$ cannot be excluded in this way since the north pole is not part of the geometry.
However, since the vector field should generate a diffeomorphism, its normal component has to vanish at the boundary.
To satisfy this, one has to take a particular linear combination of $\xi_n$ and $\xi_{-n}$, which renders the conformal group finite-dimensional, since $n \geq -1$ from regularity at the south pole.
Explicitly, the holomorphic components are
\begin{equation}
  \xi_0 = \ii z\+,\qquad
  \xi_\mathrm{R} = \frac{-1 + z^2}{2}\+,\qquad
  \xi_\mathrm{I} = \ii \frac{1 + z^2}{2}\+.
\end{equation}
Only real linear combinations of these satisfy the constraint of not moving the boundary, so the conformal group of the disk is three-dimensional.
One can show that it is isomorphic to $\mathrm{PSL}(2,\mathbb{R})$ by mapping the disk to the upper half-plane.
We have normalized the generators such that
\begin{equation}
  [\vec{\xi}_0, \vec{\xi}_\mathrm{R}] = \vec{\xi}_\mathrm{I}\+,\qquad
  [\vec{\xi}_0, \vec{\xi}_\mathrm{I}] = -\vec{\xi}_\mathrm{R}\+,\qquad
  [\vec{\xi}_\mathrm{R}, \vec{\xi}_\mathrm{I}] = - \vec{\xi}_0\+.
\end{equation}

Although the conformal group of the disk is independent of the metric, since all metrics are related by Weyl transformations, we still need the metric to compute the corresponding fluctuations of the Liouville field, $\delta\varphi = - \frac{1}{2} \nabla_{\! \ast \mu} \xi^\mu$.
When the insertion is light, the semiclassical geometry is a hyperbolic or spherical cap
\begin{equation}
  \dd s_\ast^2 = \frac{4 \gamma^2 L^2}{(1 + \eta \gamma^2 \bar{z} z)^2} \dd \bar{z} \dd z\+,
\end{equation}
without a conical singularity, here in coordinates $\abs{z} \leq 1$.
A short calculation shows that
\begin{equation} \label{eq:conf_generator_divergences}
  \nabla_{\! \ast \mu} \xi_0^\mu = 0\+,\qquad
  \nabla_{\! \ast \mu} \xi_\mathrm{R}^\mu = \frac{1 + \eta \gamma^2}{1 + \eta \gamma^2 z \bar{z}} (z + \bar{z}) \+,\qquad
  \nabla_{\! \ast \mu} \xi_\mathrm{I}^\mu = \ii \frac{1 + \eta \gamma^2}{1 + \eta \gamma^2 z \bar{z}} (z - \bar{z})\+.
\end{equation}
These are indeed the two $\abs{n} = 1$ zero modes \eqref{eq:varphi_b_solution} of §\ref{sec:bdy_fluct} with $k = 0$.
We also see that the vector field in \eqref{eq:infinitesimal_conf} is
\begin{equation}
  \xi
  = \ii z\+ \delta\alpha + \delta w - z^2\+ \delta \bar{w}
  = \xi_0\+ \delta\alpha - (\xi_\mathrm{R} + \ii \xi_\mathrm{I}) \delta w + (\xi_\mathrm{R} - \ii \xi_\mathrm{I}) \delta \bar{w}\+,
\end{equation}
which together with \eqref{eq:conf_generator_divergences} implies \eqref{eq:delta_varphi_from_w}.

\section{Zeta function regularization of boundary determinant} \label{app:bdy_det_zeta}
Here, we compute the boundary fluctuation determinant in \eqref{eq:bdy_det} by zeta function regularization for generic $k$, such that there are no zero modes.
In this scheme,
\begin{equation}
  D_\mathrm{b}(k) \equiv \prod_{n \in \mathbb{Z}} \frac{2\pi}{\Lambda_\mathrm{UV} \ell} \frac{n^2 - (1+k)^2}{\abs{n} + u - (1+k)}
  = \biggl(\frac{\Lambda_\mathrm{UV} \ell}{2\pi}\biggr)^{-\zeta_\mathrm{b}(0)} e^{-\zeta_\mathrm{b}'(0)}\+,
\end{equation}
where we have defined $u \equiv \ell^2 / (2\pi A)$,
\begin{equation}
  \zeta_\mathrm{b}(s) = \sum_{n\in\mathbb{Z}} \biggl(\frac{n^2 - (1+k)^2}{\abs{n} + u - (1+k)}\biggr)^{-s}\+,
\end{equation}
for sufficiently large $\Re(s)$ such that the sum converges, and $\zeta_\mathrm{b}$ is defined in a neighbourhood of $s=0$ through analytic continuation.

It is useful to define $\tilde{\zeta}_\mathrm{b}$ by
\begin{equation} \label{eq:helper_zeta}
  \zeta_\mathrm{b}(s) = \biggl(\frac{-(1+k)^2}{u-(1+k)}\biggr)^{-s} + 2 \tilde{\zeta}_\mathrm{b}(s)\+,\qquad\quad
  \tilde{\zeta}_b(s) = \sum_{n=1}^\infty \biggl(\frac{(n+1+k)(n-1-k)}{n + u - (1+k)}\biggr)^{-s}\+.
\end{equation}
The latter may be computed, again for sufficiently large $\Re(s)$, as
\begin{equation}
  \begin{aligned}[b]
    \tilde{\zeta}_\mathrm{b}(s)
    &= R_\mathrm{b}(s) + \sum_{n=1}^\infty (n+1+k)^{-s} + \sum_{n=1}^\infty (n-1-k)^{-s} - \sum_{n=1}^\infty (n+u-1-k)^{-s} \\
    &= R_\mathrm{b}(s) + \zeta(s, 2+k) + \zeta(s, -k) - \zeta(s, u-k)\+,
  \end{aligned}
\end{equation}
where $\zeta(s,a)$ is the Hurwitz zeta function and the residue term is
\begin{equation}
  R_\mathrm{b}(s) = \sum_{n=1}^\infty \biggl[\biggl(\frac{(n+1+k)(n-1-k)}{n + u - (1+k)}\biggr)^{-s} - (n+1+k)^{-s} - (n-1-k)^{-s} + (n+u-1-k)^{-s}\biggr],
\end{equation}
The point of this rewriting is that the terms in $R_\mathrm{b}(s)$ scale as $\mathcal{O}(n^{-s-2})$ for large $n$ and, hence, the sum converges absolutely for $\Re(s) > -1$.
The expansion in large $n$ is uniform on compact regions of $\mathbb{C} \ni s$ for fixed $u$ and $k$.
Since we are only interested in the value and first derivative of $R$ at $s = 0$, we may compute these by differentiating and evaluating before summing.
This shows that $R_\mathrm{b}(0) = R'(0) = 0$, i.e.\ there is no multiplicative anomaly.
Hence,
\begin{equation}
  \biggl(\frac{\Lambda_\mathrm{UV} \ell}{2\pi}\biggr)^{-\tilde{\zeta}_\mathrm{b}(0)} e^{-\tilde{\zeta}_\mathrm{b}'(0)}
  = \sqrt{2\pi} \biggl(\frac{\Lambda_\mathrm{UV} \ell}{2\pi}\biggr)^{-(u-k-3/2)} \frac{\Gamma(u-k)}{\Gamma(-k)\Gamma(2+k)}\+,
\end{equation}
which is consistent with the result in the Appendix of \cite{Chaudhuri:2024yau}.
As a further consistency check, it is straightforward to verify that
\begin{equation}
  \sum_{n=1}^\infty \frac{\partial^2}{\partial u^2} \log \frac{n^2 - (1+k)^2}{n + u - (1+k)} = - \frac{\partial^2 \tilde{\zeta}_\mathrm{b}'(0)}{\partial u^2}\+,
\end{equation}
where the left-hand side converges without regularization due to the $u$-derivatives.

Accounting for the single $n=0$ mode and the two-fold degeneracy of the $\abs{n} \geq 1$ eigenvalues in $\zeta_\mathrm{b}$, cf.\ \eqref{eq:helper_zeta}, we get the result \eqref{eq:bdy_det_heavy}
\begin{equation}
  D_\mathrm{b}(k)
  = -\frac{2\pi}{(1+k)^2} \biggl(\frac{\Lambda_\mathrm{UV} \ell}{2\pi}\biggr)^{-2[u - (1+k)]} \bigl(u - (1+k)\bigr) \frac{\Gamma\bigl(u - (1+k)\bigr)^2}{\Gamma(1+k)^2\Gamma(-1-k)^2}\+.
\end{equation}

\section{Hemisphere bulk determinant} \label{app:hemisphere_det}
As a consistency check, we verify the result \eqref{eq:bulk_det} of \cite{Chaudhuri:2024rgn} on the hemisphere.
The result in \eqref{eq:bulk_det} vanishes there due to the breakdown of the split into bulk and boundary fluctuations, as discussed in §\ref{sec:fluct}.
Here, we lift the zero mode by varying either the position of the boundary $\gamma$ or the mass of the fluctuation, remove it and then take limits to recover the hemisphere and $m^2 = -2/L^2$ without zero modes.

\subsubsection*{Lifting the zero mode by shifting the boundary}
We begin by deriving the hemisphere determinant with the zero mode removed by shifting the boundary, starting from \eqref{eq:bulk_det}.
Recall that the zero mode is the eigenmode
\begin{equation}
  f(\rho) = \sqrt{3}\+ \frac{1 - \rho^2}{1 + \rho^2}\+,
\end{equation}
of the Sturm--Liouville problem \eqref{eq:bulk_Sturm--Liouville}.%
\footnote{Since the mass term was absorbed into the eigenvalue in the Sturm--Liouville problem, this mode has $\lambda = 2$.}
This eigenmode comes from \eqref{eq:varphi_b_solution} with $n = 0$, here normalized such that $\int_0^\gamma \dd\rho\+ w f^2 = 1$.
It manifests as a term $\log(1-\gamma)$ when expanding $\log D_\mathrm{B}$ around $\gamma = 1$.

By taking a derivative of the Sturm--Liouville eigenvalue equation with respect to the location $\gamma$ of the boundary and integrating the result against $f$, it is straightforward to show that
\begin{equation}
  \partial_\gamma \lambda = - p(\gamma) f'(\gamma)^2 = -3\+.
\end{equation}
Hence, the eigenvalue entering in the bulk determinant is $\lambda_{\mathrm{B},0} = -3 (\gamma - 1)/L^2 + \mathcal{O}(\gamma-1)^2$ and
\begin{equation} \label{eq:hemisphere_det}
  D'_\mathrm{B}(\gamma = 1)
  = \lim_{\gamma \to 1} \frac{\Lambda_\mathrm{UV}^2 D_\mathrm{B}(\gamma)}{\lambda_{\mathrm{B},0}}
  = \frac{e^{\frac{9}{4} - 2 \zeta'(-1)}}{3 \sqrt{2\pi}} (\Lambda_\mathrm{UV} L)^{-1/3}\+.
\end{equation}

\subsubsection*{Lifting the zero mode by shifting the mass}
We will now verify \eqref{eq:hemisphere_det} by computing the hemisphere determinant for a massive scalar of generic mass and then take the limit $m^2 \to -2$, corresponding to the effective mass of the bulk fluctuation of the timelike Liouville field, see \eqref{eq:bulk_Klein--Gordon}.
Here, for convenience, we denote by $m^2$ the mass in units of the radius of the sphere, i.e.\ the physical mass is $m^2/L^2$.

Consider first the Klein--Gordon eigenvalue problem $(-\square + m^2/L^2)\phi = \lambda_\ast \phi$ on the full sphere.
Using coordinates \eqref{eq:Liouville_simple_saddle} and decomposing $\phi = \sum_{n \in \mathbb{Z}} \phi_n e^{\ii n \theta}$, the solution \eqref{eq:bulk_mode} respecting regularity at the south pole can be rewritten as
\begin{equation} \label{eq:S2_eigenmodes}
  \phi_n = \biggl(\frac{\rho}{1+\rho^2}\biggr)^{\abs{n}} \hypgeom\biggl(\Delta_\lambda + \abs{n}, 1-\Delta_\lambda + \abs{n}; 1 + \abs{n}; \frac{\rho^2}{1+\rho^2}\biggr)\+,\qquad
  \Delta_\lambda = \frac{1}{2} + \sqrt{\frac{1}{4} + \lambda}\+.
\end{equation}
Here, $\lambda$ is related to the eigenvalue of the massive Klein--Gordon operator by $\lambda_\ast = (\lambda + m^2)/L^2$.
Note that the prefactor is invariant under the reflection $\rho \to \rho^{-1}$ across the equator and, hence, smooth both at the south pole ($\rho = 0$) and the north pole ($\rho = \infty$).
The full fluctuation, including the hypergeometric function, is regular at the north pole if $1-\Delta_\lambda+\abs{n} = - \tilde{n}$ for a nonnegative integer $\tilde{n}$.
This can be understood from the fact that spherical harmonics can be constructed by restricting homogeneous polynomials in embedding space to the sphere.
It follows that the eigenvalues are the well-known $\lambda = l (l + 1)$, with $l =\abs{n}+\tilde{n}$, with corresponding multiplicities $2l+1$.

The homogeneous Dirichlet problem on the hemisphere can be solved by noting that one may extend to the full sphere by reflecting across the equator by $\phi(\rho, \theta) = -\phi(\rho^{-1}, \theta)$.
Hence, the eigenmodes on the hemisphere are the restrictions of the odd eigenmodes on the full sphere.
The expression in \eqref{eq:S2_eigenmodes} is odd under this reflection for odd $\tilde{n} = l - \abs{n}$ and even otherwise.
Hence, the eigenvalues are the same as on the full sphere, $\lambda = l (l + 1)$, but the corresponding multiplicities are $l$ instead of $2l+1$.

To be able to compare to \eqref{eq:hemisphere_det}, we have to use the same scheme as in \cite{Chaudhuri:2024rgn} since the normalization of the determinant is scheme dependent.
Hence, we employ zeta function regularization, as in Appendix~\ref{app:bdy_det_zeta}.
The eigenvalues can conveniently be written as $\lambda_\ast = [l (l+1) + m^2]/L^2 = (l + \Delta)(l+1-\Delta) / L^2$, where we parametrize the mass by $m^2 = \Delta (1-\Delta)$.
Hence, the zeta-regularized hemisphere determinant is
\begin{equation} \label{eq:hemisphere_m2_det}
  D_\mathrm{B}(m^2) = \prod_{l \in \mathbb{Z}_+} \biggl(\frac{(l+\Delta)(l+1-\Delta)}{\Lambda_\mathrm{UV}^2 L^2}\biggr)^l
  = (\Lambda_\mathrm{UV}^2 L^2)^{-\zeta_\mathrm{B}(0)} e^{-\zeta_\mathrm{B}'(0)}\+,
\end{equation}
where
\begin{subequations}
  \begin{align}
    & \zeta_\mathrm{B}(s) = \sum_{l=1}^\infty l\+ [(l+\Delta)(l+1-\Delta)]^{-s}
    = R_\mathrm{B}(s) + \zeta_\Delta(s) + \zeta_{1-\Delta}(s)\+,\\
    & \zeta_\Delta(s) = \sum_{l=1}^\infty l (l + \Delta)^{-s} = \zeta(s-1, 1+\Delta) - \Delta \zeta(s, 1+\Delta)\+,\\
    & R_\mathrm{B}(s) = \sum_{l=1}^\infty r_l(s)\+,\qquad r_l = l\+ \bigl[\bigl((l+\Delta)(l+1-\Delta)\bigr)^{-s} - (l+\Delta)^{-s} - (l+1-\Delta)^{-s}\bigr]\+.
  \end{align}
\end{subequations}
In contrast to Appendix~\ref{app:bdy_det_zeta}, the residue term does not converge in a neighbourhood of $s=0$.
However, it can be written as
\begin{equation} \label{eq:hemisphere_det_bulk_residue_term}
  R_\mathrm{B}(s) = \sum_{l=1}^{\infty} [r_l(s) - a_l(s)] + \sum_{l=1}^{\infty} a_l(s)\+,
\end{equation}
where $a_l$ is constructed from the large $l$ expansion of $r_l$,
\begin{equation}
  a_l(s) = \frac{l^{-s}}{2} \bigl[-4 l + 2 s - s (s+1)(1-2 m^2) l^{-1} + \mathcal{O}(l^{-2})\bigr] + \frac{l^{-2s}}{2} \bigl[2 l - 2 s + s (s + 1 - 2m^2) l^{-1} + \mathcal{O}(l^{-2})\bigr]\+,
\end{equation}
such that $r_l(s) - a_l(s) = \mathcal{O}(l^{-2-s}) + \mathcal{O}(l^{-2-2s})$.
Importantly, this makes the first term on the right-hand side of \eqref{eq:hemisphere_det_bulk_residue_term} converge in a neighbourhood of $s = 0$ and, much like $R_\mathrm{b}$ from Appendix~\ref{app:bdy_det_zeta}, it is $\mathcal{O}(s^2)$ and hence does not contribute to the hemisphere determinant in \eqref{eq:hemisphere_m2_det}.
Computing the last sum,
\begin{equation}
  R_\mathrm{B}(s) = \sum_{l=1}^{\infty} a_l(s) + \mathcal{O}(s^2)
  = \frac{3 m^2-1}{6} + \frac{4 m^2 - 1}{4} s + \mathcal{O}(s^2)\+.
\end{equation}
Putting the above pieces together, we get
\begin{equation}
  D_\mathrm{B}(m^2) = (\Lambda_\mathrm{UV} L)^{-\frac{1}{3} + m^2} 2\pi \Gamma(1+\Delta)^\Delta \Gamma(2-\Delta)^{1-\Delta} e^{-\frac{3}{4} - 2\zeta'(-1) - \psi^{(-2)}(1+\Delta) - \psi^{(-2)}(2-\Delta)}\+,
\end{equation}
where $\psi^{(n)}(z)$ is the polygamma function of order $n$.

Lastly, we take the limit $m^2 \to -2$.
The lowest mode, with eigenvalue $\lambda_{\ast,0} = (m^2 + 2)/L^2$, becomes a zero mode in this limit, as expected.
Removing it, we find
\begin{equation}
  D_\mathrm{B}'(m^2 = -2) = \lim_{m^2 \to -2} \frac{\Lambda_\mathrm{UV}^2 D_\mathrm{B}(m^2)}{\lambda_{\ast,0}}
  = \frac{e^{\frac{9}{4} - 2 \zeta'(-1)}}{3 \sqrt{2\pi}} (\Lambda_\mathrm{UV} L)^{-1/3}\+,
\end{equation}
which agrees with \eqref{eq:hemisphere_det} and concludes the consistency check of \eqref{eq:bulk_det} on the hemisphere.

\phantomsection
\addcontentsline{toc}{section}{\refname}
\bibliographystyle{myJHEP}
\bibliography{refs}

\end{document}